\documentclass[11pt]{article}
\usepackage{mathrsfs}
\usepackage{natbib,graphicx,setspace,lscape,longtable}
\usepackage{mathrsfs,amsmath,amsthm,amssymb,color}
\usepackage{natbib,epsfig,graphicx,pdfpages}
\usepackage{rotating}
\usepackage{subfigure}
\usepackage{graphicx}
\usepackage{algorithm}
\usepackage{algpseudocode}
\usepackage{hyperref}
\bibpunct{(}{)}{;}{a}{,}{,}

\newtheorem{theorem}{Theorem}
\newtheorem{lemma}{Lemma}
\newtheorem{proposition}{Proposition}
\newtheorem{remark}{Remark}
\usepackage{amsthm}

\newtheorem{definition}{Definition}
\newtheorem{assumption}{{Assumption}}

\def\beq{\begin{equation}}
\def\eeq{\end{equation}}
\def\beqr{\begin{eqnarray}}
\def\eeqr{\end{eqnarray}}
\def\beqrs{\begin{eqnarray*}}
\def\eeqrs{\end{eqnarray*}}
\def\bet{\begin{theorem}}
\def\eet{\end{theorem}}
\def\bel{\begin{lemma}}
\def\eel{\end{lemma}}
\def\bep{\begin{proposition}}
\def\eep{\end{proposition}}
\def\bg{\begin{figure}[tbph]\begin{center}}
\def\eg{\end{center}\end{figure}}

\def\bc{\begin{center}}
\def\ec{\end{center}}

\def\wt{\widetilde}
\def\wh{\widehat}

\def\diag{\mbox{diag}}

\numberwithin{equation}{section}

\newcommand{\Var}{\textnormal{Var}}
\newcommand{\Cov}{\textnormal{Cov}}
\newcommand{\Corr}{\textnormal{Corr}}
\newcommand{\Vc}{\textnormal{Vec}}

\newcommand{\bA}{{\mathbf A}}
\newcommand{\bB}{{\mathbf B}}

\newcommand{\bE}{{\mathbf E}}

\newcommand{\bH}{{\mathbf H}}
\newcommand{\bI}{{\mathbf I}}

\newcommand{\bL}{{\mathbf L}}

\newcommand{\bP}{{\mathbf P}}
\newcommand{\bR}{{\mathbf R}}
\newcommand{\bS}{{\mathbf S}}
\newcommand{\bU}{{\mathbf U}}
\newcommand{\bV}{{\mathbf V}}
\newcommand{\bW}{{\mathbf W}}
\newcommand{\bX}{{\mathbf X}}
\newcommand{\bY}{{\mathbf Y}}
\newcommand{\bZ}{{\mathbf Z}}

\newcommand{\bu}{{\mathbf u}}
\newcommand{\bv}{{\mathbf v}}
\newcommand{\bw}{{\mathbf w}}
\newcommand{\bx}{{\mathbf x}}
\newcommand{\by}{{\mathbf y}}
\newcommand{\bz}{{\mathbf z}}

\newcommand{\bfeta}  {\boldsymbol{\eta}}

\newcommand{\bOmega}{\boldsymbol{\Omega}}

\newcommand{\bSigma}{\boldsymbol{\Sigma}}

\newcommand{\bve}{\mbox{\boldmath$\varepsilon$}}

\newcommand{\bTheta} {\boldsymbol{\Theta}}
\newcommand{\bPhi} {\boldsymbol{\Phi}}

\newcommand{\bmu} {\boldsymbol{\mu}}

\newcommand{\bGamma} {\boldsymbol{\Gamma}}

\newcommand{\bC}{{\mathbf C}}
\newcommand{\bD}{{\mathbf D}}

\newcommand{\ve}{{\varepsilon}}
\renewcommand{\epsilon}{{\ve}}
\renewcommand{\hat}{\widehat}
\newcommand{\T}{{\rm T}}
\def\wt{\widetilde}
\renewcommand{\tilde}{\wt}
\def\JRSSB{{\sl Journal of the Royal Statistical Society}, {\bf B}}

\def\BKA{{\sl Biometrika}}
\def\JASA{{\sl Journal of the American Statistical Association}}

\begin{document}

\title{\bf Segmenting High-dimensional Matrix-valued  Time Series via Sequential Transformations}

\author{
Zhaoxing Gao\\ 
Department of Mathematics, Lehigh University\\
}

\date{}

\maketitle

\begin{abstract}
Modeling matrix-valued time series is an interesting and important research topic. In this paper, we extend the method of \cite{changguoyao2017} to matrix-valued time series. For any given $p\times q$ matrix-valued time series, we look for linear transformations to segment the matrix into many small sub-matrices for which each of them are uncorrelated with the others both contemporaneously and serially, thus they can be analyzed separately, which will greatly reduce the number of parameters to be estimated in terms of modeling. To overcome the identification issue, we propose a two-step and more structured procedure to segment the rows and columns separately. When $\max(p,q)$ is large in relation to the sample size $n$, we assume the transformation matrices are sparse and use threshold estimators for the (auto)covariance matrices. We also propose a block-wisely thresholding method to separate the columns (or rows) of the transformed matrix-valued data. The asymptotic properties are established for both fixed and diverging $\max(p,q)$. Unlike principal component analysis (PCA) for independent
data, we cannot guarantee that the required linear transformation
exists. When it does not, the proposed method provides an
approximate segmentation, which may be useful for forecasting. The proposed method is illustrated with both simulated and real data examples. We also propose a sequential transformation algorithm to segment higher-order tensor-valued time series.

\end{abstract}


\newpage

\section{Introduction}
In the era of big data, the volume, scale and structure of these contemporary data pose fundamentally new and exciting statistical challenges that cannot be tackled with traditional methods. Modern scientific studies often gather data under combinations of multiple factors. For example, neuroimaging experiments record brain activity at multiple spatial locations, at multiple time points, and under a variety of experimental stimuli. Studies of social networks record social links of a variety of types from multiple initiators of social activity to multiple receivers of the activity.  Data such as these are naturally represented not as lists or tables of numbers, but as multi-indexed arrays, or tensors. As many types of such data are collected over time, it is natural to view them as tensor-valued time series. 
The matrix-valued time series is a sequence of second-order random tensors. For example, financial and economic studies often collect data from different countries with a number of economic indicators (e.g. GDP growth, unemplyment rate, etc.) every quarter. Therefore, it is important and interesting to develop appropriate statistical methods to analyze such type of data. The most common approach to modeling such data is to stack the matrix into a large vector and then apply the standard multivariate methods. However, such approach will ignore the matrix structure of the data, it can lead to inefficient use of data, and important patterns in the data being overlooked. For example, \cite{werner2008} pointed out that after vectorizing the matrices the resulting vectors have a Kronecker structure. Ignoring this structure then means that a much larger number of parameters need to be estimated. Therefore, it is urgent to find an effective way to reduce the number of parameters especially when the dimension is large.

When modeling vector time series, the available methods to reduce the number of parameters  are in two categories: regularization and dimension reduction. The former imposes some conditions on the structure of a vector autoregressive and moving average (VARMA) model, and the later assumes there is a lower dimensional representation for the high-dimensional vector process. For the regularization method, some special structures are often imposed on the VARMA model. For example, Chapter 4 of \cite{Tsay_2014} discussed different canonical structures, see also the references therein. \cite{Davis2012} studied the VAR model with sparse coefficient matrices based on partial spectral coherence. The Lasso regularization has also been applied to VAR models, see \cite{ShojaieMichailidis_2010}, \cite{SongBickel_2011}, among others. \cite{GuoWangYao_2014} considered banded autoregressive models for vector time series, and estimated the coefficient matrices by a componentwise least squares method. For the dimension reduction method, popular ones include the canonical correlation analysis (CCA) of \cite{BoxTiao_1977}, the principle component analysis (PCA) of \cite{StockWatson_2002}, the scalar component analysis of \cite{TiaoTsay_1989}  and \cite{Huang_2014}. The factor model approach can be found in \cite{BaiNg_Econometrica_2002}, \cite{StockWatson_2005}, \cite{panyao2008}, \cite{LamYaoBathia_Biometrika_2011}, \cite{lamyao2012} and \cite{changguoyao2015}, among others. However, none of the methods mentioned above can be directly used to model matrix-valued time series if we do not vectorize it.

When the data are independent and identically distributed (i.i.d.), \cite{xueandyin2014} introduced dimension folding sufficient reduction for conditional mean functioins, \cite{li2016} proposed a dimension folding method for data with matrix-valued predictors, \cite{huagwang2012}, \cite{zhou2013} and \cite{zhouli2014} extended the generalized linear models to matrix- and tensor-valued predictors for analyzing image data. \cite{dingcook2017} studied the matrix variate regression with matrix-valued response.  An incomplete list of publications also include \cite{gupta2000}, \cite{lengtang2012}, \cite{yinli2012}, \cite{zhaoleng2014} and \cite{zhou2014}. With temporal dependence, the matrix-valued time series has not been well studied in the literature, Walden and Serroukh (2002) handled this kind of data in signal and image processing, \cite{wangetal2017} proposed a factor model for matrix-valued time series which maintains and utilize the matrix structure to achieve the dimension reduction.

In this paper, we extend the PCA approach of \cite{changguoyao2017} to matrix-valued time series  without stacking the matrix into a vector and the structure can be preserved. Our goal is as follows: let $\bY_t=(y_{ij}^t)$ be a $p\times q$ matrix valued time series, i.e. there are $pq$ recorded values at each time, for example, $p$ individuals and over $q$ indices or variables. We assume  $\bY_t$ can be represented as
\begin{equation}\label{eq:1}
  \bY_t=\bB\bW_t\bA^\T,
\end{equation}
where $\bB\in R^{p\times p}$, $\bA\in R^{q\times q}$ and $\bW_t$ is a latent $p\times q$ matrix in which the rows are divided into $p_1(\leq p)$ groups and there are no correlations across different groups at all time lags, and the columns are divided into $q_1(\leq q)$ groups and there are no correlations across different groups at all time lags either. With such a decomposition, we only need to model the small sub-matrices in $\bW$ separately and we can achieve substantial dimension reduction. As $\bB$, $\bW_t$, and $\bA$ are all latent ones and the identification is a big issue. For example, even when $\bW_t$ is observable, $(\bA,\bB)$ can be replaced by $(\bA/c,c\bB)$ for any nonzero constant $c$ without changing the relationship of (\ref{eq:1}). 

Instead of estimating them simultaneously, we propose in this paper a two-step and more structured approach: first we seek a column transformation, i.e. we transform linearly the columns of $\bY_t$ into $q$ new variables, and ideally, those $q$ new variables form $q_1$ uncorrelated groups with $q_1\leq q$. The second step applies the same segmentation method to the $p$ rows of the obtained ones in the first step and the transformation of the rows will not alter the uncorrelatedness of the column groups in the first step. In the end, this new matrix can be divided into several smaller submatrices, and those submatrices are uncorrelated with each other both contemporaneously and serially. Our method is a building block for modeling tensor-valued time series and it turns out that all tensor-valued time series can be rearranged as a matrix time series by matricization, see \cite{kolda2009}. Therefore, the proposed method can be applied sequentially  to all types of tensor-valued time series without losing the information of the structures. Unlike principal component analysis (PCA) for independent
data, we cannot guarantee that the required linear transformation
exists. When it does not, the proposed method provides an
approximate segmentation, which may be useful for forecasting. Simulation studies are carried out to assess the performance of our procedure and the proposed method is further applied to real data examples.

The rest of the paper is organized as follows. We specify the methodology in Section 2. The asymptotic properties are presented
in Section 3. A feasible approach to segmenting tensor-valued time series is given in Section 4. The numerical illustrations with both simulated and real data sets are
reported in Section 5.  All technical proofs are relegated into an Appendix. We always use the following notation, for a $p\times 1$ vector
$\bu=(u_1,..., u_p)^\T,$  $||\bu||_2 = (\sum_{i=1}^{p} u_i^2)^{1/2} $
is the Euclidean norm. $\bI_p$ denotes a $p\times p$ identity matrix. For a matrix $\bH=(h_{ij})$, $|\bH|_\infty=\max_{i,j}|h_{ij}|$,  $\|\bH
\|_2=\sqrt{\lambda_{\max} (\bH^\T \bH ) }$ is the operator norm, where
$\lambda_{\max} (\cdot) $ denotes for the largest eigenvalue of a matrix, and $\|\bH\|_F=\sqrt{\textrm{tr}(\bH\bH^\T)}$.

\section{Methodology}
\subsection{Setting and method}
Let $\bY_t=(\by_1^t,...,\by_q^t)$ be an observable $p\times q$ matrix-valued time series with $\by_i^t\in R^p$. We assume $\bY_t$ admits a latent segmentation structure:
\begin{equation}\label{seg}
\bY_t=\bX_t\bA^\T,
\end{equation}
where $\bX_t$ is an unobservable $p\times q$ matrix valued time series in which the $q$ columns can be classified into $q_1$($>1$) groups and any two groups are contemporaneously and serially uncorrelated, and $\bA\in R^{q\times q}$ is an unknown constant matrix. Before we proceed further, we give the definitions of  row- and column-covariance  matrix between two random matrices.
\begin{definition}
Let $\bU_t\in R^{s_1\times r_1}$ and $\bV_t\in R^{s_2\times r_2}$. If $r_1=r_2=r$, the  covariance matrix over the columns between $\bU_t$ and $\bV_t$ is defined as
\begin{equation}\label{covc}
\Cov_c(\bU_t,\bV_t):=\frac{1}{r}E(\bU_t-E\bU_t)(\bV_t-E\bV_t)^\T,
\end{equation}
and if $s_1=s_2=s$, the covariance matrix over the rows is defined as
\begin{equation}\label{covr}
\Cov_r(\bU_t,\bV_t):=\frac{1}{s}E(\bU_t-E\bU_t)^\T(\bV_t-E\bV_t)=\Cov_c(\bU_t^\T,\bV_t^\T).
\end{equation}
$\Var_c(\bU_t)$ and $\Var_r(\bU_t)$ can be defined in a similar way.  In particular, when $r=1$ or $s=1$, (\ref{covc}) or (\ref{covr}) reduces to the traditional case for two random vectors.
\end{definition}
In model (\ref{seg}), we assume $\bY_t$ and $\bX_t$ are both  weakly stationary in the sense that the means and the autocovariances do not vary with respect to time for any fixed $(p,q)$. The stationarity of $\bY_t$ can be inherited from $\bX_t$ through (\ref{seg}), and a sufficient condition for this is to assume $\Vc(\bX_t)$ and $\Vc(\bY_t)$ are stationary, where $\Vc(\cdot)$ is the vectorization of a matrix. Denote the segmentation of $\bX_t$ by
\begin{equation}\label{seg:x}
\bX_t=(\bx_1^t,...,\bx_q^t)=(\bX_t^{(1)},...,\bX_t^{(q_1)})
\end{equation}
with $\Cov_r(\bX_t^{(i)},\bX_s^{(j)})=\bf{0}$ for all $t,s$ and $i\neq j$. Therefore, all the autocovariances of $\bX_t^\T$ are of the same block-diagonal structure with $q_1$ blocks and $\bX_t^{(1)},...,\bX_t^{(q_1)}$ can be modelled or forecasted separately as far as their linear dynamic structure is concerned. See Remark 4 in Section 4 for details.

Now we spell out how to find the segmentation transformation under (\ref{seg}) and (\ref{seg:x}). Without loss of generality, we assume
\begin{equation}\label{std}
  \Var_r(\bY_t)=\bI_q\quad\text{and}\quad \Var_r(\bX_t)=\bI_q.
\end{equation}
The first equation in (\ref{std}) is implied by replacing $\bY_t$ by $\bY_t\hat{\bS}_{y,0}^{-1/2}$, where $\hat{\bS}_{y,0}$ is a consistent estimator of $\Var_r(\bY_t)$. The second equation is to conceptually replace $\bX_t$ by $\bX_t\wh\bS_{x,0}^{-1/2}$ where $\wh\bS_{x,0}^{-1/2}$ is a consistent estimator for $\Var_r(\bX_t)$, and it will not alter the fact that there are no correlations across different groups. As both $\bA$ and $\bX_t$ are unobservable, (\ref{std}) implies that we can view $\wh\bS_{y,0}^{-1/2}\bA\wh\bS_{x,0}^{1/2}$ as $\bA$.
As a consequence of (\ref{std}), the transformation matrix $\bA$ in (\ref{seg}) is orthogonal. Let $l_j$ be the number of columns of $\bX_t^{(j)}$ with $l_1+\cdots+l_{q_1}=q$. Write $\bA=(\bA_1,...,\bA_{q_1})$, where $\bA_j\in R^{q\times l_j}$. It follows from (\ref{seg}) and (\ref{seg:x}) that

\begin{equation}\label{pt}
\bX_t^{(j)}=\bY_t\bA_{j},\quad j=1,...,q_1.
\end{equation}

However, similar to that in \cite{changguoyao2017}, $\bA$ and $\bX_t$ are not uniquely identified in (\ref{seg}), even with additional assumption in (\ref{std}). For example, let $\bH_j$ be any $l_j\times l_j$ orthogonal matrix, and $\bH=\diag(\bH_1,...,\bH_{q_1})$. Then $(\bA,\bX_t)$ in (\ref{seg}) can be replaced  by $(\bA\bH,\bX_t\bH)$ while (\ref{seg:x}) still holds.   In fact, only $\mathcal{M}(\bA_1),...,\mathcal{M}(\bA_{q_1})$ are uniquely defined by (\ref{seg}), where $\mathcal{M}(\bA_j)$ denotes the linear space spanned by the columns of $\bA_j$. As a result, $\bY_t\bGamma_j$ can be taken as $\bX_t^{(j)}$ for any $q\times l_j$ matrix $\bGamma_j$ as long as $\bGamma_j^\T\bGamma_j=\bI_{l_j}$ and $\mathcal{M}(\bGamma_j)=\mathcal{M}(\bA_j)$. Thus, to estimate $\bA=(\bA_1,...,\bA_{q_1})$, it is sufficient to estimate the linear spaces $\mathcal{M}(\bA_1),...,\mathcal{M}(\bA_{q_1})$.

To discover the latent segmentation, we introduce some notation first. We denote $\by_{i:}^t$ and $\bx_{i:}^t$, respectively,  the row vectors of $\bY_t$ and $\bX_t$. For any integer $k$, let $\bSigma_y(k)=\Cov_r(\bY_{t+k},\bY_{t})$, $\bSigma_x(k)=\Cov_r(\bX_{t+k},\bX_{t})$, $\bSigma_{y,i,j}(k)=\Cov(\by_{i:}^{t+k},\by_{j:}^{t})$ and $\bSigma_{x,i,j}(k)=\Cov(\bx_{i:}^{t+k},\bx_{j:}^t)$. By (\ref{std}), we have $\bSigma_y(0)=\bSigma_x(0)=\bI_q$. For a pre-specified integer $k_0$, define
\begin{equation}\label{wy}
\bW_y=\sum_{k=0}^{k_0}\bSigma_y(k)\bSigma_y(k)^\T=\bI_q+\sum_{k=1}^{k_0}\bSigma_y(k)\bSigma_y(k)^\T
\end{equation}
and
\begin{equation}\label{wx}
\bW_x=\sum_{k=0}^{k_0}\bSigma_x(k)\bSigma_x(k)^\T=\bI_q+\sum_{k=1}^{k_0}\bSigma_x(k)\bSigma_x(k)^\T.
\end{equation}
It follows from (\ref{seg}) and (\ref{seg:x}) that both $\bSigma_x(k)$ and $\bW_x$ are block-diagonal, and
\begin{equation}\label{wyx}
\bW_y=\bA\bW_x\bA^\T.
\end{equation}
\begin{remark}
There are other ways to obtain the relationship (\ref{wyx}). For example,
we may also define
\begin{equation}\label{wyp}
\bW_y^{(1)}=\frac{1}{p^2}\sum_{k=0}^{k_0}\sum_{i=1}^p\sum_{j=1}^p\bSigma_{y,i,j}(k)\bSigma_{y,i,j}(k)^\T,
\end{equation}
and
\begin{equation}\label{wxp}
\bW_x^{(1)}=\frac{1}{p^2}\sum_{k=0}^{k_0}\sum_{i=1}^p\sum_{j=1}^p\bSigma_{x,i,j}(k)\bSigma_{x,i,j}(k)^\T,
\end{equation}
then (\ref{wyx}) still holds as
\begin{equation}\label{wyxp}
\bW_{y}^{(1)}=\bA \bW_{x}^{(1)}\bA^\T. 
\end{equation}
Under some regularity conditions, by Lemma 5, the asymptotic properties can also be established with higher estimation errors since there are more covariance matrices to be estimated in (\ref{wyp}). For simplicity, we only deal with (\ref{wyx}) as (\ref{wyxp}) can be analyzed in a similar way.
\end{remark}
Note that both $\bW_y$ and $\bW_x$ are positive definite matrices, therefore we have the following decomposition
\begin{equation}\label{wxdc}
\bW_x\bGamma_x=\bGamma_x\bD,
\end{equation}
where $\bGamma_x$ is a $q\times q$ orthogonal matrix with the columns being the orthonormal eigenvectors of $\bW_x$, and $\bD$ is a diagonal matrix with the corresponding eigenvalues as the elements on the main diagonal. By (\ref{wyx}) and (\ref{wxdc}), $\bW_y\bA\bGamma_x=\bA\bGamma_x\bD$ and hence the columns of $\bGamma_y:=\bA\bGamma_x$ are the orthonormal eigenvectors of $\bW_y$. Consequently,
\begin{equation}\label{gam:yx}
\bY_t\bGamma_y=\bY_t\bA\bGamma_x=\bX_t\bGamma_x,
\end{equation}
where the last equality follows from (\ref{seg}). Let
\begin{equation}\label{wxdiag}
\bW_x=\diag(\bW_{x,1},...,\bW_{x,q_{1}}),
\end{equation}
where $\bW_{x,j}$ is an $l_j\times l_j$ positive definite matrix, and the eigenvalues of $\bW_{x,j}$ are also the eigenvalues of $\bW_x$. Suppose that $\bW_{x,i}$ and $\bW_{x,j}$ do not share the same eigenvalues for any $i\neq j$. Then if we line up the eigenvalues of $\bW_x$ (i.e. the eigenvalues of $\bW_{x,1},...,\bW_{x,q_1}$ combining together) in the main diagonal of $\bD$ according to the order of the blocks in $\bW_x$, $\bGamma_x$ must be a block-diagonal orthogonal matrix of the same shape as $\bW_x$; see Proposition 1(i). However the order of the eigenvalues is latent, and any $\bGamma_x$ defined by (\ref{wxdc}) is nevertheless a column-permutation (i.e. a matrix consisting of the same column vectors but arranged in a different order) of such a block-diagonal orthogonal matrix; see Proposition 1(ii). By Proposition 1(i), write $\bGamma_x=\diag(\bGamma_{x,1},...,\bGamma_{x,q_1})$, it follows from (\ref{seg:x}) and (\ref{gam:yx}) that
\begin{equation}\label{gam:yxpt}
  \bY_t\bGamma_y=\bX_t\bGamma_x=(\bX_t^{(1)}\bGamma_{x,1},...,\bX_t^{(q_1)}\bGamma_{x,q_1}),
\end{equation}
hence $\bX_t\bGamma_x$ does not alter the fact that there  are no correlations between different groups in $\bX_t$ and $\bGamma_y$ can be regarded as $\bA$ as long as the eigenvalues of $\bW_x$ are ordered appropriately. As they are all latent, $\bY_t\bGamma_y$ can be taken as a permutation of $\bX_t$, and $\bGamma_y$ can be viewed as a column-permutation of $\bA$; see the discussion below (\ref{pt}). This leads to the following three-step estimation for $\bA$ and $\bX_t$:
\begin{description}
  \item[Step 1.] \textit{Let $\hat{\bS}_{y,0}$ be a consistent estimator for $\Var_r(\bY_t)$. Replace $\bY_t$ by $\bY_t\hat{\bS}_{y,0}^{-1/2}$ .}
  \item[Step 2.] \textit{Let $\hat{\bS}$ be a consistent estimator for $\bW_y$. Calculate a $q\times q$ orthogonal matrix $\hat{\bGamma}_y$ with columns being the orthonormal eigenvectors of $\hat{\bS}$.}
  \item[Step 3.] \textit{The columns of $\hat{\bA}=(\hat{\bA}_1,...,\hat{\bA}_{q_1})$ are a permutation of the columns of $\hat{\bGamma}_y$ such that $\hat{\bX}_t=\bY_t\hat{\bA}$ is segmented into $q_1$ uncorrelated sub-matrix series $\hat{\bX}_t^{(j)}=\bY_t\hat{\bA}_j$, $j = 1,...,q_1$.}
\end{description}
In Steps  1 and 2, the estimators $\hat{\bS}_{y,0}$ and $\hat{\bS}$ should be consistent, and will be constructed under various scenarios in Section 3 below. The permutation in Step 3 can be carried out by grouping the columns of $\hat{\bZ}_t:=\bY_t\hat{\bGamma}_y$.

We now state a proposition that demonstrate the assertion after (\ref{wxdiag}), the proof is similar to Proposition 1 in \cite{changguoyao2017} and we therefore omit it.
\begin{proposition}
 (i) The orthogonal matrix $\bGamma_x$ in (\ref{wxdc}) can be taken as a block-diagonal orthogonal matrix with the same block structure as $\bW_x$. (ii) An orthogonal matrix $\bGamma_x$ satisfied (\ref{wxdc}) if and only if its columns are a permutation of the columns of a block-diagonal orthogonal matrix described in (i), provided that any two different blocks $\bW_{x,i}$ and $\bW_{x,j}$ do not share the same eigenvalues.
\end{proposition}

From Proposition 1, we can see that the proposed method will not be able to separate $\bX_t^{(i)}$ and $\bX_t^{(j)}$ if $\bW_{x,i}$ and $\bW_{x,j}$ share one or more common eigenvalues. But it does not rule out the possibility that each block $\bW_{x,j}$ may have multiple eigenvalues.
\subsection{Permutation}
\subsubsection{Permutation rule.}
According to the discussion in Section 2.1, $\hat{\bA}$ is a permutation of the columns of $\hat{\bGamma}_y$ and the permutation can be carried out by grouping the columns of $\hat{\bZ}_t:=\bY_t\hat{\bGamma}_y$ into $q_1$ groups, where $q_1$ and the number of columns $l_j$ ($1\leq j\leq q_1$) are unknown. Let $\hat{\bZ}_t=(\hat{\bz}_1^t,...,\hat{\bz}_{q}^t)$, $\bZ_t=\bY_t\Gamma_y=(\bz_1^t,...,\bz_q^t)$, and $\bGamma_{i,j}(h)$ denote the covariance matrix between two series $\hat{\bz}_i^t$ and $\hat{\bz}_j^t$ at lag $h$, i.e. $\bGamma_{i,j}(h)=\Corr(\hat{\bz}_i^{t+h},\hat{\bz}_j^t)$. We say that $\hat{\bz}_i^t$ and $\hat{\bz}_j^t$ are connected if the multiple null hypothesis
\begin{equation}\label{hypo}
  H_0:\bGamma_{i,j}(h)=\mathbf{0}\quad \text{for any}\quad h=0,\pm 1,\pm 2,...,\pm m
\end{equation}
is rejected, where $m\geq 1$ is a prescribed integer. We should mention that the true $\bGamma_{i,j}(h)$ is not known since $\hat{\bz}_i^t$ is also one estimator for $\bz_i^t$, but it will be asymptotically equivalent to $\Corr(\bz_i^{t+h},\bz_j^t)$ as long as $\hat{\bGamma}_y$ is consistent to $\bGamma_y$. Given the structure of $\bW_x$, this can be done under some regularity conditions and therefore we also denote the true $\bGamma_{i,j}(h)=\Corr(\bz_i^{t+h},\bz_j^{t})$, and the estimator $\wh \bGamma_{i,j}(h)=\wh\Corr(\wh\bz_i^{t+h},\wh\bz_j^{t})$ which  will be specified in Section 3. See the proofs of Theorems 1-3 in Section 3.  The permutation in Step 3 in Section 2.1 can be performed as follows.
\begin{center}
\begin{enumerate}
\item[i.] \textit{Start with the $q$ groups with each group containing  one column of $\hat{\bZ}_t$ only.}
  \item[ii.] \textit{Combine two groups together if one connected pair are found.}
  \item[iii.] \textit{Repeat Step ii above until all connected pairs are within one group.}
\end{enumerate}
\end{center}
We introduce below one way to identify the connected pairs of the transformed matrix $\hat{\bZ}_t$.
\subsubsection{Maximum cross correlation method.}
Similar to Chang et al. (2017), one natural way to test hypothesis $H_0$ in (\ref{hypo}) is to use the maximum cross correlation over all elements of $\bGamma_{i,j}(h)$ and all the lags between $-m$ to $m$:
\begin{equation}\label{maxcor}
  \hat{L}_n(i,j)=\max_{|h|\leq m}|\hat{\bGamma}_{i,j}(h)|_{\infty},
\end{equation}
where $\hat{\bGamma}_{i,j}(h)$ is a sample correlation matrix between $\hat{\bz}_i^t$ and $\hat{\bz}_j^t$ at lag $h$ when the dimension $p$ and $q$ are fixed, and it is a block-wisely thresholded sample correlation matrix when $p$ and $q$ are moderately high, and it will be constructed under different scenarios in Section 3. We would reject $H_0$ for the pair $(\hat{\bz}_i^t,\hat{\bz}_j^t)$ if $\hat{L}_n(i,j)$ is greater than an appropriate threshold value.

For the $q_0=q(q-1)/2$ pairs of $\hat{\bZ}_t$, we propose a ratio based method to those pairs for which $H_0$ will be rejected. We re-arrange the $q_0$ pairs obtained $\hat{L}_n(i,j)$'s in descending order: $\hat{L}_1\geq\cdots\geq \hat{L}_{q_0}$. Define
\begin{equation}\label{rhat}
  \hat{d}=\arg\max_{1\leq j<c_0q_0}\hat{L}_j/\hat{L}_{j+1},
\end{equation}
where $c_0\in(0,1)$ is a prescribed constant. Similar ideas can be found in \cite{changguoyao2017} and \cite{lamyao2012}.

To state the asymptotic property of the above approach, similar to Chang et al. (2017), we use a graph representation. define
\begin{equation}\label{graph}
  E=\left\{(i,j):\max_{|h|\leq m}|\bGamma_{i,j}(h)|_{\infty}>0\right\}
\end{equation}
Each $(i,j)$ can be viewed as an edge. For the presentation of the theoretical results and to avoid the case of $``0/0"$, we modify (\ref{rhat}) as
\begin{equation}\label{rhat:md}
\hat{d}=\arg\max_{1\leq j<q_0}(\hat{L}_j+C\delta_n)/(\hat{L}_{j+1}+C\delta_n),
\end{equation}
where $C>0$ and $\delta_n\rightarrow 0$ as $n\rightarrow\infty$. We will specify $\delta_n$ in Section 3 below under different scenarios. To make $E$ in (\ref{graph}) be identified, we further assume
\begin{equation*}
  \min_{(i,j)\in E}\max_{|h|\leq m}|\bGamma_{i,j}(h)|_\infty\geq \epsilon_n
\end{equation*}
for some $\epsilon_n>0$ and $n\epsilon_n^2\rightarrow\infty$. The consistency of our permutation method is stated in Section 3 under different settings of dimensionality.

\section{ Theoretical properties}
In this section, we will show that, under some regularity conditions, there exists a matrix $\bA$ that transforms $\bY_t$ into several smaller submatrices, and the estimator $\hat{\bA}=(\hat{\bA}_1,...,\hat{\bA}_{q_1})$, is an adequate estimator for $\bA$ in (\ref{seg}) in the sense that $\mathcal{M}(\hat{\bA}_j)$ is consistent to $\mathcal{M}(\bA_j)$ for each $j=1,...,q_1$.  From Section 2, the estimators $\wh\bS_{y,0}$ for $\Var_r(\bY_t)$ in (\ref{std}), $\hat{\bS}$ for $\bW_y$ and $\hat{\bGamma}_{i,j}(h)$ in (\ref{maxcor}) for $\bGamma_{i,j}(h)$ are important and our goal is to show that $\hat{\bGamma}_y$ is a valid estimator for $\bA$ up to a column permutation. Similar to Chang et al. (2017), we establish the consistency under three different asymptotic modes: (i) the dimensions $\max(p,q)$ is fixed, (ii) $\max(p,q)=o(n^c)$, and (iii) $\log \{\max(p,q)\}=o(n^c)$, as the sample size $n\rightarrow\infty$, where $c>0$ is a small constant.

As the choice of $\bA$ in model (\ref{seg}) is not unique, we consider the error in estimating $\mathcal{M}(\bA_j)$ instead of a particular $\bA_j$. To this end, we first extend the discrepancy measure
used by \cite{panyao2008} to a more general form below. Let $\bH_i$ be a
$p\times r_i$ matrix with rank$(\bH_i) = r_i$, and $\bP_i =
\bH_i(\bH_i'\bH_i)^{-1} \bH_i'$, $i=1,2$. Define
\begin{equation}\label{dmeasure}
D(\mathcal{M}(\bH_1),\mathcal{M}(\bH_2))=\sqrt{1-
\frac{1}{\min{(r_1,r_2)}}\textrm{tr}(\bP_1\bP_2)}.
\end{equation}

Then $D \in [0,1]$. Furthermore,
$D(\mathcal{M}(\bH_1),\mathcal{M}(\bH_2))=0$ if and only if
either $\mathcal{M}(\bH_1)\subset \mathcal{M}(\bH_2)$ or
$\mathcal{M}(\bH_2)\subset \mathcal{M}(\bH_1)$, and 1 if and only if
$\mathcal{M}(\bH_1) \perp \mathcal{M}(\bH_2)$.
When $r_1 = r_2=r$ and $\bH_i'\bH_i= \bI_r$,
$D(\mathcal{M}(\bH_1),\mathcal{M}(\bH_2))
$ is the same as that in \cite{panyao2008}.

We always assume that the weakly stationary process $\Vc(\bY_t)$ is $\alpha$-mixing in the sense that the mixing coefficients $\alpha_{p,q}(k)\rightarrow 0$ as $k\rightarrow\infty$, where
\begin{equation}\label{amix}
  \alpha_{p,q}(k)=\sup_{i}\sup_{A\in\mathcal{F}_{-\infty}^i,B\in\mathcal{F}_{i+k}^\infty}|P(A\cap B)-P(A)P(B)|,
\end{equation}
and $\mathcal{F}_i^j$ is the $\sigma$-field generated by $\{\Vc(\bY_t):i\leq t\leq j\}$. In the sequel, we denote by $\sigma_{i,j}^{(k)}$ the $(i,j)$-th element of $\bSigma_y(k)$ and $\gamma_{ij,kl}^{(h)}$ the $(k,l)$-th element of $\bSigma_{y,i,j}{(h)}$ . We also define $\bmu_i=E(\by_i^t)$ and $\bmu=E\bY_t=(\bmu_1,...,\bmu_q)$, and
let
\begin{equation}\label{sigyhat}
  \hat{\bSigma}_y(k)=\frac{1}{np}\sum_{t=1}^{n-k}(\bY_{t+h}-\bar{\bY})^\T(\bY_t-\bar{\bY}),\quad \text{where}\quad \bar{\bY}=\frac{1}{n}\sum_{t=1}^n\bY_t,
\end{equation}
\begin{equation}\label{siggamma}
  \hat{\bSigma}_{y,i,j}{(h)}=\frac{1}{n}\sum_{t=1}^{n-h}(\by_{i:}^{t+h}-\bar{\by}_{i:})^\T(\by_{j:}^t-\bar{\by}_{j:}),\quad \text{where}\quad \bar{\by}_{i:}=\frac{1}{n}\sum_{t=1}^n\by_{i:}^t,
\end{equation}
and other sample estimators can be defined in a similar way.

To show the consistency of the maximum cross correlation method, we introduce some additional notation here.
Let
\begin{equation}\label{eigen:df}
  \varpi_n=\min_{1\leq i<j\leq q_1}\min_{\lambda\in\sigma(\bW_{x,i}),\mu\in\sigma(\bW_{x,j})}|\lambda-\mu|,
\end{equation}
where $\bW_{x,i}$ is defined in (\ref{wxdiag}) and $\sigma(\bW_{x,i})$ denotes the set containing all the eigenvalues of $\bW_{x,i}$.  
The true maximum cross correlations of $\bZ_t=\bY_t\bGamma_y$ in the descending order are denoted by $L_1\geq\cdots\geq L_{q_0}$.  Define
$$\chi_n=\max_{1\leq j<d-1}L_j/L_{j+1},$$
where $d=|E|$ is the true number of connected pairs in the graph $E$.

\subsection{Asymptotic properties when $n\rightarrow\infty$ and $p,q$ are fixed}

When the dimension is fixed, the sample estimators for $\bW_y$ and ${\bGamma}_{i,j}(h)$ are defined as
\begin{equation}\label{shat:fixed}
  \hat{\bS}=\bI_q+\sum_{k=1}^{k_0}\hat{\bSigma}_y(k)\hat{\bSigma}_y(k)^\T\quad\text{and}\quad \hat{\bGamma}_{i,j}(h)=(\hat{\delta}_{i,k}^{-1}\hat{\delta}_{j,l}^{-1}\hat{\delta}_{ij,kl}^{(h)}),
\end{equation}
respectively, where  $\hat{\bSigma}_y(k)$ is defined in (\ref{sigyhat}), $\hat{\delta}_{ij,kl}^{(h)}=\hat{\bv}_i^\T \hat{\bSigma}_{y,k,l}{(h)}\hat{\bv}_j$, $\hat{\delta}_{i,k}^2=\hat{\delta}_{ii,kk}^{(0)}$,
 $\hat{\bv}_i$ is the $i$-th column of $\hat{\bGamma}_y$ and $\hat{\bSigma}_{y,k,l}{(h)}$ is the sample estimator for ${\bSigma}_{y,k,l}{(h)}$ defined in (\ref{siggamma}). In addition, $\hat{\bS}_{y,0}=\wh\bSigma_y(0)$.

We introduce some assumptions first.
\begin{assumption}\label{a1}
It holds that $\sup_t\max_{1\leq i\leq p,1\leq j\leq q}E|y_{ij}^t-\mu_{ij}|^{2\gamma}\leq K_1$ for some constants $\gamma>2$ and $K_1>0$.
\end{assumption}

\begin{assumption}\label{a2}
The mixing coefficients $\alpha_{p,q}(k)$ defined in (\ref{amix}), satisfy the condition that $\sum_{k=1}^\infty \alpha_{p,q}^{1-2/\gamma}<\infty$ for some $\gamma$ defined in Assumption 1.
\end{assumption}

\begin{theorem}\label{thm1}
(i) Under Assumptions \ref{a1} and \ref{a2}, if $\chi_n/\epsilon_n=o_p(n^{1/2})$, $\varpi_n$ is positive and the singular values of $\bSigma_{y,i,j}{(h)}$ are uniformly bounded away from $\infty$ for all $|h|\leq m$. Then for $\hat{d}$ defined in (\ref{rhat:md}), we have $P(\hat{E}=E)\rightarrow 1$.

\noindent(ii)  Under Assumptions \ref{a1} and \ref{a2},  $p$ and  $q$ are fixed and $\varpi_n$ is positive. Then
  $$\max_{1\leq j\leq q_1}D(\mathcal{M}(\hat{\bA}_j),\mathcal{M}(\bA_j))=O_p(n^{-1/2}),$$
  where $\hat{\bA}=(\hat{\bA}_1,...,\hat{\bA}_{q_1})$ is a permutation of the $q$ orthogonal eigenvectors of $\hat{\bS}$ defined in (\ref{shat:fixed}).
\end{theorem}
\subsection{Asymptotic properties when $n\rightarrow\infty$ and $\max(p,q)=o(n^c)$}
In order to deal with large $p$, we impose a sparsity condition on the transformation matrix $\bA$ first. Similar assumptions can be found in \cite{changguoyao2017}.

\begin{assumption}\label{a3}
   For $\bA=(a_{i,j})$ in (\ref{seg}), we assume that $\max_{1\leq j\leq q}\sum_{i=1}^q|a_{i,j}|^\iota\leq s_1$ and $\max_{1\leq j\leq q}\sum_{j=1}^q|a_{i,j}|^\iota\leq s_2$, for some constant $\iota\in[0,1)$, where $s_1$ and $s_2$ may diverge together with $q$.
\end{assumption}

It is well known that if $p$ and $q$ diverge faster than $n^{1/2}$, the sample autocoariance matrices $\hat{\bSigma}_y(k)=(\hat{\sigma}_{i,j}^{(k)})$ and $\hat{\bSigma}_{y,i,j}{(h)}=(\hat{\gamma}_{ij,kl}^{(h)})$ are not consistent estimators for $\bSigma_y(k)$ and $\bSigma_{y,i,j}{(h)}$, respectively. Under the sparsity assumption, here we adopt the thresholded estimator  for large covariance matrix by \cite{bickel2008},
\begin{equation}\label{threshold}
  T_u(\hat{\bSigma}_y(k))=\{\hat{\sigma}_{i,j}^{(k)}I(|\hat{\sigma}_{i,j}^{(k)}|\geq u)\}\quad\text{and}\quad T_v(\hat{\bSigma}_{y,i,j}{(h)})=\{\hat{\gamma}_{ij,kl}^{(h)}I(|\hat{\gamma}_{ij,kl}^{(h)}|\geq v\},
\end{equation}
where $I(\cdot)$ is the indicator function, $u$ and $v$ are, respectively, the threshold level for $\hat{\bSigma}_y(k)$ and $\hat{\bSigma}_{y,i,j}{(h)}$. By Lemma \ref{lm3}, under the setting of $pq=o(n^{(\beta-1)/2})$, we can show that $\max_{1\leq i,j\leq q}|\hat{\sigma}_{i,j}^{(k)}-{\sigma}_{i,j}^{(k)}|=O_p(\vartheta_n)$ and $\max_{1\leq i,j\leq p,1\leq k,l\leq q}|\hat{\gamma}_{ij,kl}^{(h)}-{\gamma}_{ij,kl}^{(h)}|=O_p(\theta_n)$, where
\begin{equation}\label{th:u}
  \vartheta_n=q^{2/\beta}n^{-(\beta-1)/\beta}\quad\text{and}\quad \theta_n=(pq)^{2/\beta}n^{-(\beta-1)/\beta},
\end{equation}
and $\vartheta_n,\theta_n\rightarrow0$ as $n\rightarrow\infty$.
Hence we set the threshold level at $u=M\vartheta_n$ and $v=M\theta_n$,
where $M>0$ is a constant and $\beta$ is defined in Lemma \ref{lm5}. By an abuse of notation and (\ref{threshold}), we sometimes also write $\hat{\bSigma}_{y,k,l}{(h)}=(\hat{\gamma}_{kl,st}^{(h)})_{1\leq s,t\leq q}$, and hence
\begin{equation}\label{th:est}
  \hat{\bS}=\bI_q+\sum_{k=1}^{k_0}T_u(\hat{\bSigma}_y(k))T_u(\hat{\bSigma}_y(k))^\T
\end{equation}
and
\begin{equation}\label{th:block}
 \hat{\bGamma}_{i,j}(h)=\{[\hat{\bv}_i^\T T_v(\hat{\bSigma}_{y,k,k}{(h)})\hat{\bv}_i\hat{\bv}_j^\T T_v(\hat{\bSigma}_{y,l,l}{(h)})\hat{\bv}_j]^{-1/2}\hat{\bv}_i^\T T_v(\hat{\bSigma}_{y,k,l}{(h)})\hat{\bv}_j)\},
\end{equation}
where $\hat{\bv}_i$ is the $i$-th column of $\hat{\bGamma}_y$ and $\hat{\bGamma}_{i,j}(h)$ is a block-wisely thresholded estimator for $\bGamma_{i,j}$ with $T_v(\hat{\bSigma}_{y,k,l}{(h)})$ being a thresholded estimator for $\bSigma_{y,i,j}{(h)}$. In particular, $\hat{\bS}_{y,0}=T_u(\wh\bSigma_y(0))$.

\begin{assumption}\label{a4}
The mixing coefficients $\alpha_{p,q}(k)$ given in (\ref{amix}) satisfy $\sup_{p,q}\alpha_{p,q}(k)=O\{k^{-a}\}$ as $k\rightarrow\infty$ for some constant $a>\gamma/(\gamma-2)$, where $\gamma$ is given in Assumption \ref{a1}.
\end{assumption}

\begin{assumption}\label{a5}
  $\max_{1\leq k\leq k_0}|\bSigma_x(k)|_\infty<\infty$ and $\max_{1\leq k\leq k_0}\max_{1\leq i,j\leq p}|\bSigma_{x,i,j}(k)|_\infty<\infty$, where $\bSigma_{x,i,j}(k)=\Cov(\bx_{i:}^{t+k},\bx_{j:}^t)$.
\end{assumption}
Assumptions 1 and 4 together ensure the Fuk-Nagaev type inequalities for $\alpha$-mixing processes with power-type rates. Assumption 5 is used to establish Lemma 4. Write
\begin{equation}\label{min:eigen}
\rho_j=\min_{i\neq j}\min_{\lambda\in\sigma(\bW_{x,i}),\mu\in\sigma(\bW_{x,j})}|\lambda-\mu|,\quad j=1,...,q,
\end{equation}
\begin{equation}\label{max:group}
  \delta=s_1s_2\max_{1\leq j\leq q_1}l_j,\quad \kappa=\max_{1\leq k\leq k_0}\|\bSigma_x(k)\|_2, \quad \nu_n=\max_{|h|\leq m}\max_{1\leq i,j\leq p}\|\bSigma_{i,j}^{(h)}\|_2,
\end{equation}
\begin{equation}\label{d1d2}
  d_{1n}= \nu_n\varpi_n^{-1}[\kappa\vartheta_n^{1-\iota}\delta+\vartheta_n^{2(1-\iota)}\delta^2] \quad\text{and}\quad
  d_{2n}=\theta_n^{1-\iota}\delta.
\end{equation}
\begin{theorem}(i). Under Assumptions \ref{a1}, \ref{a3}-\ref{a5}, $pq=o(n^{(\beta-1)/2})$ and $\chi_n/\epsilon_n=o_p(\delta_n^{-1})$ where $\delta_n=d_{1n}+d_{2n}$. Then for $\hat{d}$ defined in (\ref{rhat:md}), we have $P(\hat{E}=E)\rightarrow 1$ with the threshold level $v\asymp \theta_n$.

 (ii). If Assumptions \ref{a1}, \ref{a3}-\ref{a5} hold, $pq=o\{n^{(\beta-1)/2}\}$, and $\min_{1\leq j\leq q_1}\rho_j>0$, then
  \begin{equation*}
    \max_{1\leq j\leq q_1}\rho_jD(\mathcal{M}(\hat{\bA}_j),\mathcal{M}(\bA_j))=O_p(\kappa\vartheta_n^{1-\iota}\delta+\vartheta_n^{2(1-\iota)}\delta^2),
  \end{equation*}
  where $\hat{\bA}=(\hat{\bA}_1,...,\hat{\bA}_{q_1})$ is a permutation of the $q$ orthogonal eigenvectors of matrix $\hat{\bS}$ defined in (\ref{th:est}) with the threshold $u\asymp\vartheta_n$ given in (\ref{th:u}).
\end{theorem}
\subsection{Asymptotic properties for $n\rightarrow\infty$ and $\log\{\max(p,q)\}=o(n^c)$}
To handle the ultra-high-dimensional case where $p$ and $q$ grow at an exponential rate of $n$, we need some stronger assumptions on the tail probabilities of $\by_i^t$ and the mixing coefficients $\alpha_{p,q}(k)$ defined in (\ref{amix}).
\begin{assumption}\label{a6}
  For any $x>0$ and diagonal matrix $\bD_{p}=\diag(d_1,...,d_p)$ with $\sum_{i=1}^pd_i^2=1$, $\sup_{t}\max_{1\leq j\leq q}P(\|\bD_p(\by_j^t-\bmu_j)\|_2>x)\leq K_2\exp(-K_3x^{r_1})$, where $K_2,K_3>0$ and $r_1\in(0,2]$ are constants.
\end{assumption}

\begin{assumption}\label{a7}
  For all $k\geq 1$, $\sup_{p,q}\alpha_{p,q}(k)\leq\exp(-K_4k^{r_2})$, where $K_4>0$ and $r_2\in(0,1]$ are some constants.
\end{assumption}
Assumption \ref{a6} requires the tail probabilities of linear combinations of $\by_j$ decay exponentially fast uniform for $p$ and $q$. when $r_1=2$ and $d_l=1$ for some $1\leq l\leq p$, $y_{l,j}$ is sub-Gaussian for all $1\leq j\leq q$. The restriction of $r_1\leq 2$ and $r_2\leq 1$ are only for the convenience of presentation, and Theorem 3 below holds for the ultra high-dimensional cases with
\begin{equation}\label{logq}
  \log (pq)=o(n^{\gamma_1/(2-\gamma_1)}),\,\, \text{where}\,\,\gamma_1^{-1}=2r_1^{-1}+r_2^{-1},
\end{equation}
see Lemma 7 for details.
Write
\begin{equation}\label{deltan}
    \delta_{1n}= \nu_n\varpi_n^{-1}[\kappa(n^{-1}\log q)^{(1-\iota)/2}\delta+(n^{-1}\log q)^{1-\iota}\delta^2] \,\,\text{and}\,\,
  \delta_{2n}=(n^{-1}\log pq)^{(1-\iota)/2}\delta.
\end{equation}
  \begin{theorem}
  (i). Under Assumptions \ref{a3} and \ref{a5}-\ref{a7}, $\chi_n/\epsilon_n=o_p(\delta_n^{-1})$, where $\delta_n=\delta_{1n}+\delta_{2n}$. Then for $\hat{d}$ defined in (\ref{rhat:md}), we have $P(\hat{E}=E)\rightarrow 1$ the threshold level $v\asymp(n^{-1}\log pq)^{1/2}$.

   (ii). If Assumptions \ref{a3} and \ref{a5}-\ref{a7} hold, $\min_{1\leq j\leq q_1}\rho_j>0$ and $(p,q)$ satisfy (\ref{logq}), then
     \begin{equation*}
    \max_{1\leq j\leq q_1}\rho_jD(\mathcal{M}(\hat{\bA}_j),\mathcal{M}(\bA_j))=O_p(\kappa(n^{-1}\log q)^{(1-\iota)/2}\delta+(n^{-1}\log q)^{1-\iota}\delta^2),
  \end{equation*}
  where $\hat{\bA}=(\hat{\bA}_1,...,\hat{\bA}_{q_1})$ is a permutation of the $q$ orthogonal eigenvectors of matrix $\hat{\bS}$ defined in (\ref{th:est}) with the threshold level $u\asymp(n^{-1}\log q)^{1/2}$.
  \end{theorem}
  \subsection{Choice of the threshold}
 When the dimension is large, the performance of the proposed method depends critically on the choices of two tuning
parameters: $u$ and $v$, defined in (\ref{threshold}).
  In practice, we adopt the cross-validation method of Bickel and Levina (2008), who proposed to select the threshold by minimizing the Frobenius
norm of the difference between the estimator after thresholding and the sample covariance matrix computed from the independent data. For high-dimensional dependent data, our numeric
experiments show that the cross-validation based method also has a reasonably
good performance. However, we are unable to provide a theoretical justification of
this method, and pose it as an open problem. We illustrate this method as follows.

For a given data set $\{\bY_1,...,\bY_n\}$ and each $0\leq k\leq k_0$, we denote $\wh u_k$  the choice of the threshold for $\bSigma_y(k)$. Let $n_1=[n(1-1/\log(n))]$ and $n_2=n-n_1$. For $s=1,...,N_1$, we sample a subset $\{\bY_{s_{1,1}},...,\bY_{s_{1,n_1}}\}$, and the rest of the data are denoted as $\{\bY_{s_{2,1}},...\bY_{s_{2,n_2}}\}$. Let
$$\wh \bSigma_{y,s_1}(k)=\frac{1}{n_1}\sum_{t=1}^{n_1}(\bY_{s_{1,t}+k}-\bar{\bY}_{s_1})^\T(\bY_{s_{1,t}}-\bar{\bY}_{s_1}),$$
$$\wh \bSigma_{y,s_2}(k)=\frac{1}{n_2}\sum_{t=1}^{n_2}(\bY_{s_{2,t}+k}-\bar{\bY}_{s_2})^\T(\bY_{s_{2,t}}-\bar{\bY}_{s_2}),$$
where $\bar\bY_{s_i}$ is the sample mean of the $n_i$ data matrices for $i=1,2$, and the terms with $s_{i,t}+k$ exceeding $n$ are set to be $0$. We select the parameters $\wh u_k$
by minimizing
$$R_{1,k}(u)=\frac{1}{N_1}\sum_{s=1}^{N_1}\|T_u(\bSigma_{y,s_1}(k))-\bSigma_{y,s_2}(k)\|_F^2.$$

For $0\leq |h|\leq m$, the choice of  $\wh v_h$ can be calculated as follows. Note that
$$\Vc(\wh\Cov(\bz_i^{t+h},\wh\bz_j^{t}))=\Vc\{\frac{1}{n}\sum_{t=1}^{n-h}\bY_{t+h}\wh\bv_i\wh\bv_j^\T\bY_t^\T\}=\frac{1}{n}\sum_{t=1}^{n-h}\bY_t\otimes\bY_{t+h}\Vc(\wh\bv_i\wh\bv_j^\T).$$
Thus, we define
$$\wh \bOmega_{y,s_1}(h)=\frac{1}{n_1}\sum_{t=1}^{n_1}\bY_{s_{1,t}}\otimes\bY_{s_{1,t}+h}\quad \text{and}\quad \wh \bOmega_{y,s_2}(h)=\frac{1}{n_2}\sum_{t=1}^{n_2}\bY_{s_{2,t}}\otimes\bY_{s_{2,t}+h},$$
and  select the parameters  $\wh v_h$ by minimizing
$$ R_{2,h}(v)=\frac{1}{N_1}\sum_{s=1}^{N_1}\|T_v(\bOmega_{y,s_1}(h))-\bOmega_{y,s_2}(h)\|_F^2.$$

\section{Segmenting tensor-valued time series}
Let $\{\mathbb{Y}_t\}_{t=1}^n$ be observations of tensor-valued time series. We use the notation in Kolda and Bader (2009) and assume $\mathbb{Y}_t\in R^{p_1\times\cdots\times p_r}$. We expect that there exist transformation matrices $\bC_1,...,\bC_r$ where $\bC_i\in R^{p_i\times p_i}$ for $1\leq i\leq r$, such that
\begin{equation}\label{tensor}
  \mathbb{Y}_t=\mathbb{X}_t\times_1\bC_1\times_2\cdots\times_r\bC_r,
\end{equation}
where $\mathbb{X}_t$ is also a tensor with the same size as $\mathbb{Y}_t$, and can be divided into many sub-tensors for which any two different tensors have no correlations at all time lags. When $r=2$, (\ref{tensor}) reduces to the form of (\ref{eq:1}) as $\mathbb{Y}_t=\bC_1\mathbb{X}_t\bC_2^\T$ and $\mathbb{X}_t$ is a matrix containing many uncorrelated submatrices. As we have seen that it is difficult to estimate $\bC_1,...,\bC_r$ and $\mathbb{X}_t$ simultaneously and the identifications are not clear. We propose an $r$-step procedure to estimate them.
Denote $\mathbb{Y}_t^{(m)}$ the $m$-mode matricization of $\mathbb{Y}_t$, which is obtained by arranging the mode-$m$ fibers to be the columns of the resulting matrix. Let $\wp_m=\Pi_{i=1,i\neq m}^rp_i$. Then $\mathbb{Y}_t^{(m)}\in R^{p_m\times\wp_m}$. For each $1\leq m\leq r$, we write
\begin{equation}\label{est:tensor}
  \mathbb{Y}_t^{(m)}=\bC_m\mathbb{X}_t^{(m)},
\end{equation}
where we still use $\bC_i$ even though they are not identical to those in (\ref{tensor}).
For example, when $r=2$ and hence $\mathbb{Y}_t$ is a matrix, and $\mathbb{Y}_t^{(1)}=\mathbb{Y}_t$ and $\mathbb{Y}_t^{(2)}=\mathbb{Y}_t^\T$. Then (\ref{est:tensor}) reduces to
\begin{equation}\label{matrix:ts}
  \mathbb{Y}_t=\bC_1\mathbb{X}_t\quad\text{and}\quad\mathbb{Y}_t=\mathbb{X}_t\bC_2^\T,
\end{equation}
and both $\bC_1$ and $\bC_2$ can be estimated by the method proposed in Section 2. Note that (\ref{est:tensor}) can be written as $ \mathbb{Y}_t^{(m),\T}=\mathbb{X}_t^{(m),\T}\bC_m^\T$. In view of this, when $m\geq2$, Algorithm 1 should be able to estimate the transformation matrices $\bC_i$ and $\mathbb{X}_t$ sequentially.

\begin{algorithm}[t]
\caption{Sequential transformation \label{alg:st}}
{\bf Require} $\{\mathbb{Y}_t\}_{t=1}^n$

1: {\bf Repeat}\\
2: \quad{\bf For} {$m=1:r$} {\bf do}\\
3:\quad\quad Standardize $\mathbb{Y}_t^{(m),\T}$ as (\ref{std})\\
4:\quad\quad Perform the procedure in Section 2 and obtain $\mathbb{X}_t^{(m)}$ \\
5:\quad\quad Rearrange the transformed matrix $\mathbb{X}_t^{(m)}$  as a tensor $\mathbb{X}_t$ \\
6:\quad\quad Let $\mathbb{Y}_t=\mathbb{X}_t$ and $m=m+1$\\
7: \quad {\bf End For}\\
8: {\bf Until} {$m=r+1$}
\end{algorithm}
After $r$ steps, we obtain a tensor which contains many uncorrelated subtensors, and they can be modelled separately as far as their linear dynamic structure is concerned.

\begin{remark}
We have not discussed the way to build a dynamic model for matrix- or tensor-valued time series in terms of the predictions. For example, once we have divided the original matrix into several uncorrelated sub-matrices, each of them possesses a lower-dimensional structure and they can be modelled as a matrix autoregressive model:
\begin{equation}\label{mm}
\bW_t^{(i,j)}=\bL_1\bW_{t-1}^{(i,j)}\bR_1^\T+\cdots+\bL_s\bW_{t-s}^{(i,j)}\bR_s^\T+\bE_t^{(i,j)},
\end{equation}
where $\bW_t^{(i,j)}$ is a sub-matrix of $\bW_t$ in model (\ref{eq:1}), $\bE_t^{(i,j)}$ is a noise matrix, and $\bL_k$ and $\bR_k$ ($1\leq k\leq s$) are the coefficient matrices. This is beyond the scope of this paper and research in this direction is on-going.
\end{remark}
\section{Numerical results}
\subsection{Simulation}
In this section, we illustrate the finite sample
properties of the proposed methodology using simulated data. We only study the performance of the column transformation in (\ref{seg}) since the row transformation in the second step is essentially the same. As the estimated $\wh \bA$ is an orthogonal matrix for the `normalized' model in which $\Var_r(\bY_t) = \Var_r(\bX_t) = \bI_q$. We should use $\wh \bS_{y,0}^{-1/2} \bA\wh\bS_{x,0}^{1/2}$ instead of $\bA$ in computing estimation error (\ref{dmeasure}), see the discussion after (\ref{std}). Let $\bA^* = \{\bSigma_y(0)\}^{-1/2} \bA \{\bSigma_x(0)\}^{1/2} \equiv  (\bA_1^*, \ldots, \bA_{q_1}^*)$, and $\{\bSigma_y(0)\}^{-1/2} \bA = (\bH_1, \ldots, \bH_{q_1})$. Since $\{\bSigma_x(0)\}^{1/2}$ is a block-diagonal matrix, it holds that $\mathcal{M}(\bH_j) = \mathcal{M}(\bA_j^*)$ for $1\le j \le q_1$. Therefore, we only need to replace $\bA$ by $\wh \bS_{y,0}^{-1/2} \bA$.

Since the goal is to specify (via estimation) the $q_1$ linear spaces
$\mathcal{M}(\bA_j), \; j=1, \ldots, q_1$, simultaneously, we first introduce the
concept of a `correct' specification. We call $\wh \bA= ( \wh \bA_1,
\ldots, \wh \bA_{\wh q_1})$ a {\sl correct specification} for $\bA$
if (i) $\wh q_1 = q_1$, and (ii) rank$(\wh \bA_j) = {\rm
 rank}(\bA_j)$ for $j=1, \ldots, q_1$,
 after re-arranging the order of $\wh \bA_1,
\ldots, \wh \bA_{\wh q_1}$ (we still denote the rearranged submatrices as $\wh \bA_1,
\ldots, \wh \bA_{\wh q_1}$ for the simplicity in notation).
When more than one $\bA_j$ have the same rank, we pair each those $\bA_j$
with the $\wh \bA_j$ for which
\[
 D ( \mathcal{M}(\bA_j), \mathcal{M}(\wh \bA_j) ) = \min_{ {\rm rank}(\wh \bA_i) = {\rm rank}(\bA_j) }
 D ( \mathcal{M}(\bA_j), \mathcal{M}(\wh \bA_i) ).
\]
Note that a correct specification for $\bA$ implies
a {\sl structurally correct segmentation} for $\bX_t$, which will be
abbreviated as `correct segmentation' hereafter.
For a correct segmentation, we report the estimation
error defined as
\begin{equation} \label{barD}
\bar D(\wh \bA, \bA) = {1 \over q_1} \sum_{j=1}^{q_1}  D ( \mathcal{M}(\bA_j), \mathcal{M}(\wh \bA_j) ).
\end{equation}

In addition to the correct segmentations, we also report the proportions of the near-complete segmentations with $\wh q_1=q_1-1$ in all the following examples.

\begin{figure}
\begin{center}
{\includegraphics[width=0.8\textwidth]{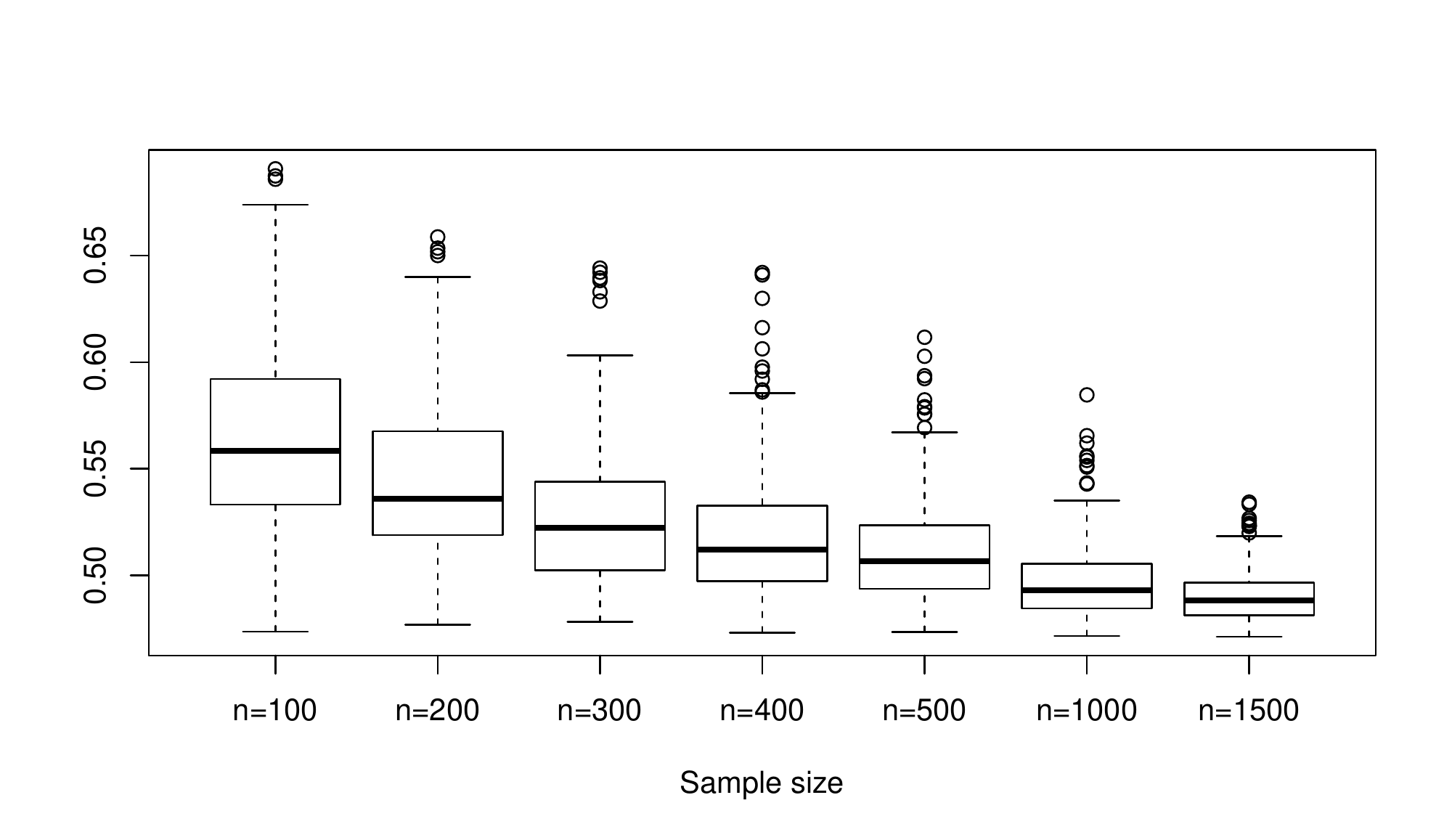}}
\caption{The boxplots of estimation errors $\bar D(\wh \bA, \bA)$ in example 1.}\label{fig1}
\end{center}
\end{figure}

{\bf Example 1.}
We consider the model (\ref{seg}) with $p=3$ and $q=6$. The columns of $\bX_t$ are generated as follows:
$$\bx_i^t=\bfeta_{t+i-1}^{(1)}\,(i=1,2,3),\quad \bx_i^{t}=\bfeta_{t+i-4}^{(2)}\,(i=4,5)\quad\text{and}\quad \bx_6^t=\bfeta_t^{(3)},$$
where
$$\bfeta_t^{(j)}=\bPhi^{(j)}\bfeta_{t-1}^{(j)}+\bve_t^{(j)}-\bTheta^{(j)}\bve_{t-1}^{(j)},\quad j=1,2,3.$$
The elements of $\bPhi^{(j)}$ are drawn independently from $U(-3,3)$ and then normalized by $0.9\times\bPhi^{(j)}/\|\bPhi^{(j)}\|_2$ so that $\bfeta^{(j)}$ is stationary, and the elements of $\bTheta^{(j)}$  are drawn independently from $U(-1,1)$. Meanwhile, the elements of the transformation matrix $\bA$ are also drawn independently from $U(-3,3)$. Thus $\bX_t$ consists of three independent submatrices with, respectively, $3$, $2$ and $1$ columns. In the experiments, we choose $c_0=0.75$ in (\ref{rhat}) and $k_0=2$ in (\ref{wy}), and the other $k_0$'s give similar results. The sample sizes $(n)$ are $100$, $200$, $300$, $400$, $500$, $1000$ and $1500$, and the number of replications is $500$ for each case. The proportions of the correct, incorrect and near-complete segmentations are reported in Table \ref{Table1}. From Table \ref{Table1}, we can see that the proposed method improves as the sample size increases. With a dimension of $pq=18$, we can see that the performance is reasonably well even for a small sample size and the sum of the proportions of the complete and the near complete ones more than $90\%$ for a small sample size $n=100$, from which we can see that we have achieved sufficient dimension reduction. We next study the estimation errors (\ref{barD}), and the box plots of the errors in the complete segmentations are shown in Figure  \ref{fig1}. From Figure \ref{fig1}, we can see that the proposed method improves as the sample size increases.

\begin{figure}
\begin{center}
{\includegraphics[width=0.8\textwidth]{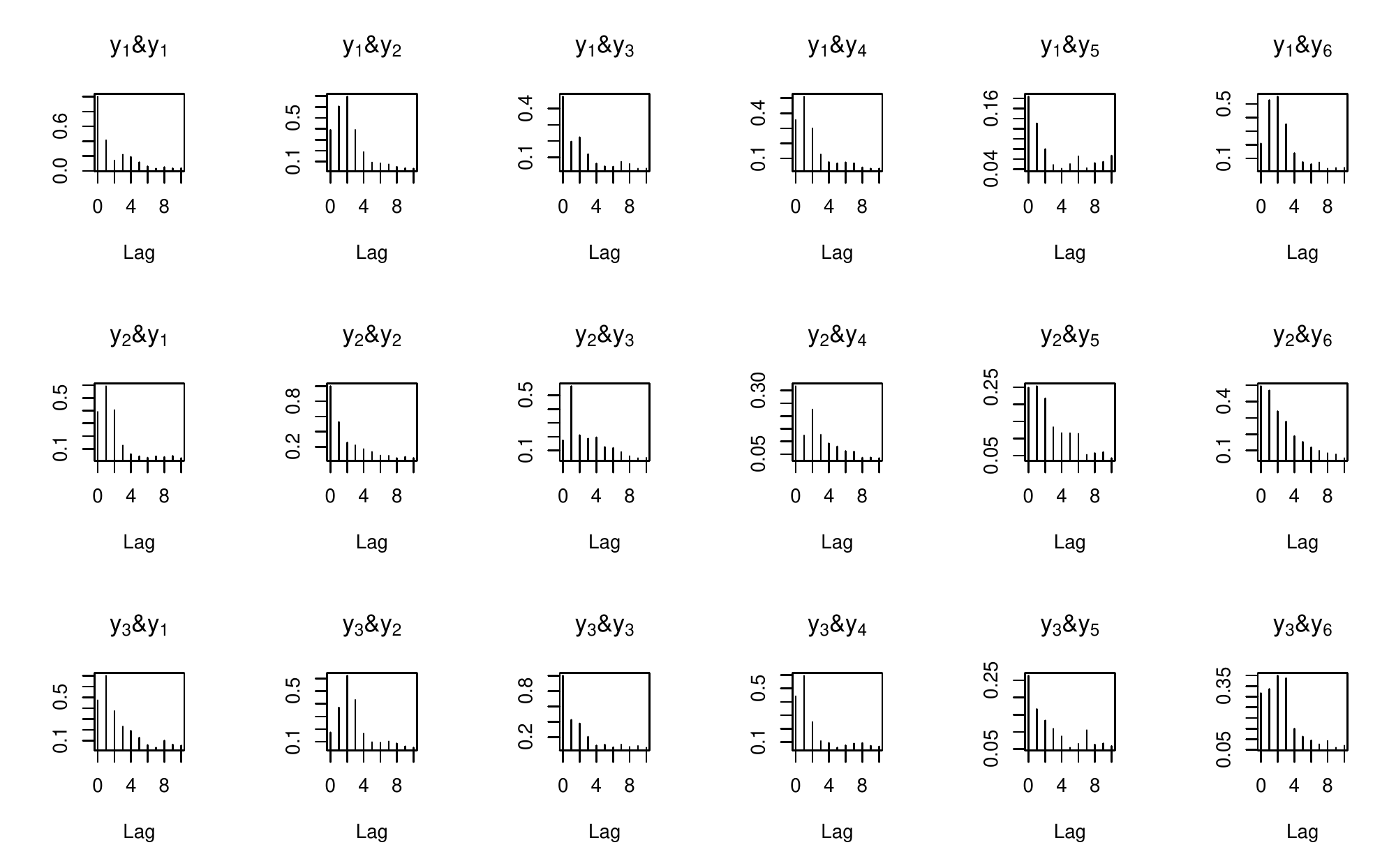}}
{\includegraphics[width=0.8\textwidth]{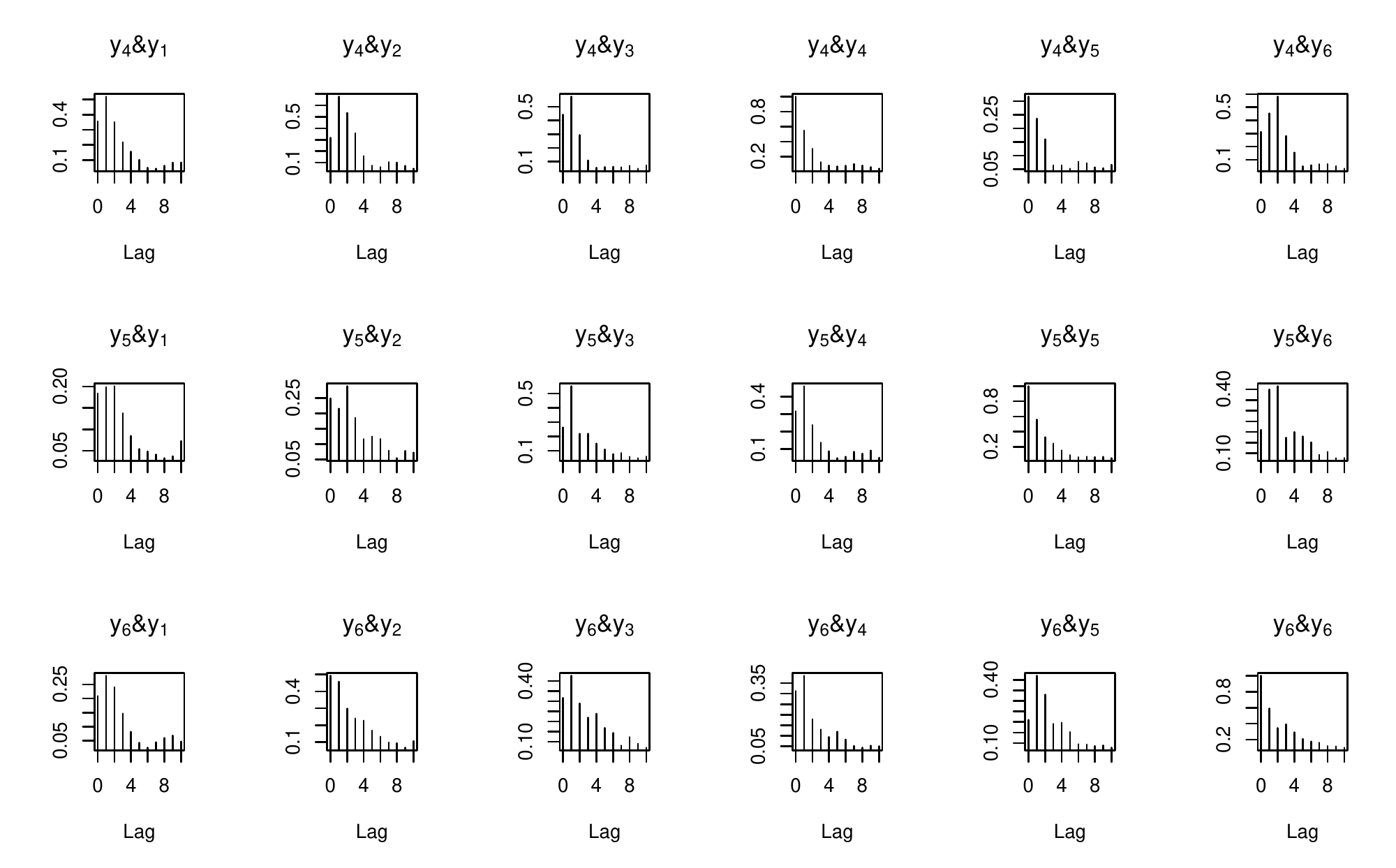}}
\caption{Cross correlogram of the $\by_i^t$ in Example 1, $n=1500$.}\label{fig2}
\end{center}
\end{figure}

\begin{figure}
\begin{center}
{\includegraphics[width=0.8\textwidth]{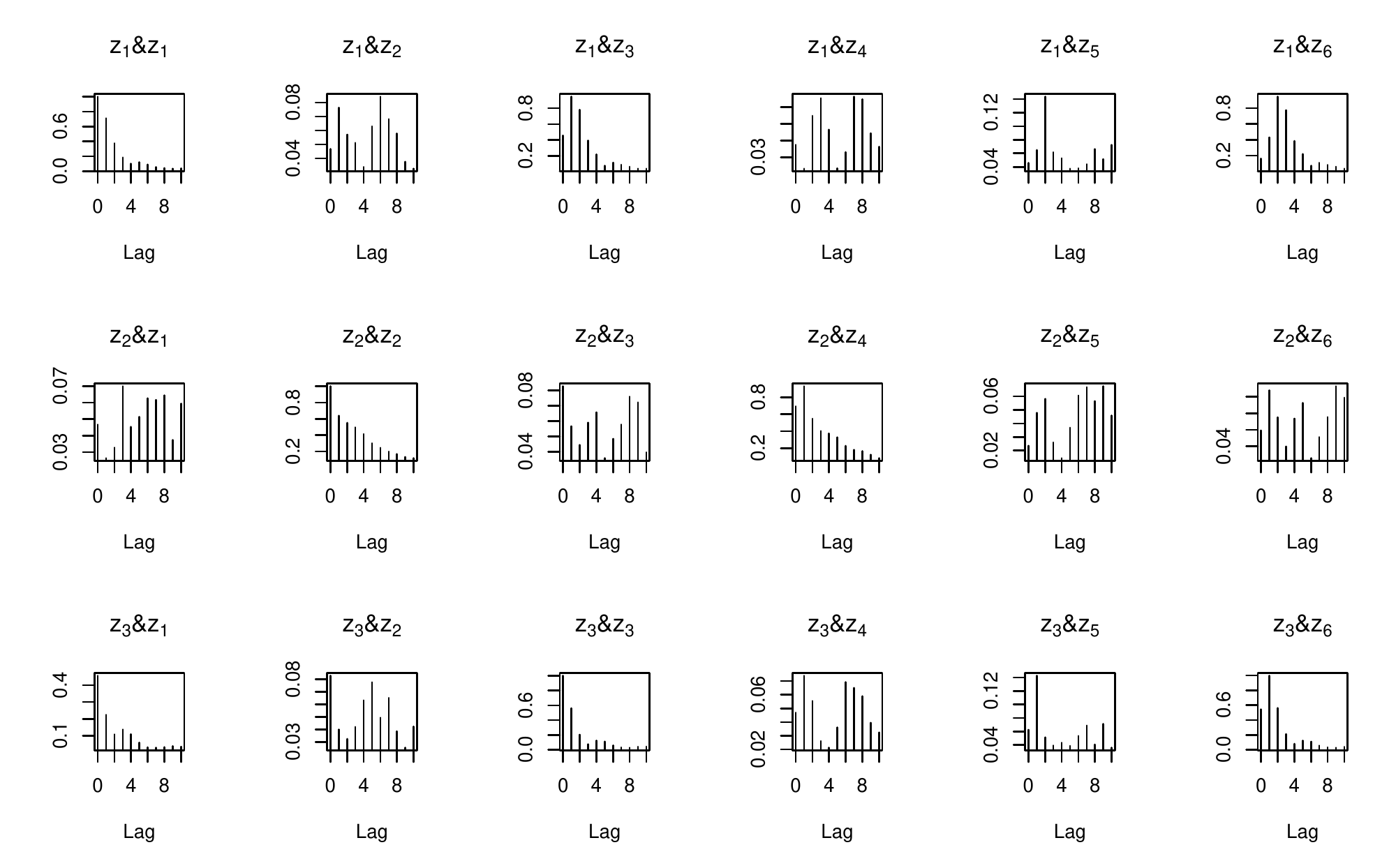}}
{\includegraphics[width=0.8\textwidth]{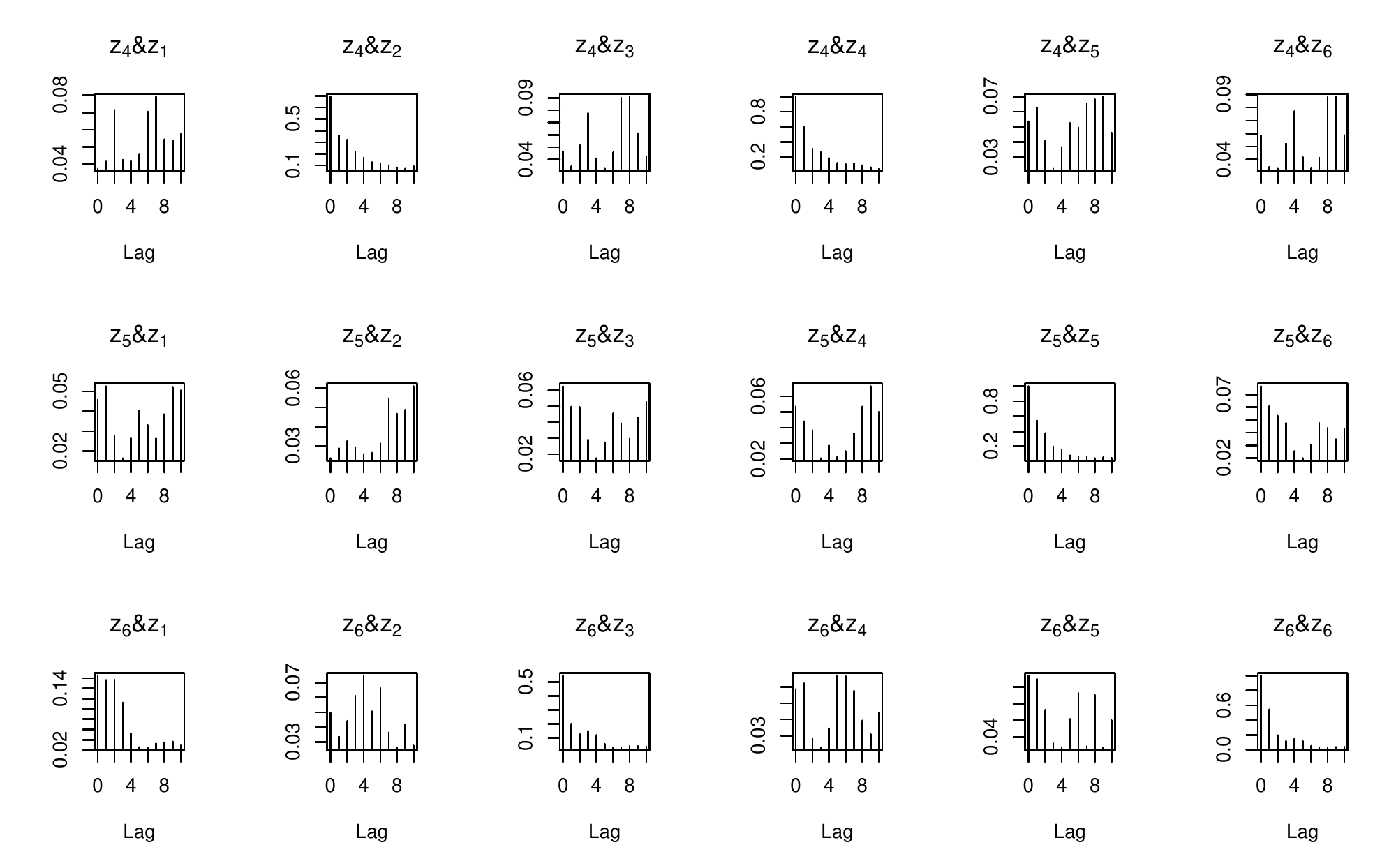}}
\caption{Cross correlogram of the transposed series $\wh \bz_i^t=\bY_t{\wh \bv_i}$ in Example 1, $n=1500$. The components of $\wh \bZ_t$ can be segmented into 3 groups: $\{1,3,6\}$, $\{2,4\}$ and $\{5\}$.}\label{fig3}
\end{center}
\end{figure}

As a concrete example, we report the correlogram of $\bY_t$ for one replication. We choose $m=10$ and for $0\leq k\leq m$ in Figure \ref{fig2}, the correlation between $\by_i^t$ and $\by_j^t$ are computed by
$$\wh \Corr(\by_i^{t+k},\by_j^t)=|\diag\{\wh\bSigma_{y,i,i}(0)\}^{-1/2}\wh\bSigma_{y,i,j}(k)\diag\{\wh\bSigma_{y,j,j}(0)\}^{-1/2}|_\infty.$$
From Figure \ref{fig2}, we can see that each of the columns of $\bY_t$ are highly correlated with the others. We then apply our method to $\bY_t$ and Figure \ref{fig3} depicts the cross correlogram of the transformed matrix $\bZ_t=\bY_t\wh\bGamma_y$, and the scales of the $y$-axes are not the same. We can see from Figure \ref{fig3} that the columns of $\bZ_t$ can be divided into $3$ groups: $\{1,3,6\}$, $\{2,4\}$ and $\{5\}$.
 
\begin{table}
\begin{center}
 \setlength{\abovecaptionskip}{0pt}
\setlength{\belowcaptionskip}{3pt}
 \caption{The proportions of correct and incomplete segmentations in Example
1. $p=3,q=6$} \centering
          \label{Table1}
\begin{tabular}{l|ccccccc}
\hline \hline
\qquad \qquad  $n$  & 100 & 200 & 300  & 400 & 500 &1000 & 1500 \\
\hline {\footnotesize Correct segmentation} & 0.608  & 0.668  &
0.748 & 0.806 & 0.864 & 0.948 & 0.980
\\
{\footnotesize Incorrect segmentation} & 0.392  & 0.332  & 0.252
& 0.294 & 0.246 & 0.052 & 0.020\\
{\footnotesize Near-Complete segmentation with $\wh q_1=2$} & 0.298  & 0.306  & 0.250
& 0.190 & 0.136 & 0.052 & 0.020\\
   \hline \hline
\end{tabular}
          \end{center}
\end{table}

{\bf Example 2.} In this example, we slightly increase the dimension as $p=q=6$, and the data generating processes are the same as those in Example 1. We observe that the performance of the proposed method is not as good as that in Example 1 when the dimension is higher with $pq=36$. Nevertheless, similar conclusions can also be obtained from Table \ref{Table2} such as the proportion of the complete segmentation increases as the sample size becomes larger, and the sum of the proportions of the complete and the near complete ones is more than $90\%$ even for a small sample size. The box plots of the estimation errors are similar to that in Example 1 and hence we do not report it here. From this example, we can see that when the dimension is higher, the proposed method without thresholding may not work well, and we will illustrate this point in the next example.
\begin{table}
\begin{center}
 \setlength{\abovecaptionskip}{0pt}
\setlength{\belowcaptionskip}{3pt}
 \caption{The proportions of correct and incomplete segmentations in Example
2. $p=6,q=6$.} \centering
          \label{Table2}
\begin{tabular}{l|ccccccc}
\hline \hline
\qquad \qquad  $n$  & 100 & 200 & 300  & 400 & 500 &1000 & 1500 \\
\hline {\footnotesize Correct segmentation} & 0.504  & 0.620  &
0.648 & 0.670 & 0.732 & 0.766 & 0.780
\\
{\footnotesize Incorrect segmentation} & 0.496  & 0.380  & 0.352
& 0.330 & 0.268 & 0.234 & 0.220\\
{\footnotesize Near-Complete segmentation with $\wh q_1=2$} & 0.354  & 0.320  & 0.312
& 0.308 & 0.264 & 0.230 & 0.210\\
   \hline \hline
\end{tabular}
          \end{center}
\end{table}

\begin{figure}
\begin{center}
{\includegraphics[width=0.8\textwidth]{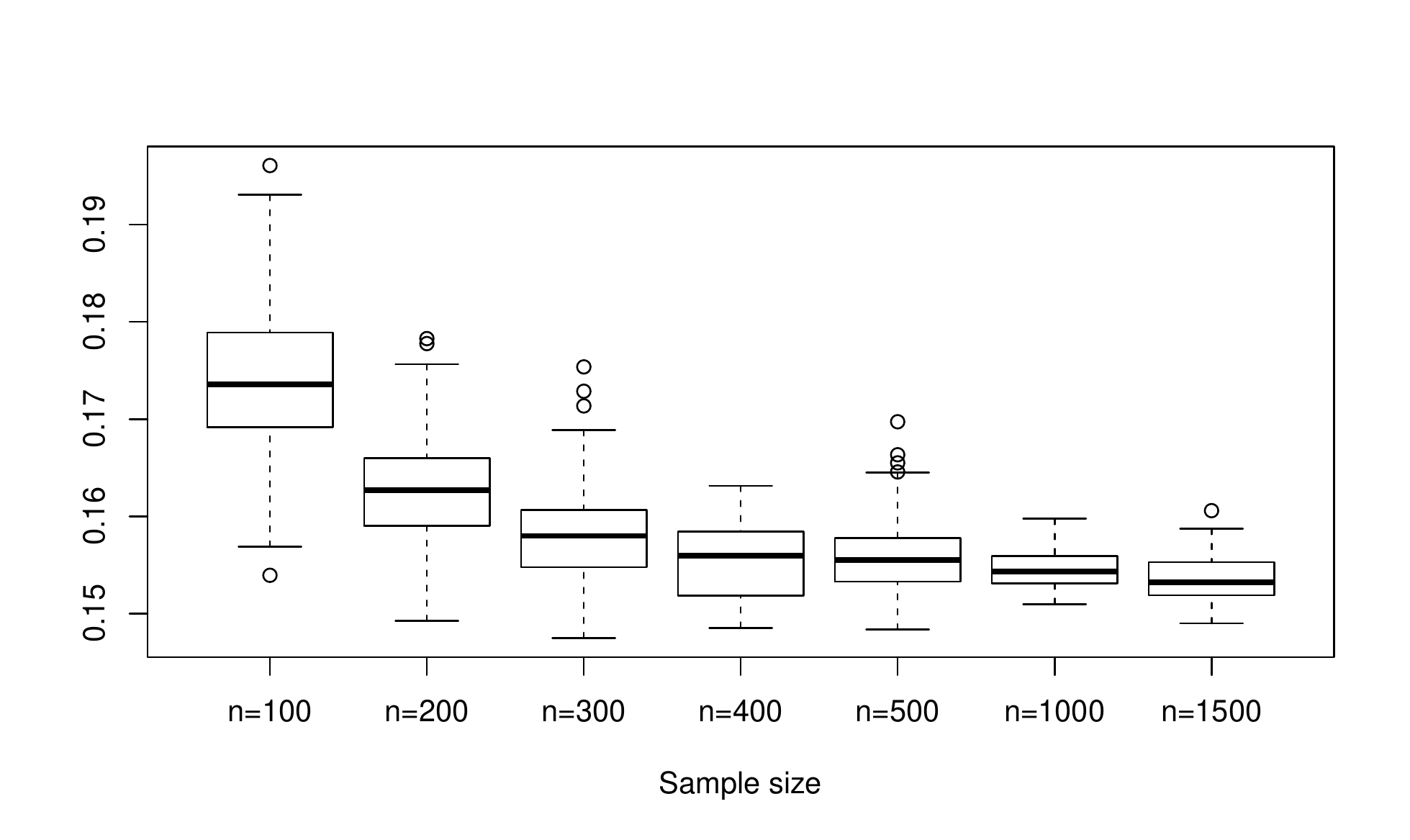}}
\caption{The boxplots of estimation errors $\bar D(\wh \bA, \bA)$ in Example 3.}\label{fig4}
\end{center}
\end{figure}

\begin{figure}
\begin{center}
{\includegraphics[width=0.8\textwidth]{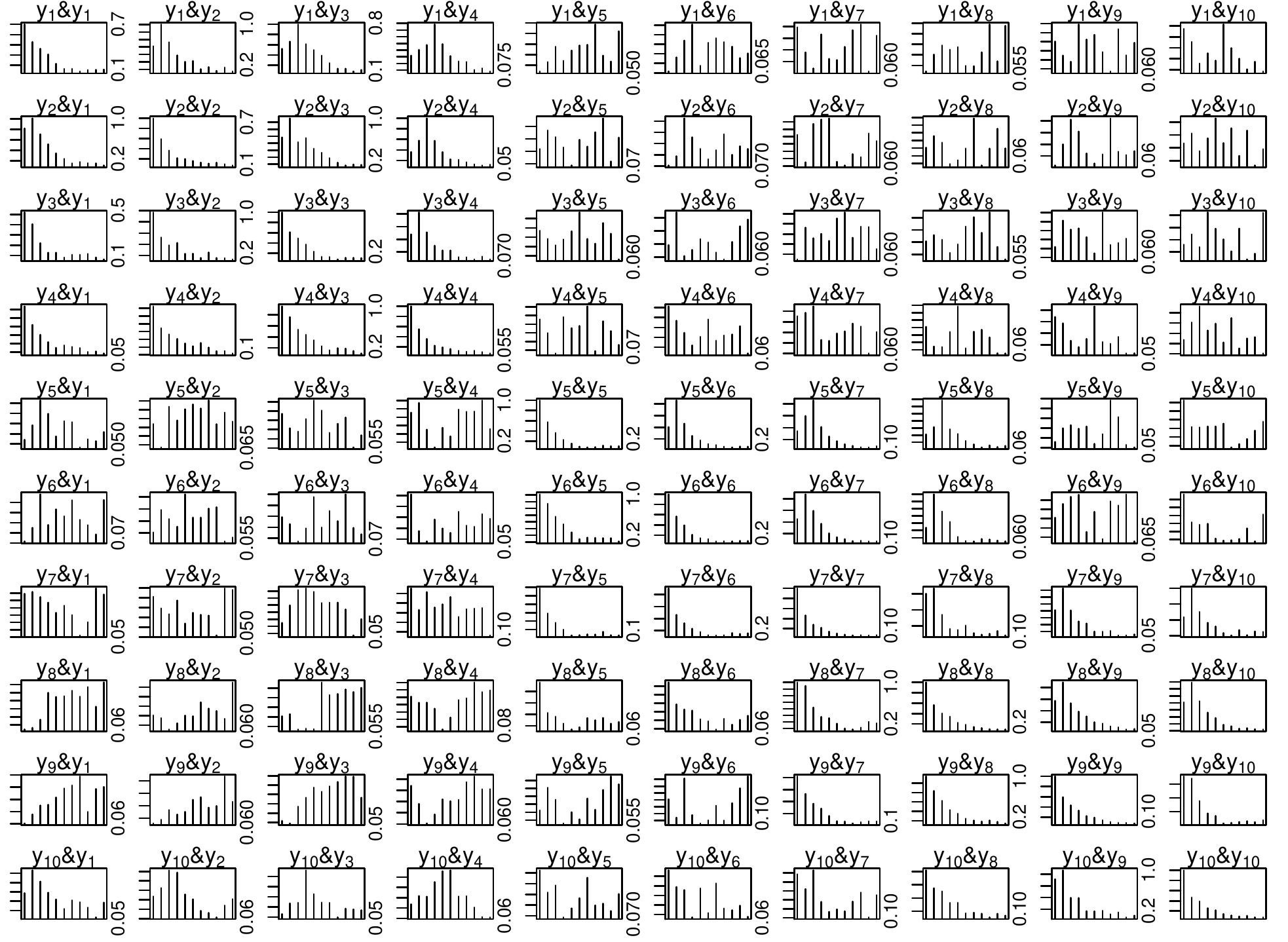}}
\caption{Cross correlogram of the $\by_i^t$ in Example 3, $n=1500$.}\label{fig5}
\end{center}
\end{figure}

\begin{figure}
\begin{center}
{\includegraphics[width=0.8\textwidth]{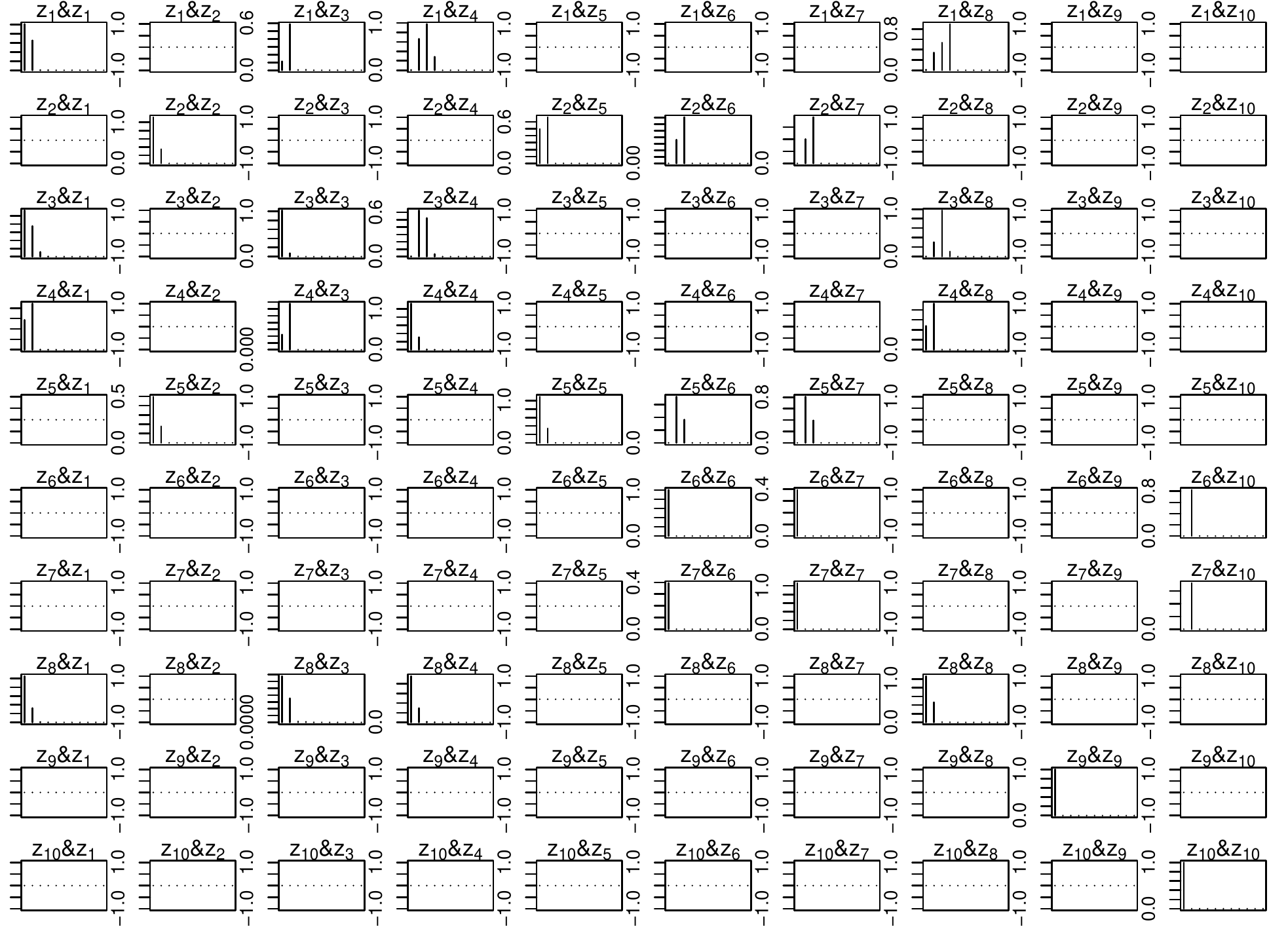}}
\caption{Cross correlogram of the transposed series $\wh \bz_i^t=\bY_t{\wh \bv_i}$ in Example 3, $n=1500$. The components of $\wh \bZ_t$ can be segmented into 4 groups: $\{1,3,4,8\}$, $\{2,5,7\}$, $\{6,10\}$ and $\{9\}$.}\label{fig6}
\end{center}
\end{figure}

{\bf Example 3.} We consider model (\ref{seg}) with $p=q=10$ and the dimension is $pq=100$. The columns of $\bX_t$ are generated as follows:
$$\bx_i^t=\bfeta_{t+i-1}^{(1)}\,(i=1,2,3,4),\,\, \bx_i^{t}=\bfeta_{t+i-5}^{(2)}\,(i=5,6,7),\,\,\bx_i^t=\bfeta_{t+i-8}^{(3)}\,(i=8,9)\,\,\text{and}\quad\bx_{10}^{t}=\bfeta_{t}^{(4)},$$
where
$$\bfeta_t^{(j)}=\bPhi^{(j)}\bfeta_{t-1}^{(j)}+\bve_t^{(j)}-\bTheta^{(j)}\bve_{t-1}^{(j)},\quad j=1,2,3,4,$$
and $\bPhi^{(j)}$ and $\bTheta^{(j)}$ are generated in the same way as Example 1. To fulfill the assumption 3, let 

$$\bB_i=\left(\begin{array}{cc}
\cos(\theta_i\pi)&\sin(\theta_i\pi)\\
-\sin(\theta_i\pi)&\cos(\theta_i\pi)
\end{array}\right),\quad \bA=\diag(\bB_1,\bB_2,\bB_3,\bB_4,\bB_5),$$
where $\theta_1=\pi/5$, $\theta_2=\pi/6$, $\theta_3=\pi/7$, $\theta_4=\pi/8$, $\theta_5=\pi/9$. Thus, $\bA$ is a sparse orthogonal transformation matrix. Since the covariance of $\bX_t$ and $\bY_t$ are all block-diagonal by the data generating process, hence $\wh \bS_{y,0}^{-1/2}\bA\bS_{x,0}^{1/2}$ is also block-diagonal and hence satisfies Assumption 3.
If we apply our methodology to this model without thresholding the covariance matrices, the results are reported in Table \ref{Table3}. From Table \ref{Table3}, we can see that the proportions  of the complete segmentations are pretty low for all the sample sizes and it does not necessarily improve as the sample size increases. Now we adopt the thresholding technique as discussed in Section 3, and the choice of the threshold is computed by a cross-validation method as Section 3.4. Table \ref{Table4} presents the proportions of the correct, incorrect and near-complete segmentations for model (\ref{seg}) with $pq=100$ dimensions. From Table \ref{Table4}, we can see that the thresholding technique works reasonably well for even small sample sizes. Figure \ref{fig4} reports the boxplots of the estimation errors, from which we can see that the estimation is very accurate especially when the sample size is large. Figure \ref{fig5} displays the cross correlogram of $\bY_t$ for one instance of the $500$ replications and the correlations are calculated in the same way as that in Example 1 with thresholded estimators for the (auto)covariance matrices and Figure \ref{fig6} gives the correlogram of the transformed data matrix $\bZ_t=\bY_t\wh \bGamma_y$. We can see from Figure \ref{fig5} that all the columns of $\bY_t$ are highly correlated while those of $\bZ_t$ can be separated into $4$ groups: $\{1,3,4,8\}$, $\{2,5,7\}$, $\{6,10\}$ and $\{9\}$. The threshold of the correlation is $0.39$ according to the rule of (\ref{rhat:md}), and those pairs will be treated as uncorrelated ones if their maximal correlation is below this threshold level. Together with Tables \ref{Table3}-\ref{Table4} and Figures \ref{fig4}-\ref{fig5}, we can achieve substantial dimension reduction using our proposed method.
\begin{table}
\begin{center}
 \setlength{\abovecaptionskip}{0pt}
\setlength{\belowcaptionskip}{3pt}
 \caption{The proportions of correct and incomplete segmentations in Example
3 without thresholding. $p=10,q=10$.} \centering
          \label{Table3}
\begin{tabular}{l|ccccccc}
\hline \hline
\qquad \qquad  $n$  & 100 & 200 & 300  & 400 & 500 &1000 & 1500 \\
\hline {\footnotesize Correct segmentation} & 0.148  & 0.184  &
0.186 & 0.170 & 0.216 & 0.174 & 0.164
\\
{\footnotesize Incorrect segmentation} & 0.852  & 0.816  & 0.814
& 0.830 & 0.784 & 0.826 & 0.836\\
{\footnotesize Near-Complete segmentation with $\wh q_1=3$} & 0.216  & 0.310  & 0.388
& 0.470 & 0.462 & 0.674 & 0.744\\
   \hline \hline
\end{tabular}
          \end{center}
\end{table}

\begin{table}
\begin{center}
 \setlength{\abovecaptionskip}{0pt}
\setlength{\belowcaptionskip}{3pt}
 \caption{The proportions of correct and incomplete segmentations in Example
3 by thresholding. $p=10,q=10$.} \centering
          \label{Table4}
\begin{tabular}{l|ccccccc}
\hline \hline
\qquad \qquad  $n$  & 100 & 200 & 300  & 400 & 500 &1000 & 1500 \\
\hline {\footnotesize Correct segmentation} & 0.440  & 0.572  &
0.654 & 0.666 & 0.840 & 0.910 & 0.936
\\
{\footnotesize Incorrect segmentation} & 0.560  & 0.428  & 0.346
& 0.334 & 0.260 & 0.080 & 0.064\\
{\footnotesize Near-Complete segmentation with $\wh q_1=3$} & 0.490  & 0.294  & 0.282
& 0.286 & 0.084 & 0.062 & 0.042\\
   \hline \hline
\end{tabular}
          \end{center}
\end{table}

\subsection{Real data example}
{\bf Example 4}.
In this section, we illustrate our methodology by using the $10\times 10$ Portfolios which are the intersections of $10$ portfolios formed on size (market equity) and investment. The data can be downloaded at \url{http://mba.tuck.dartmouth.edu/pages/faculty/ken.french/data_library.html}. We collect the data from January 1964 to December 2015 with  624 months and overall 62400 observations. The $100$ series are shown in Figure 7.

\begin{figure}
\begin{center}
{\includegraphics[width=0.9\textwidth]{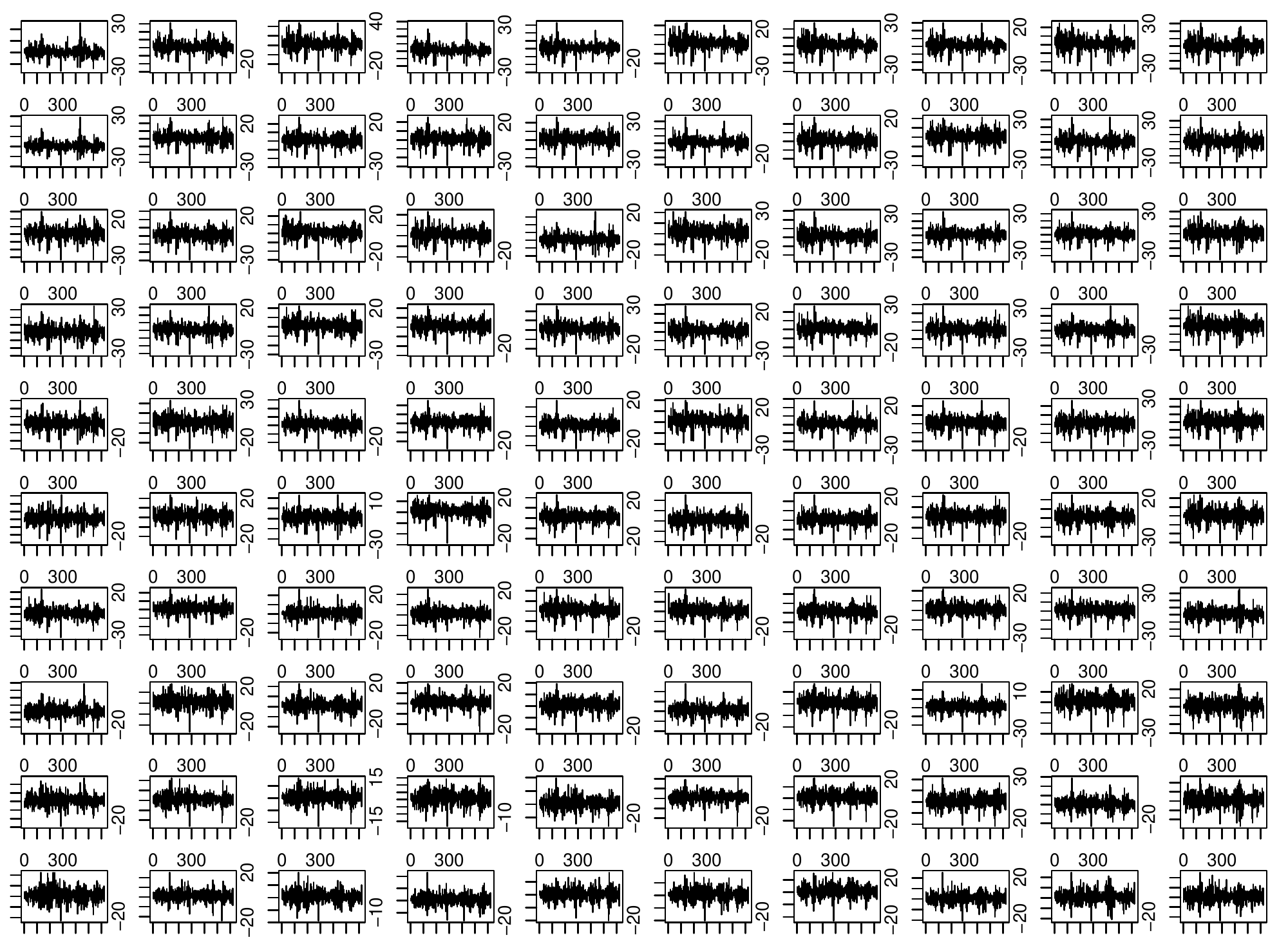}}
\caption{Plots of the $10\times 10$ series in Example 4.}\label{fig7}
\end{center}
\end{figure}

We denote the original series as $\{\bY_t\}_{t=1}^{624}$ with $\by_i^t$ as the $i$-th column of $\bY_t$. For $m=10$, Figure 8 shows the cross correlogram  of $\bY_t$ across $m$ lags, and the computing method is similar to that in the simulation study. From Figure 8, we can see that all columns are highly correlated. If we build a dynamic model for $\bY_t$ directly, using the model (\ref{mm}), there will be too many parameters to be estimated. In the following analysis, we will apply our method to this series.

First, we apply our method to the $10\times 10$ portfolios directly without the thresholding technique, and the cross correlogram of $\bZ_t=\bY_t\wh\bGamma_y$ is shown in Figure 9. According to the criterion in (\ref{rhat:md}), the threshold is chosen as $0.339$ and those pairs with a maximum correlation below $0.339$ will be treated as uncorrelated ones. From Figure 9, we can see that the columns of $\bZ_t$ can be segmented into $3$ groups: $\{1,2,3,4,5,7,9,10\}$, $\{6\}$,and $\{8\}$. 

Next, we denote $\wt\bZ_t$ as the standardized data of $\bZ_t^\T$ and perform the second step of the sequential transformation algorithm. Figure 10 shows the plots of the cross correlogram of the transformed data $\bW_t=\wt\bZ_t\wh \bGamma_{\wt z}$. The threshold in (\ref{rhat:md}) is $0.489$ and the columns of $\bW_t$ can be divided into $2$ groups: $\{1,2,3,5,6,7,8,9,10\}$ and $\{4\}$. Together with the first step of the transformation, we can see that, for a series with dimension $100$ and sample size $624$, we can only segment the matrix into $6$ groups in total, and many correlations in Figures 9-10 may not be significant enough. According to our simulation study in section 5.1, it will be useful if we adopt the thresholding technique under the sparsity assumption.

\begin{figure}
\begin{center}
{\includegraphics[width=0.48\textwidth]{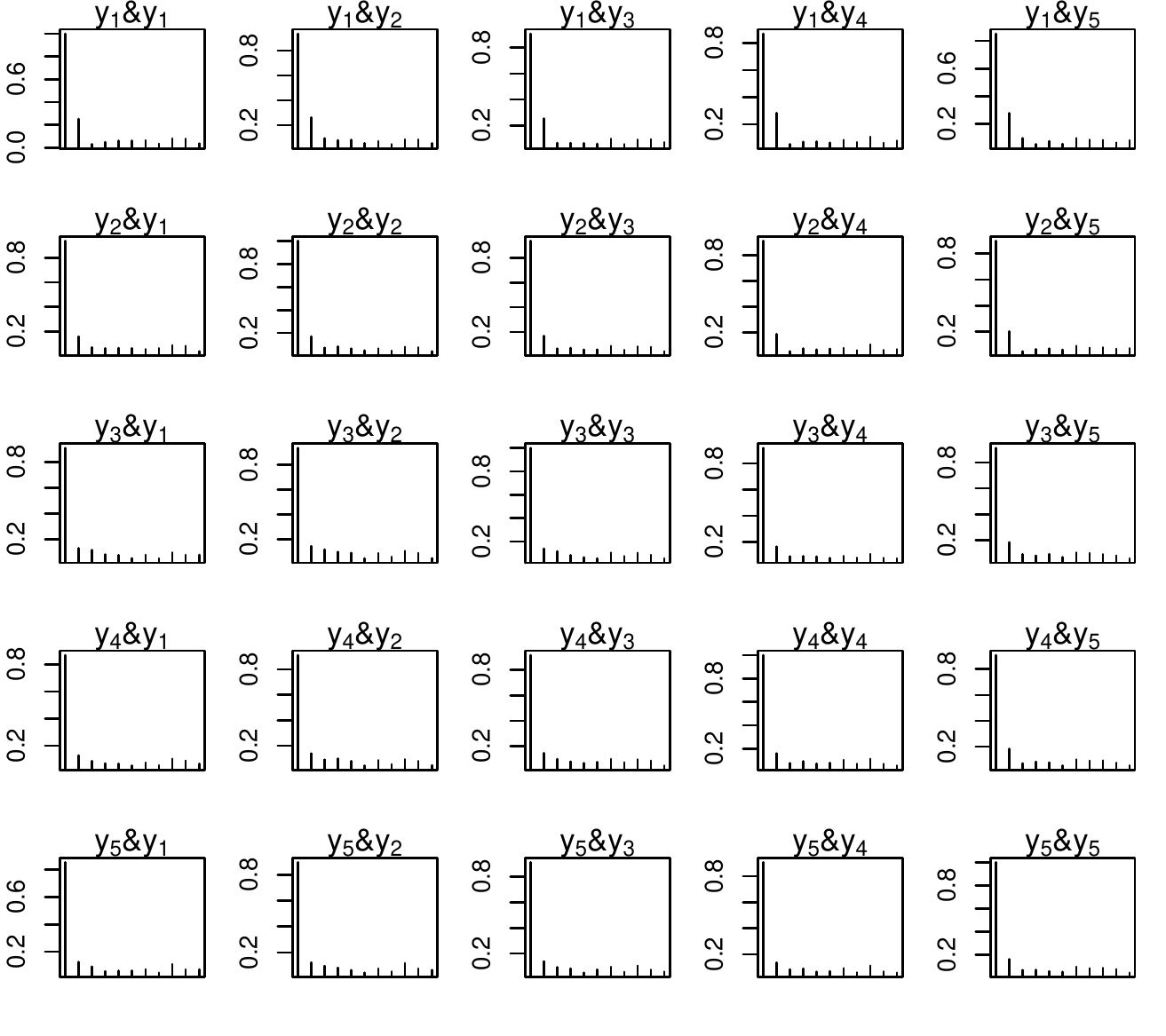}}
{\includegraphics[width=0.48\textwidth]{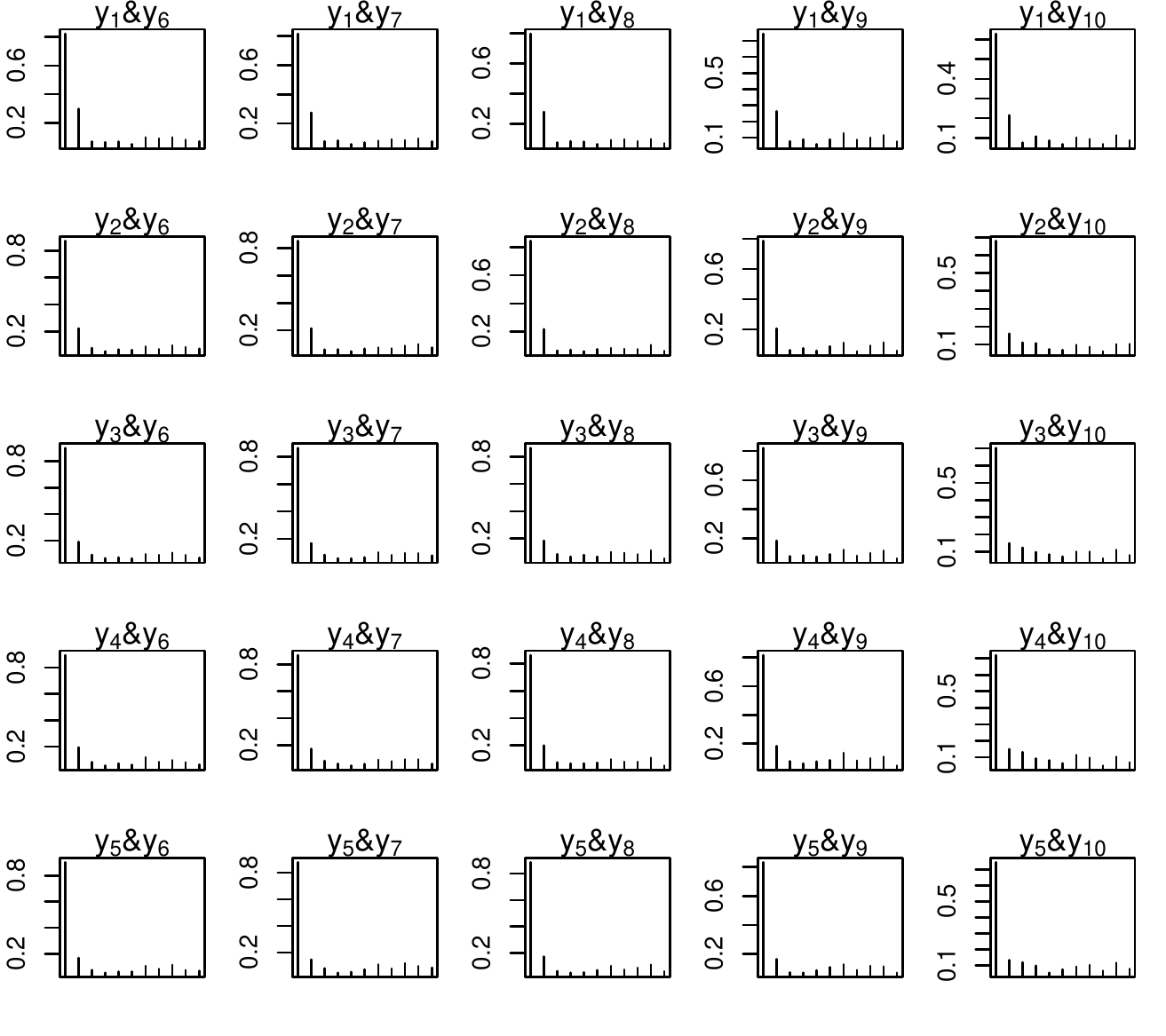}}
{\includegraphics[width=0.48\textwidth]{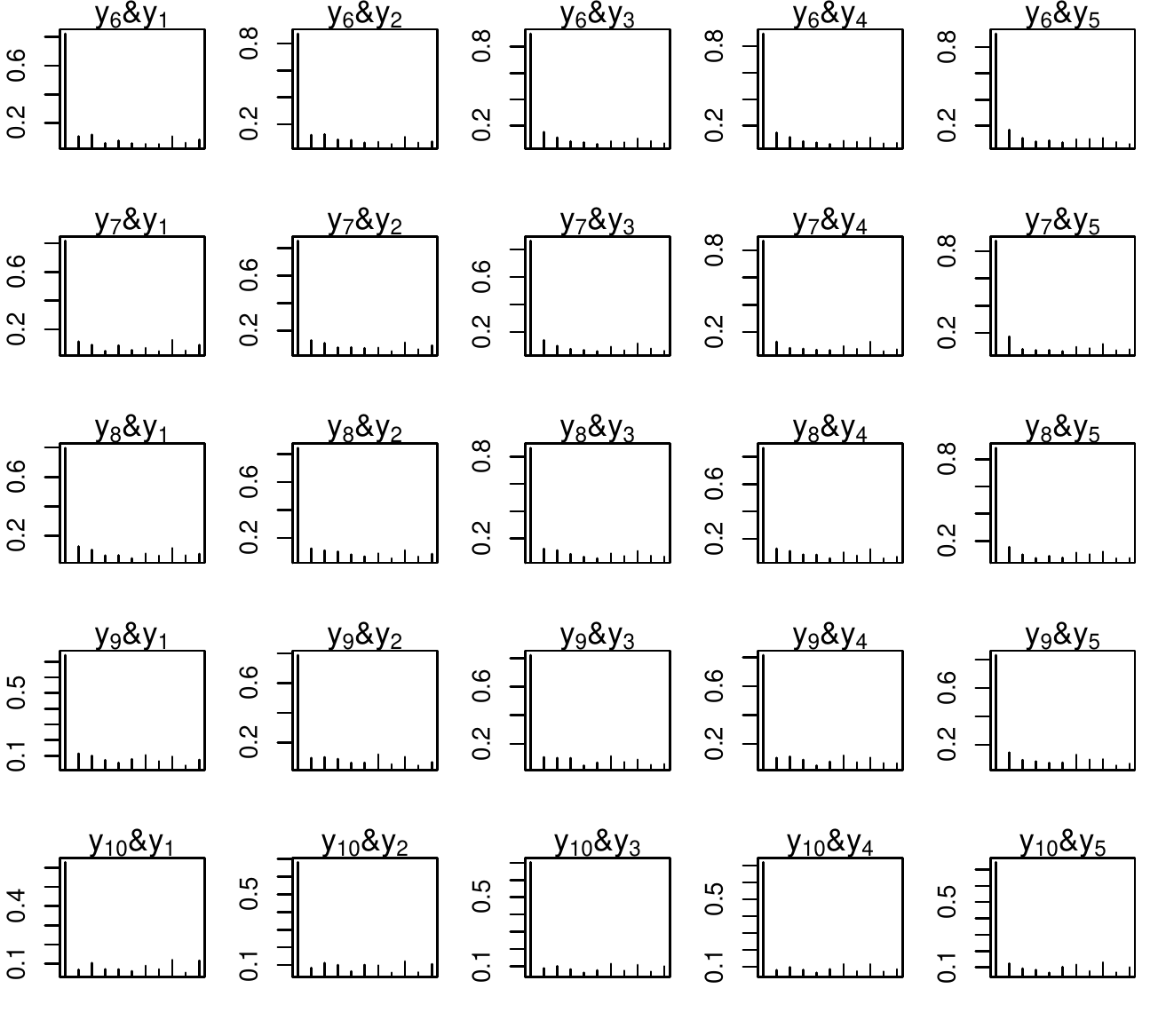}}
{\includegraphics[width=0.48\textwidth]{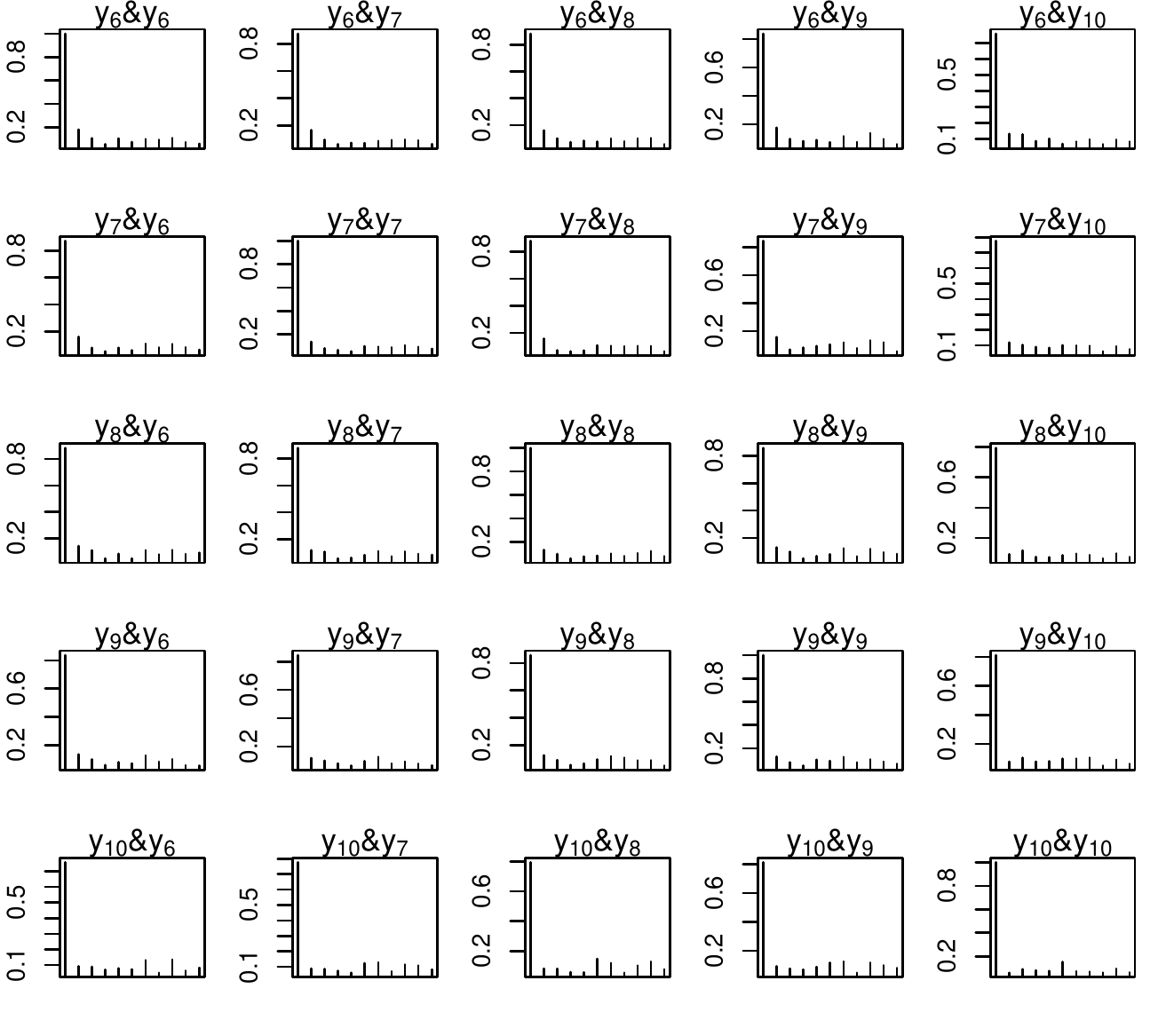}}
\caption{Cross correlogram of $\bY_t$ in Example 4.}\label{fig8}
\end{center}
\end{figure}

\begin{figure}
\begin{center}
{\includegraphics[width=0.48\textwidth]{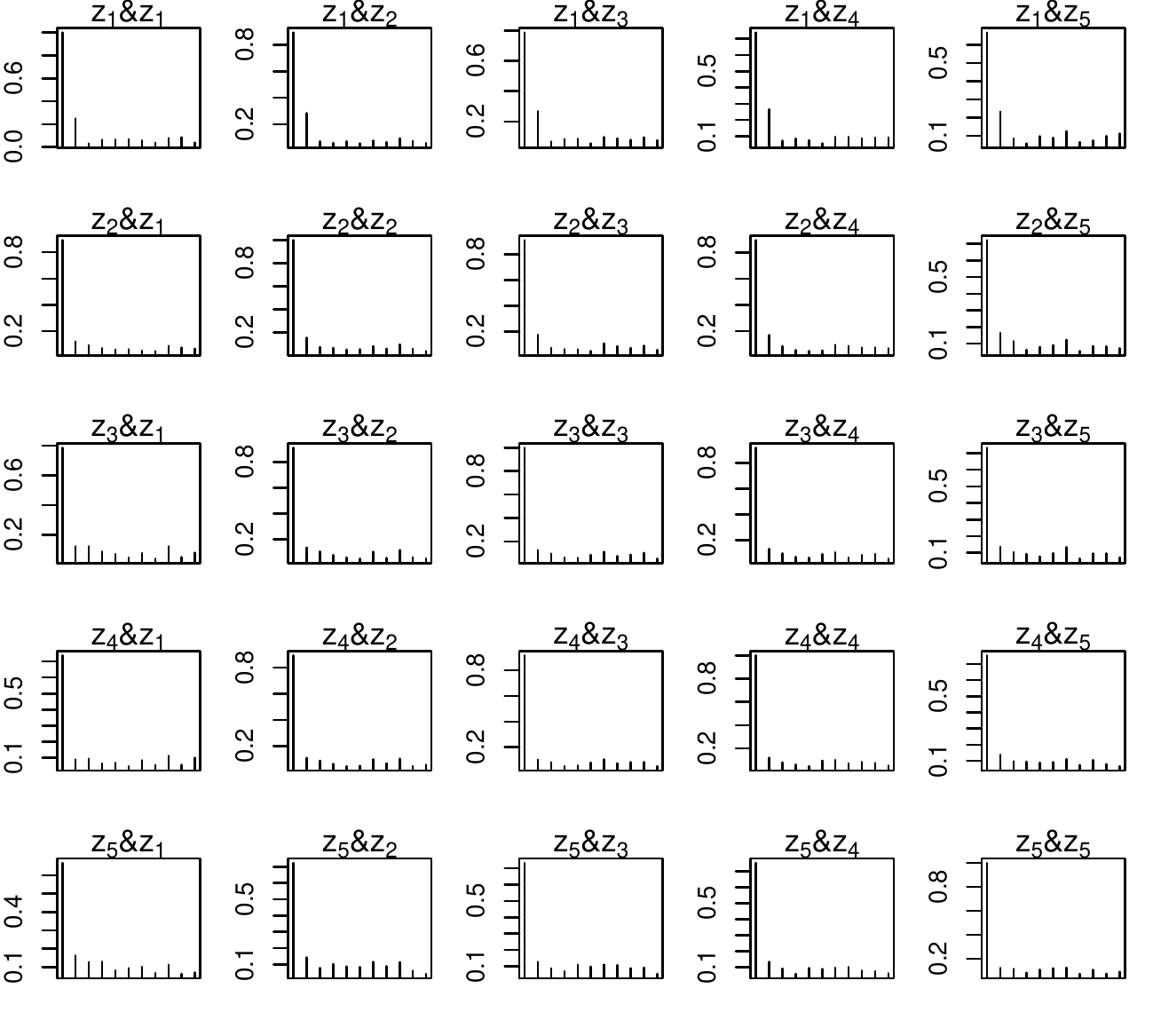}}
{\includegraphics[width=0.48\textwidth]{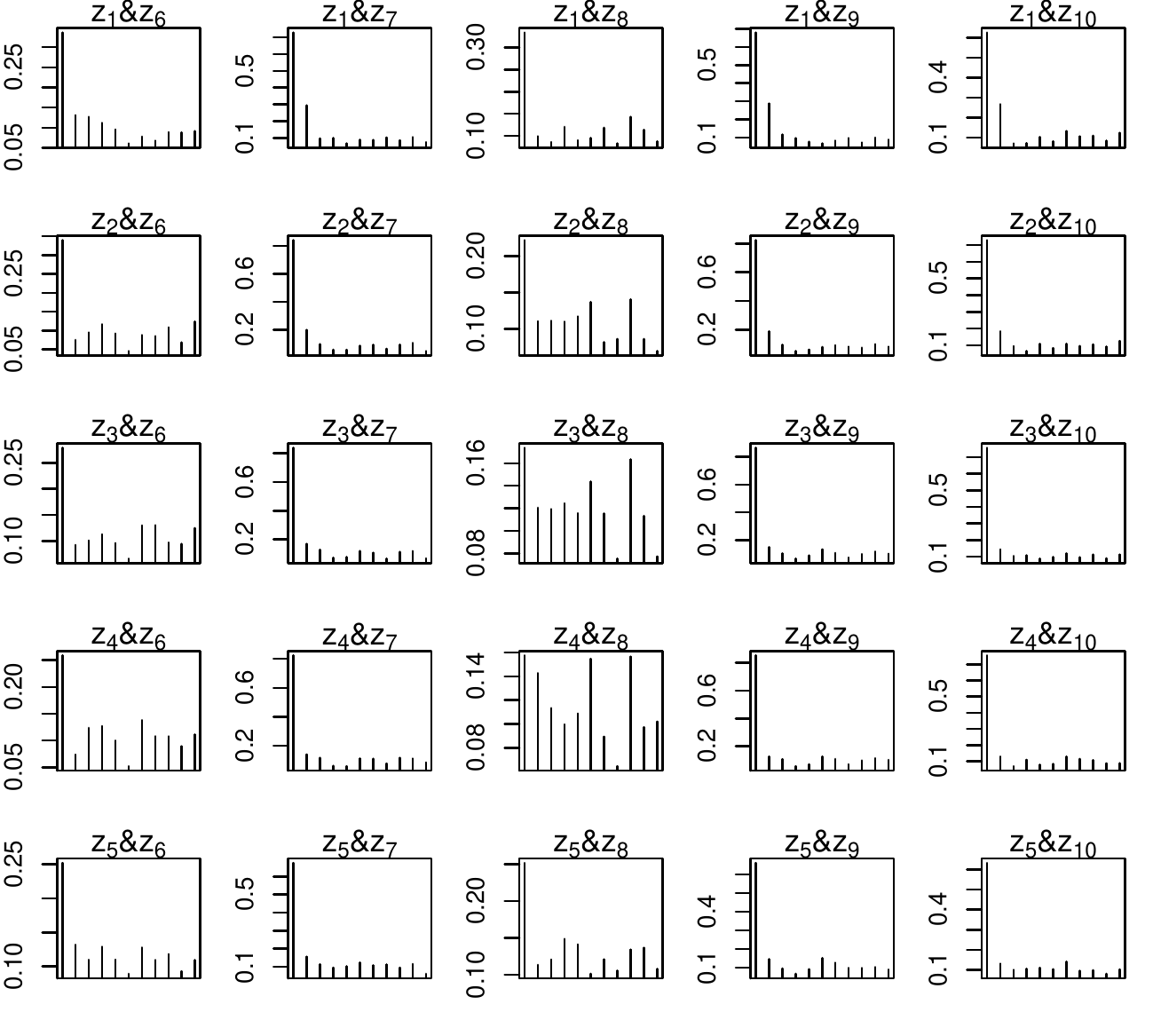}}
{\includegraphics[width=0.48\textwidth]{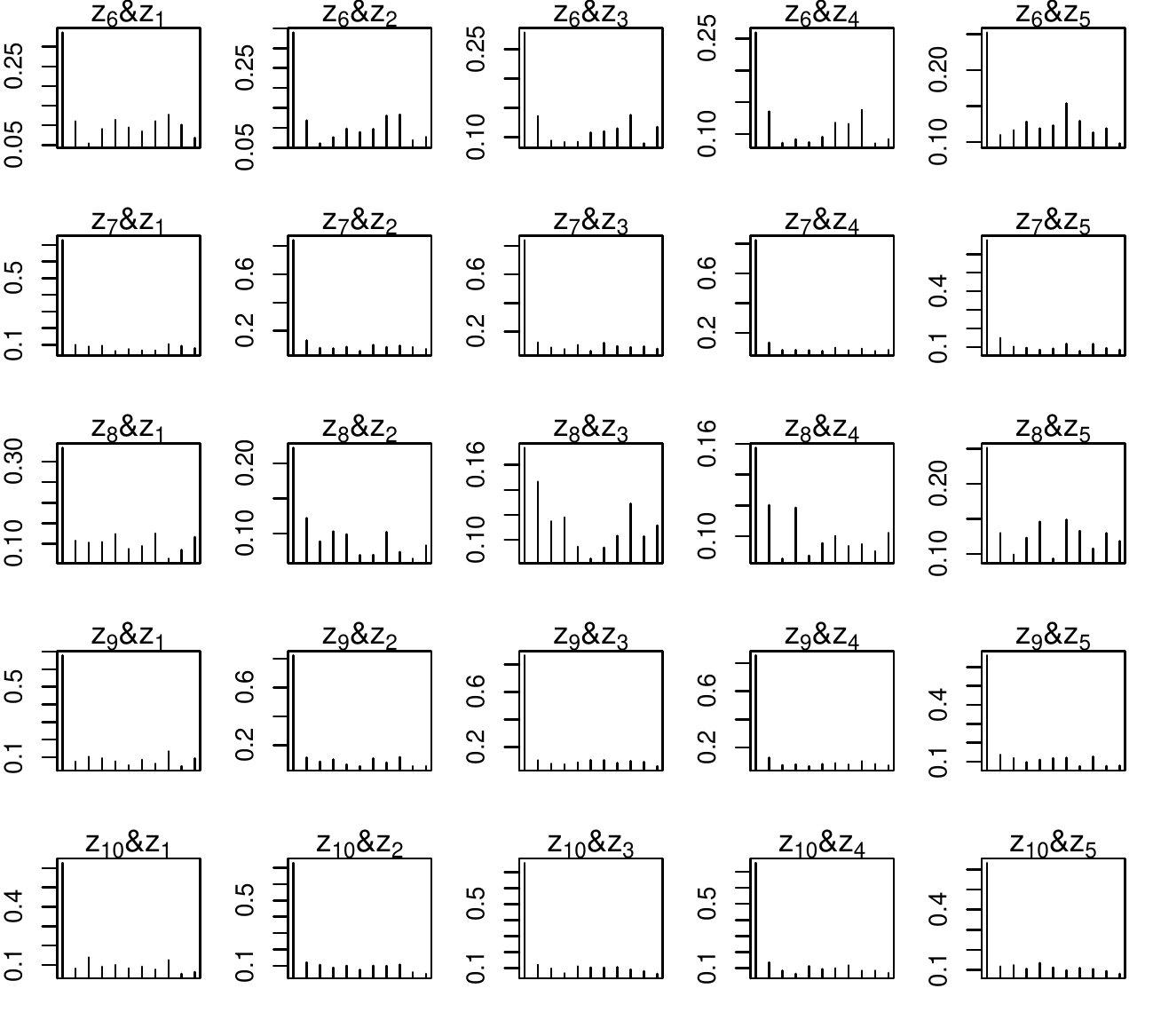}}
{\includegraphics[width=0.48\textwidth]{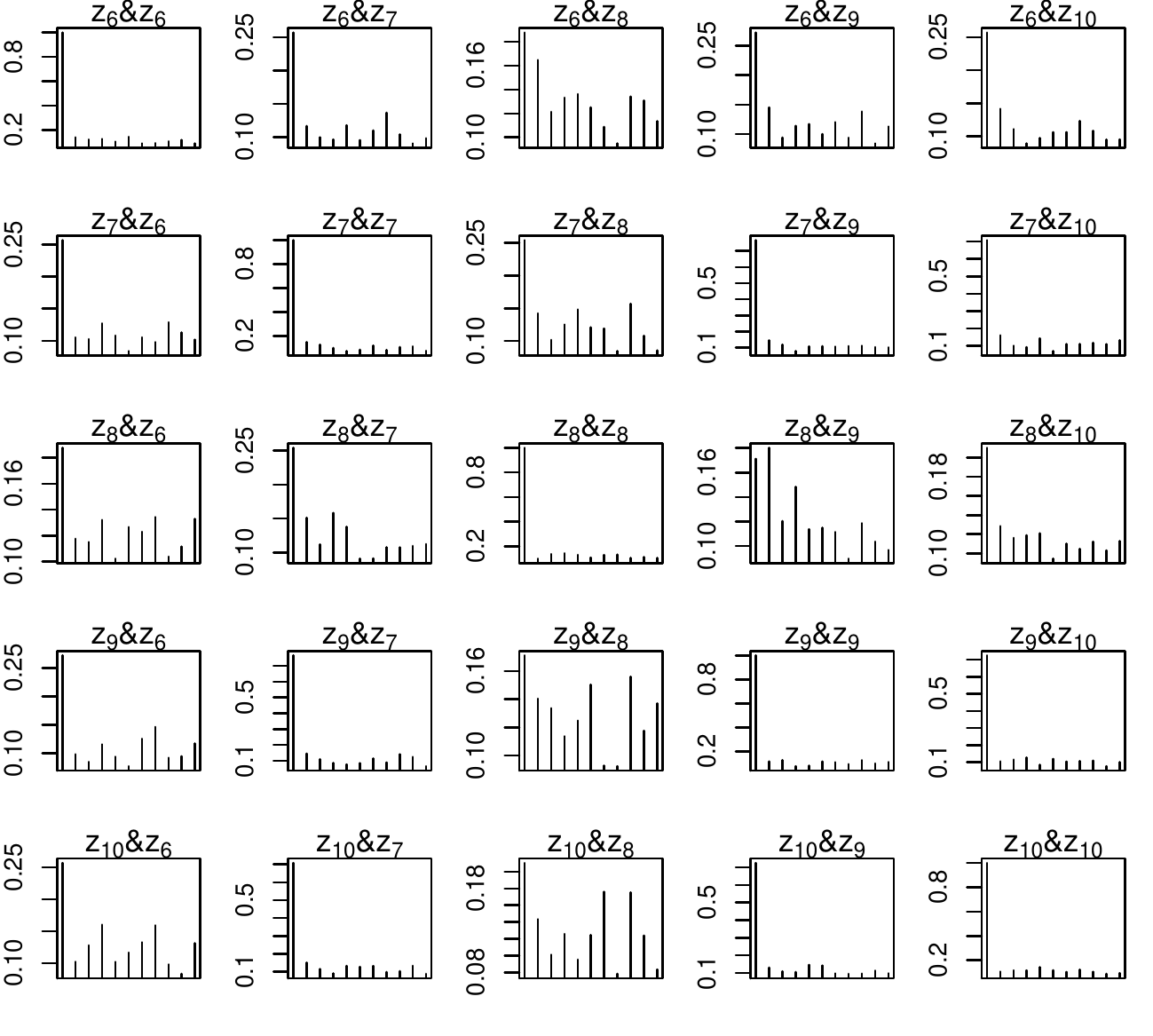}}
\caption{Cross correlogram of the transposed series $\wh \bz_i^t=\bY_t{\wh \bv_i}$ without thresholding in Example 4. The components of $\wh \bZ_t$ can be segmented into 3 groups: $\{1,2,3,4,5,7,9,10\}$, $\{6\}$,and $\{8\}$.}\label{fig9}
\end{center}
\end{figure}
\begin{figure}
\begin{center}
{\includegraphics[width=0.48\textwidth]{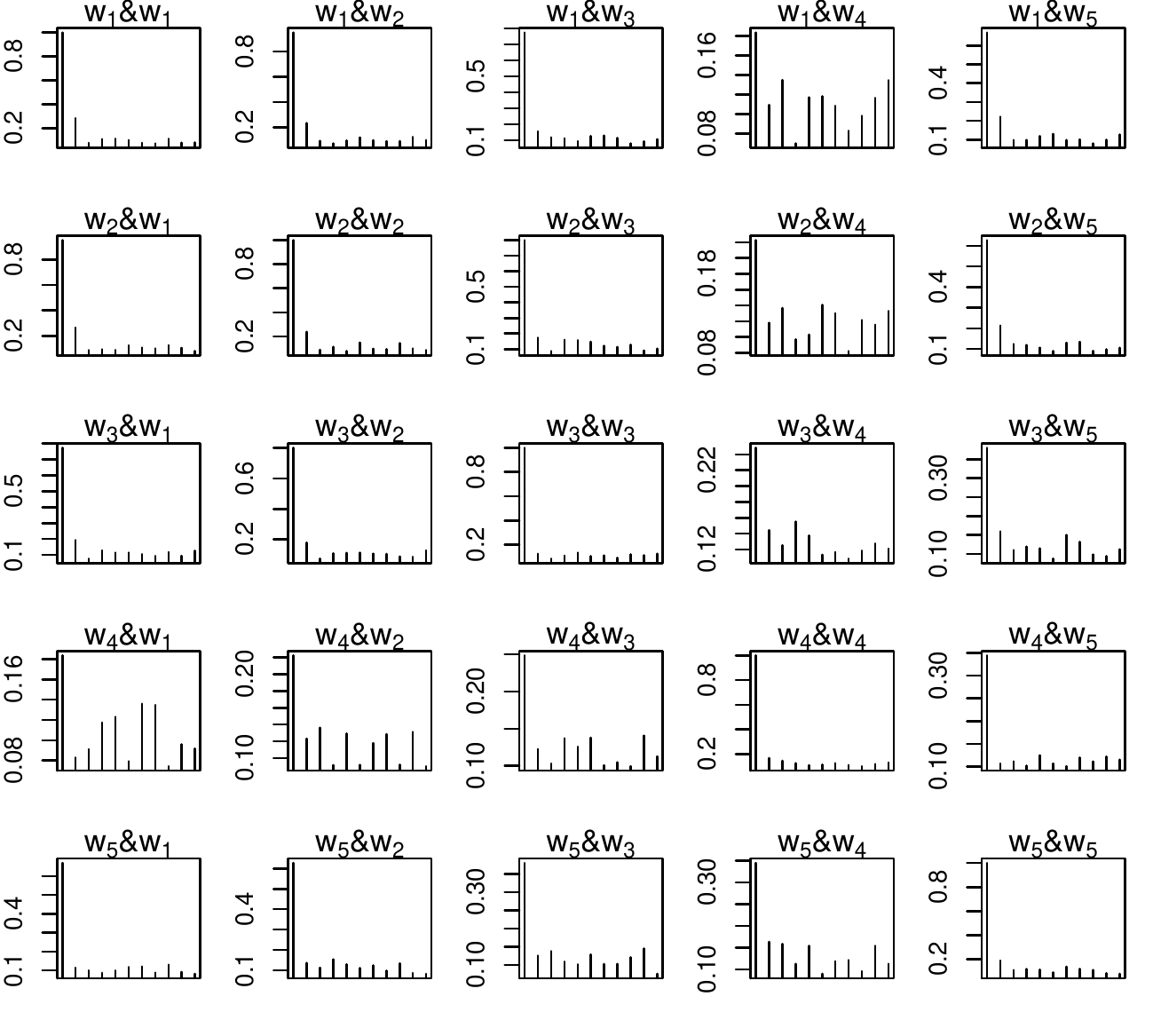}}
{\includegraphics[width=0.48\textwidth]{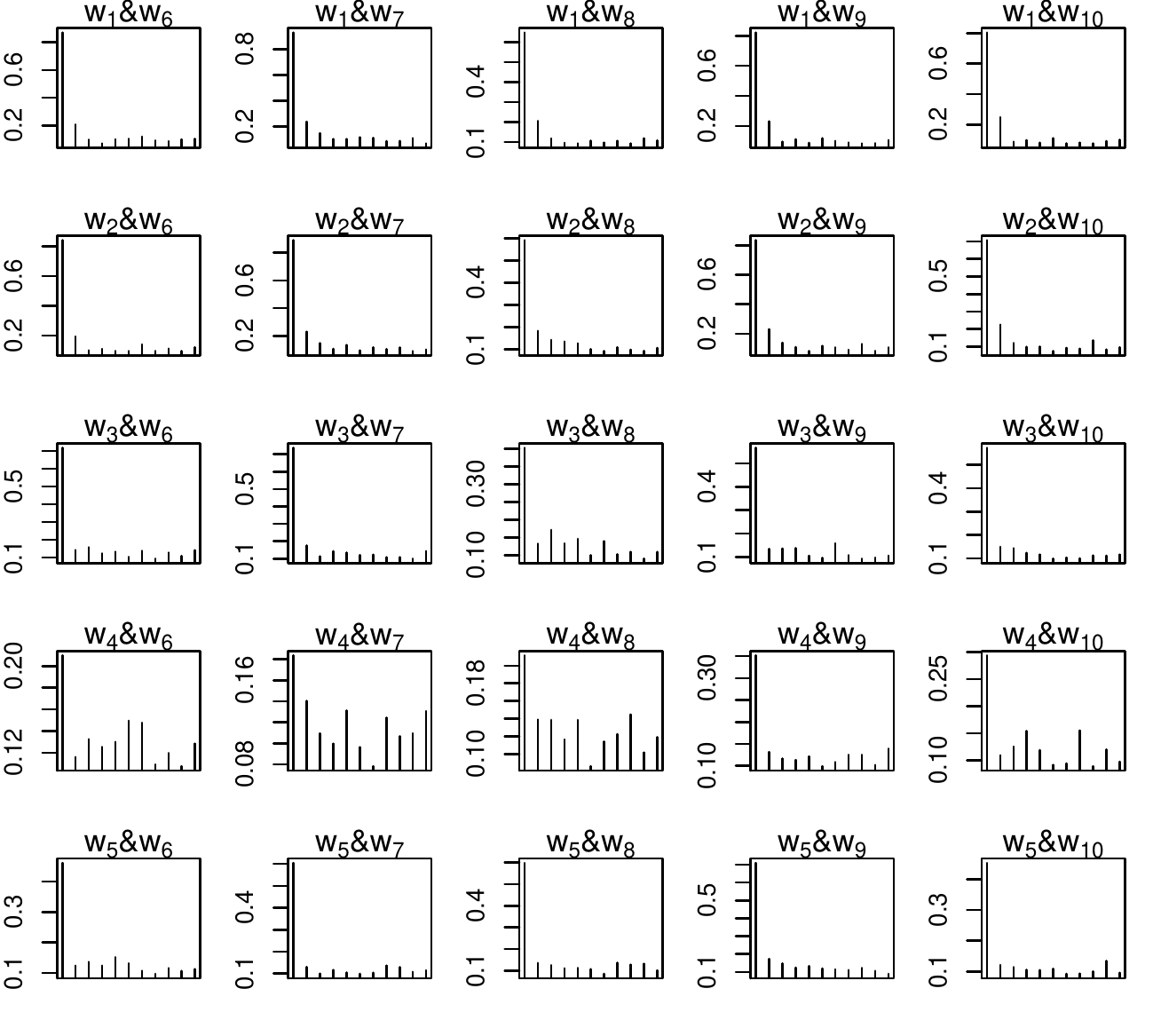}}
{\includegraphics[width=0.48\textwidth]{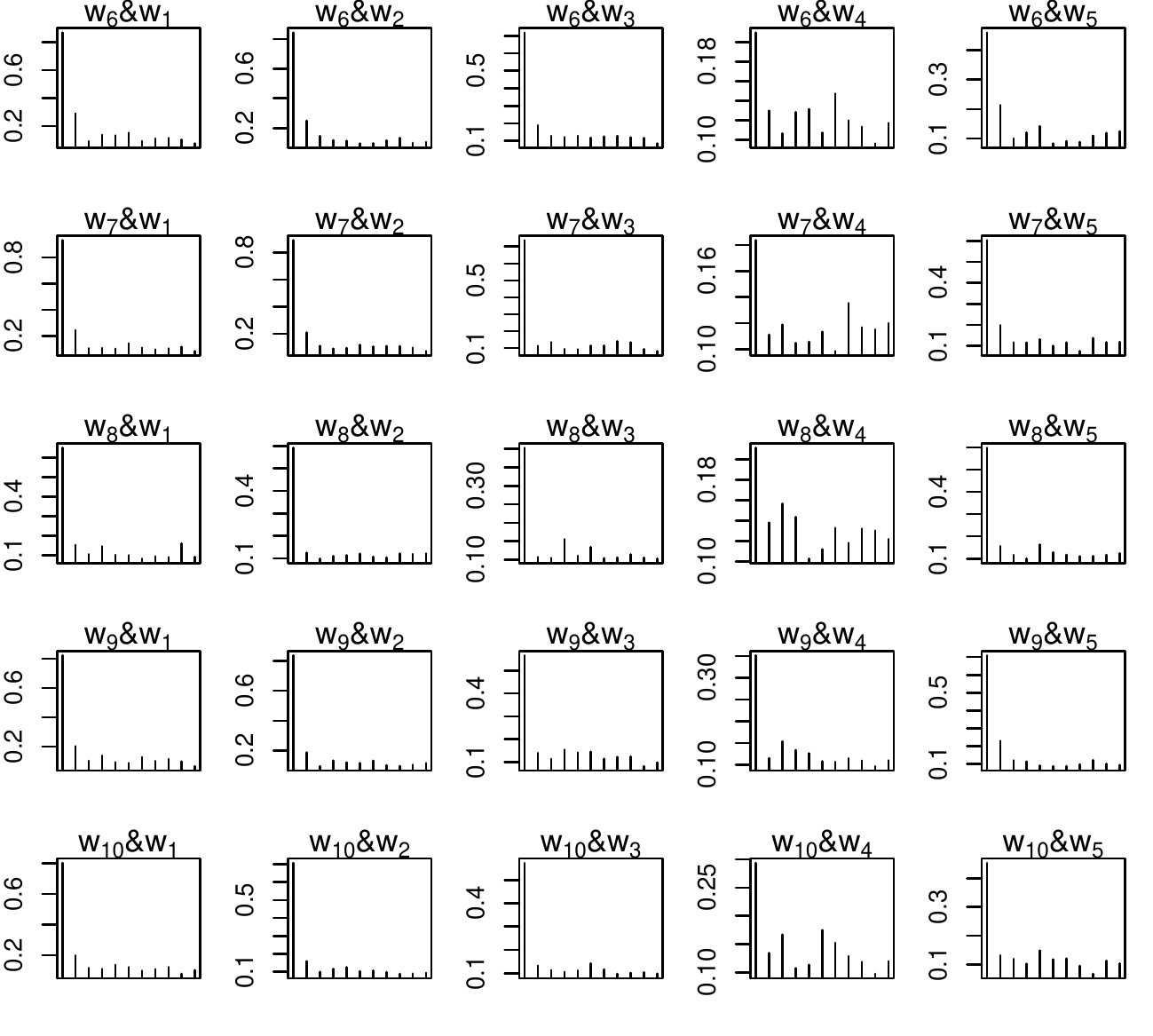}}
{\includegraphics[width=0.48\textwidth]{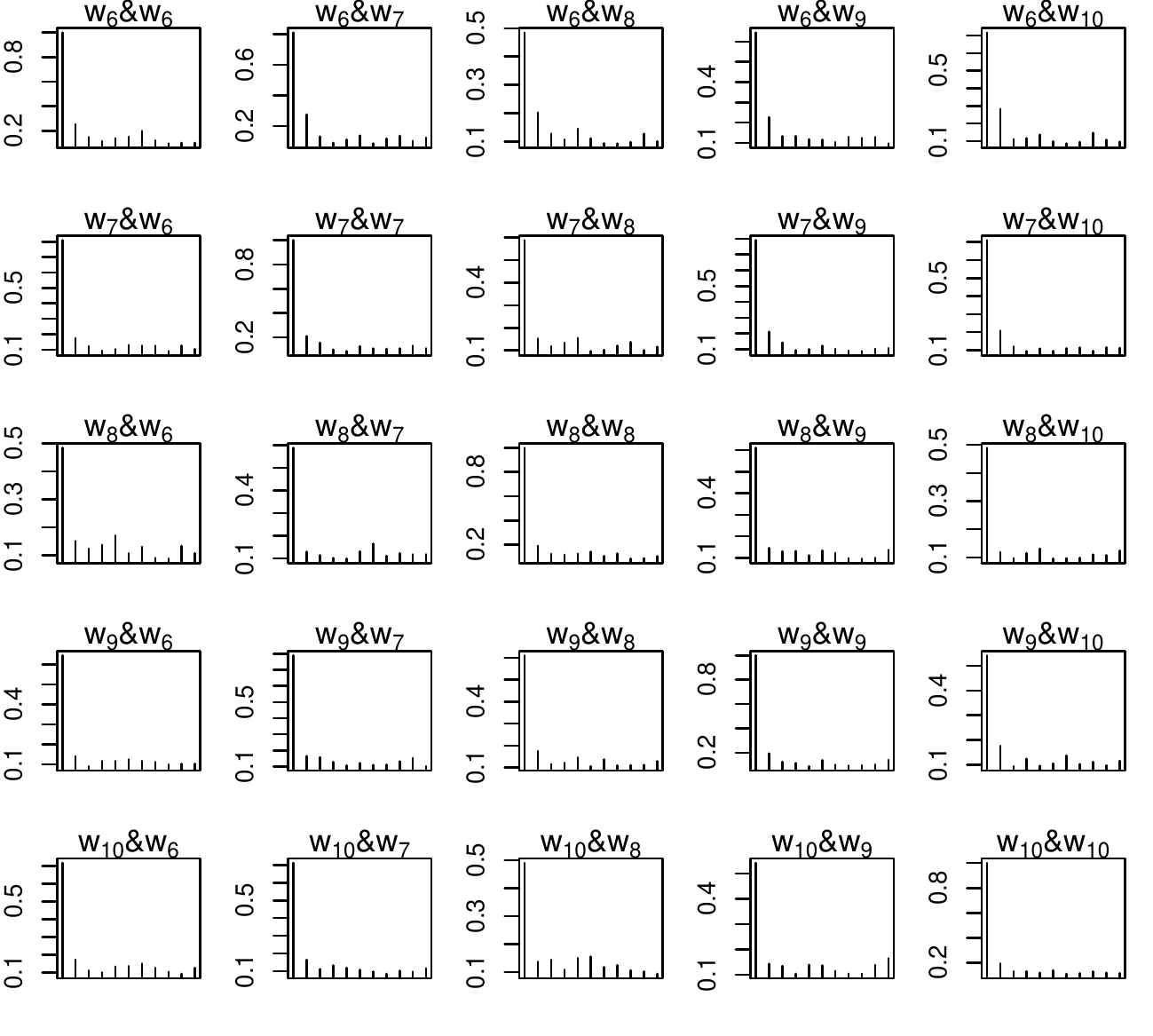}}
\caption{Cross correlogram of the transposed series $\wh \bw_i^t=\bZ_t{\wh \bv_i}$ without thresholding in Example 4. The components of $\wh \bW_t$ can be segmented into 2 groups: $\{1,2,3,5,6,7,8,9,10\}$ and $\{4\}$.}\label{fig10}
\end{center}
\end{figure}
We now apply our method in section 2 to $\bY_t$ using the thresholded estimators, and the correlations of the columns of the transformed data $\bZ_t=\bY_t\hat{\Gamma}_y$ are shown in Figure 11. By adopting the criterion in (\ref{rhat:md}), we find that the threshold is $0.35$ and those with maximum correlations less than $0.35$ are taken to be $0$. From Figure 11, we can see that the columns of $\wh \bZ_t$ can be divided into $8$ groups: $\{1,4,9\}$, $\{2\}$, $\{3\}$, $\{5\}$, $\{6,\}$, $\{7\}$, $\{8\}$ and $\{10\}$. Then, $\wh\bX_t$ is of the form
\begin{equation}\label{bx}
\wh\bX_t=\{(\wh\bz_1^t,\wh\bz_4^t,\wh\bz_9^t),(\wh\bz_2^t),(\wh\bz_3^t),(\wh\bz_5^t),(\wh\bz_6^t),(\wh\bz_7^t),(\wh\bz_8^t),(\wh\bz_{10}^t)\}.
\end{equation}

We next apply our method to the data of $\bZ_t^\T$, as indicated by the sequential transformation algorithm in section 4. By an abuse of notation, we still denote the standardized data as $\wt\bZ_t$, and we note that the correlation structure of the rows are essentially the same with (\ref{bx}). Figure 12 shows the correlations of the columns of $\bW_t=\wt\bZ_t\wh \bGamma_{\wt z}$ and we can see all the correlations are very small or close to $0$. In fact, there are only $14$ pairs which have nonzero correlations and the maximum is $0.1937$. If we choose $0.1937$ as a threshold, then every column can be treated as uncorrelated with each other and they can be segmented into $10$ groups. Instead we still use the criterion of (\ref{rhat:md}) and find the threshold level as $0.1481$, then they can be divided into $6$ groups: $\{1,3,8,10\}$, $\{2,6\}$, $\{4\}$, $\{5\}$, $\{7\}$ and $\{9\}$. In the end, we combine the results of Figure 11 and Figure 12, and rearrange the columns of $\bW_t$, then the segmented matrix $\wh\bW_t$ is of the form

\begin{figure}
\begin{center}
{\includegraphics[width=0.48\textwidth]{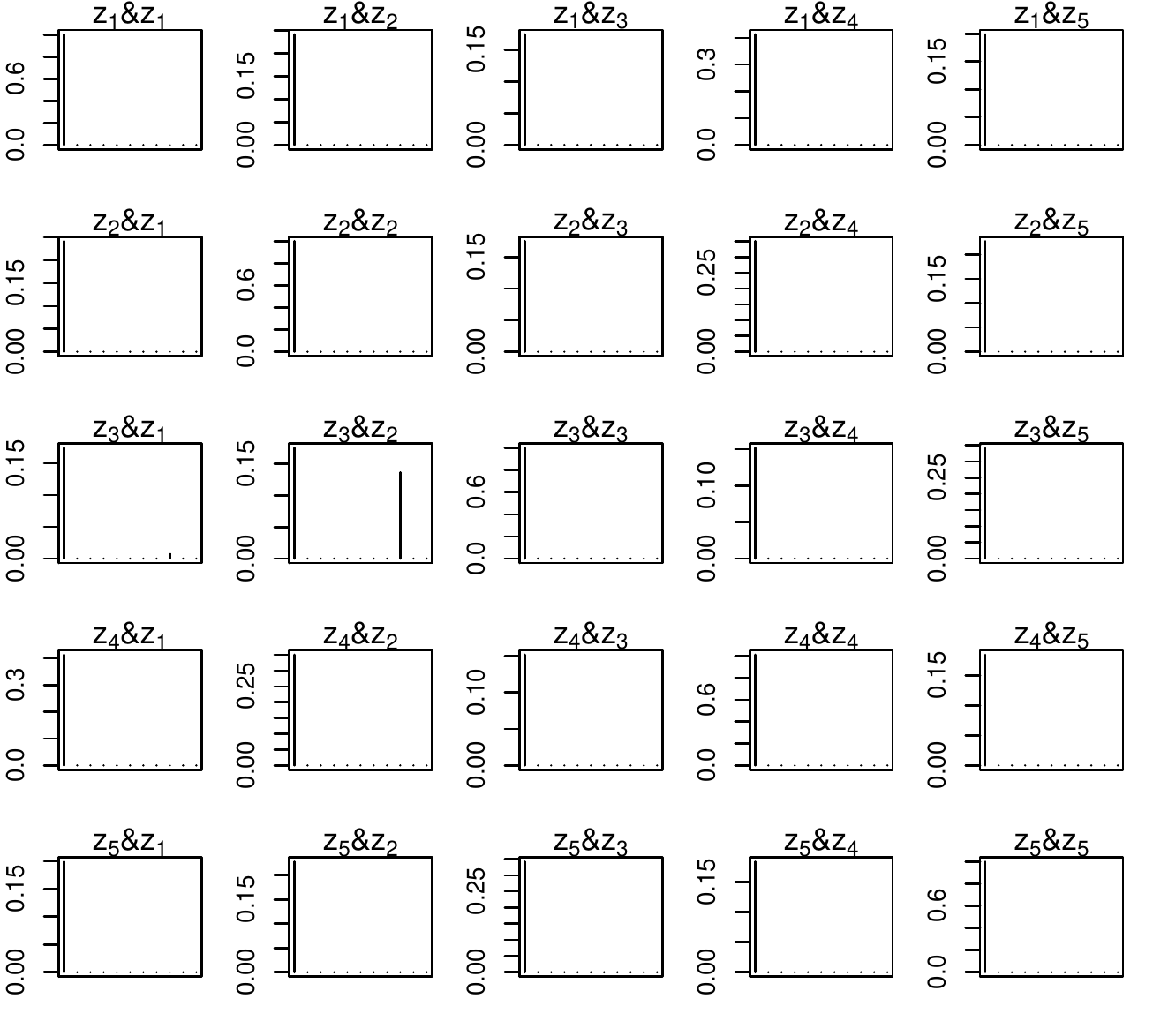}}
{\includegraphics[width=0.48\textwidth]{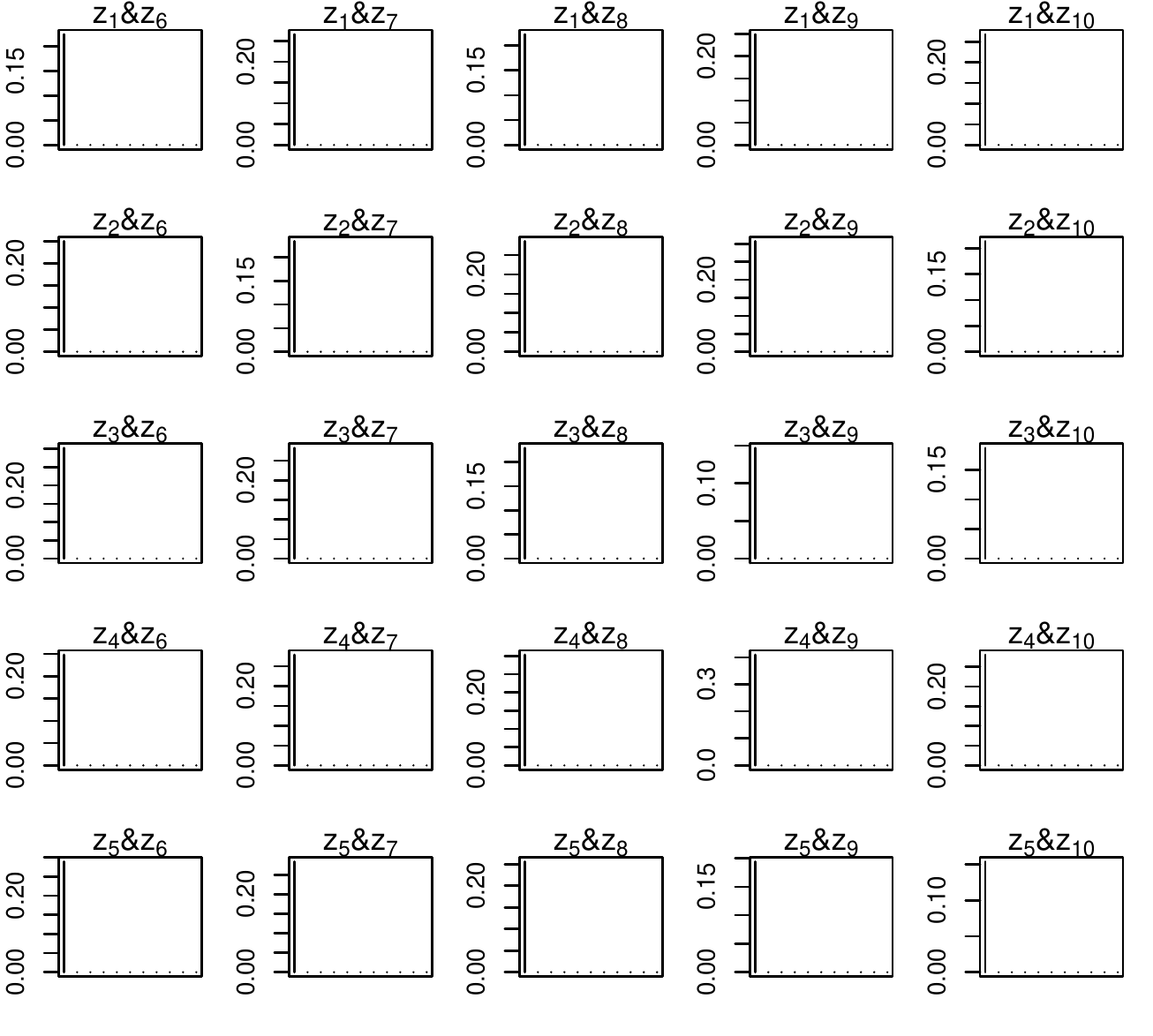}}
{\includegraphics[width=0.48\textwidth]{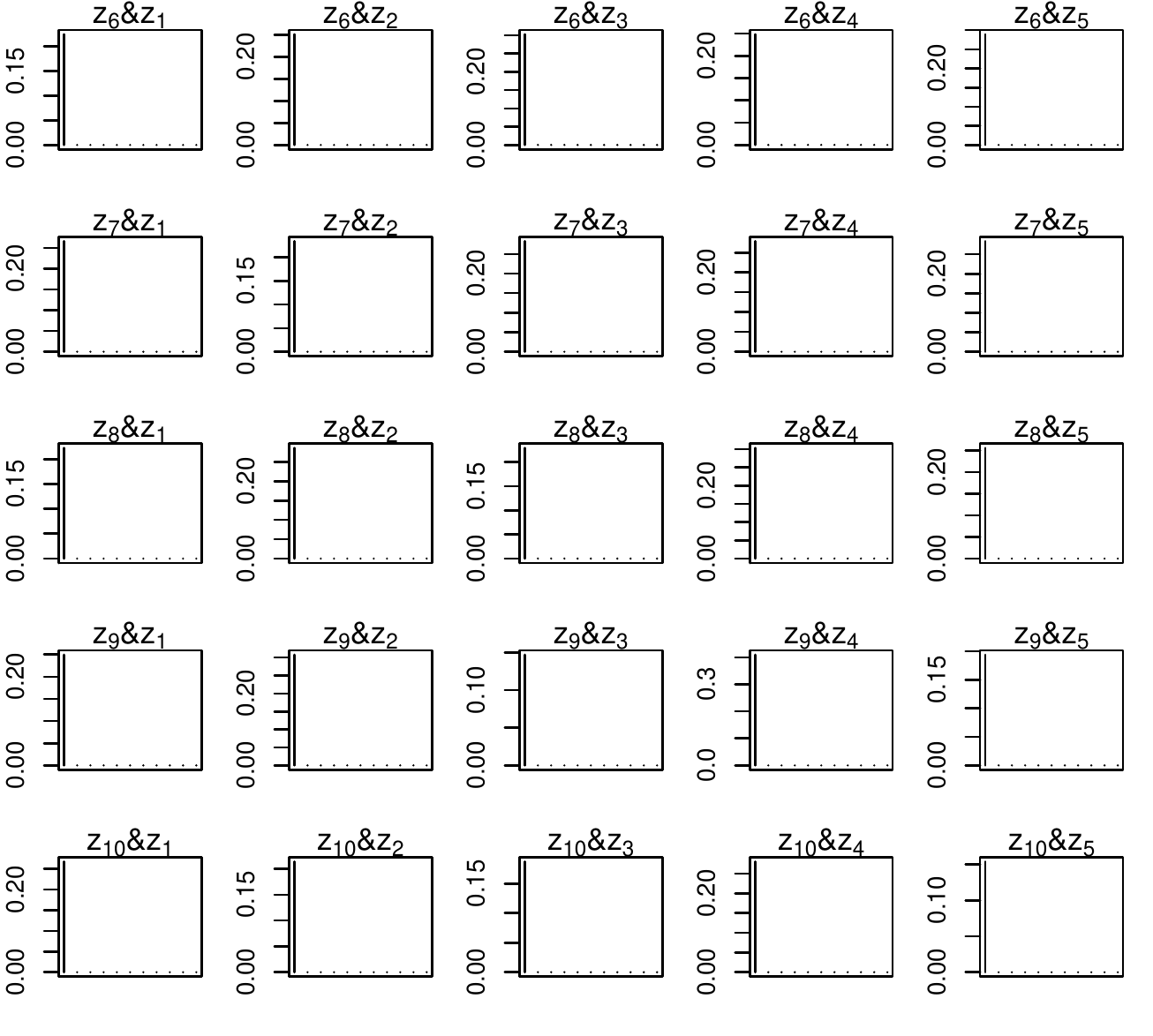}}
{\includegraphics[width=0.48\textwidth]{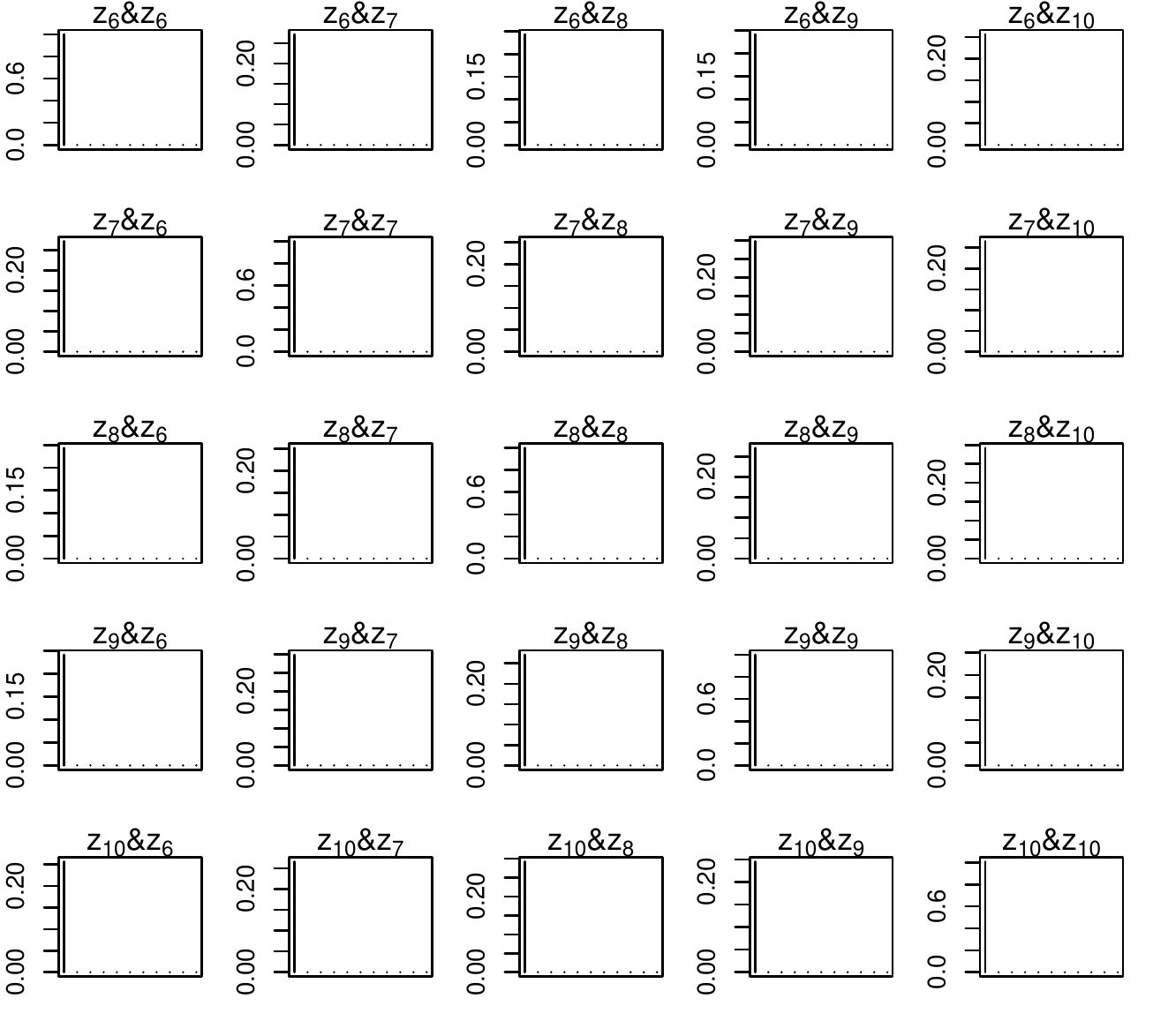}}
\caption{Cross correlogram of the transposed series $\wh \bz_i^t=\bY_t{\wh \bv_i}$ with thresholding in Example 4. The components of $\wh \bZ_t$ can be segmented into 8 groups: $\{1,4,9\}$, $\{2\}$, $\{3\}$, $\{5\}$, $\{6\}$, $\{7\}$, $\{8\}$ and $\{10\}$.}\label{fig11}
\end{center}
\end{figure}
\begin{figure}
\begin{center}
{\includegraphics[width=0.48\textwidth]{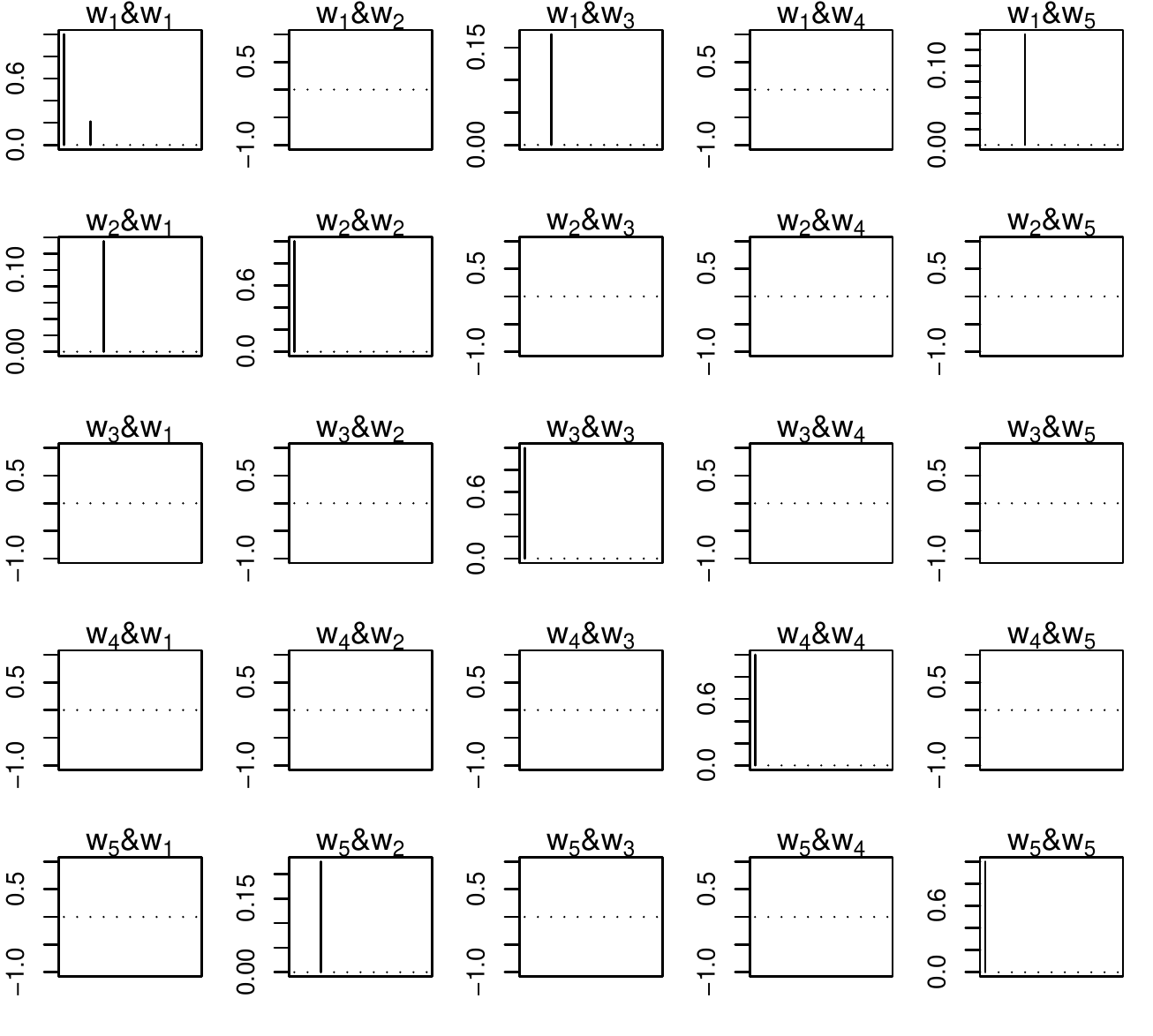}}
{\includegraphics[width=0.48\textwidth]{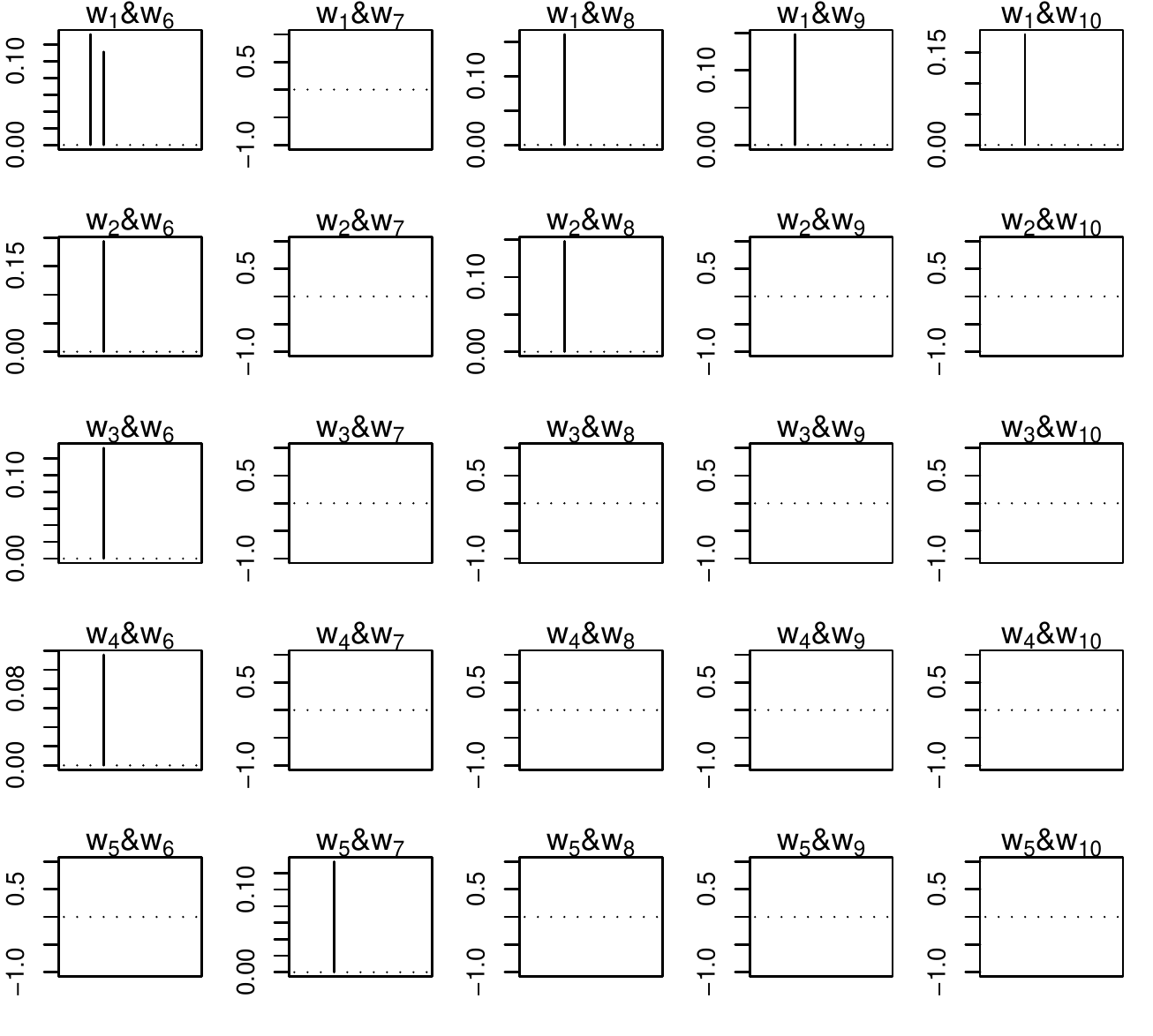}}
{\includegraphics[width=0.48\textwidth]{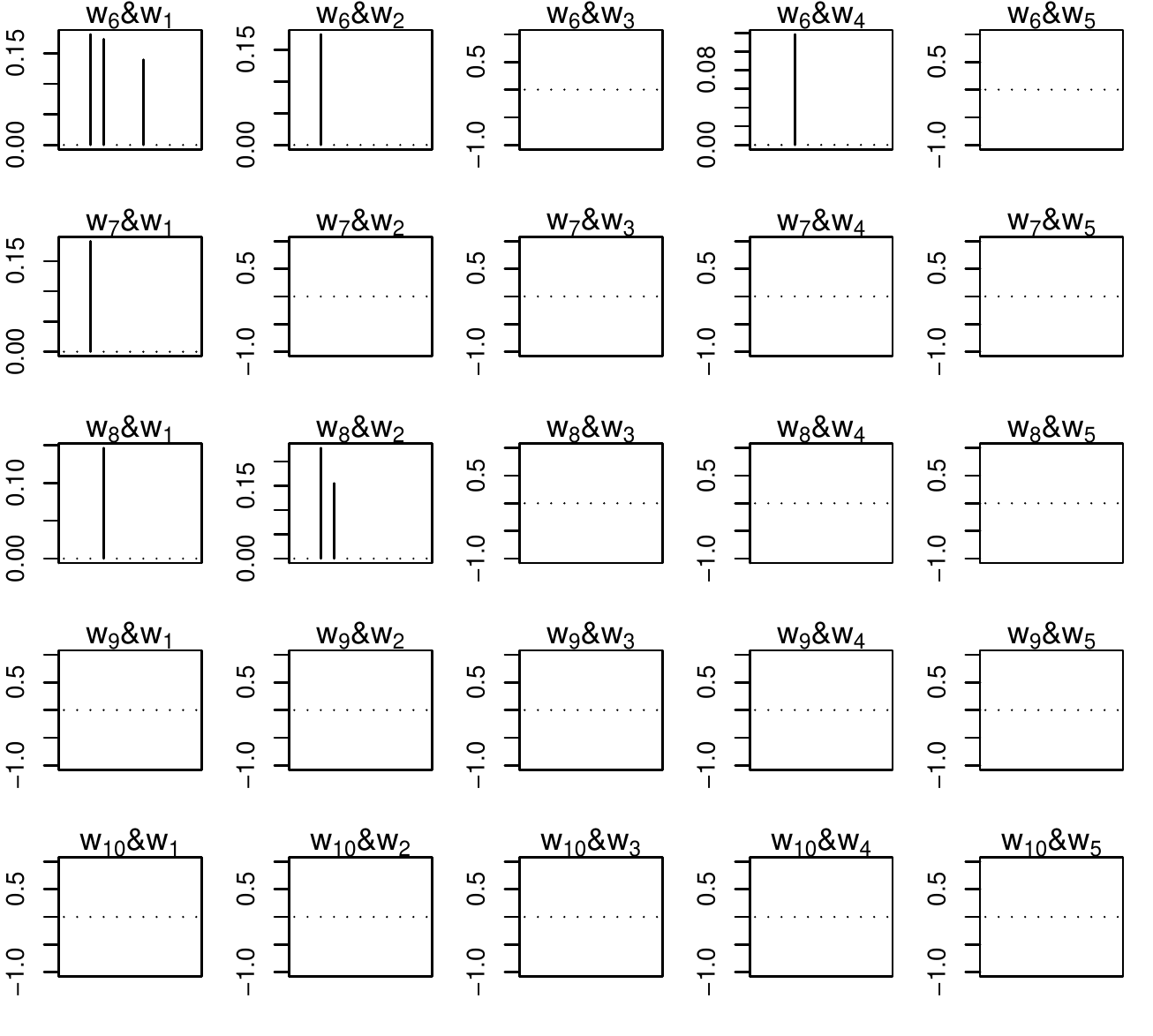}}
{\includegraphics[width=0.48\textwidth]{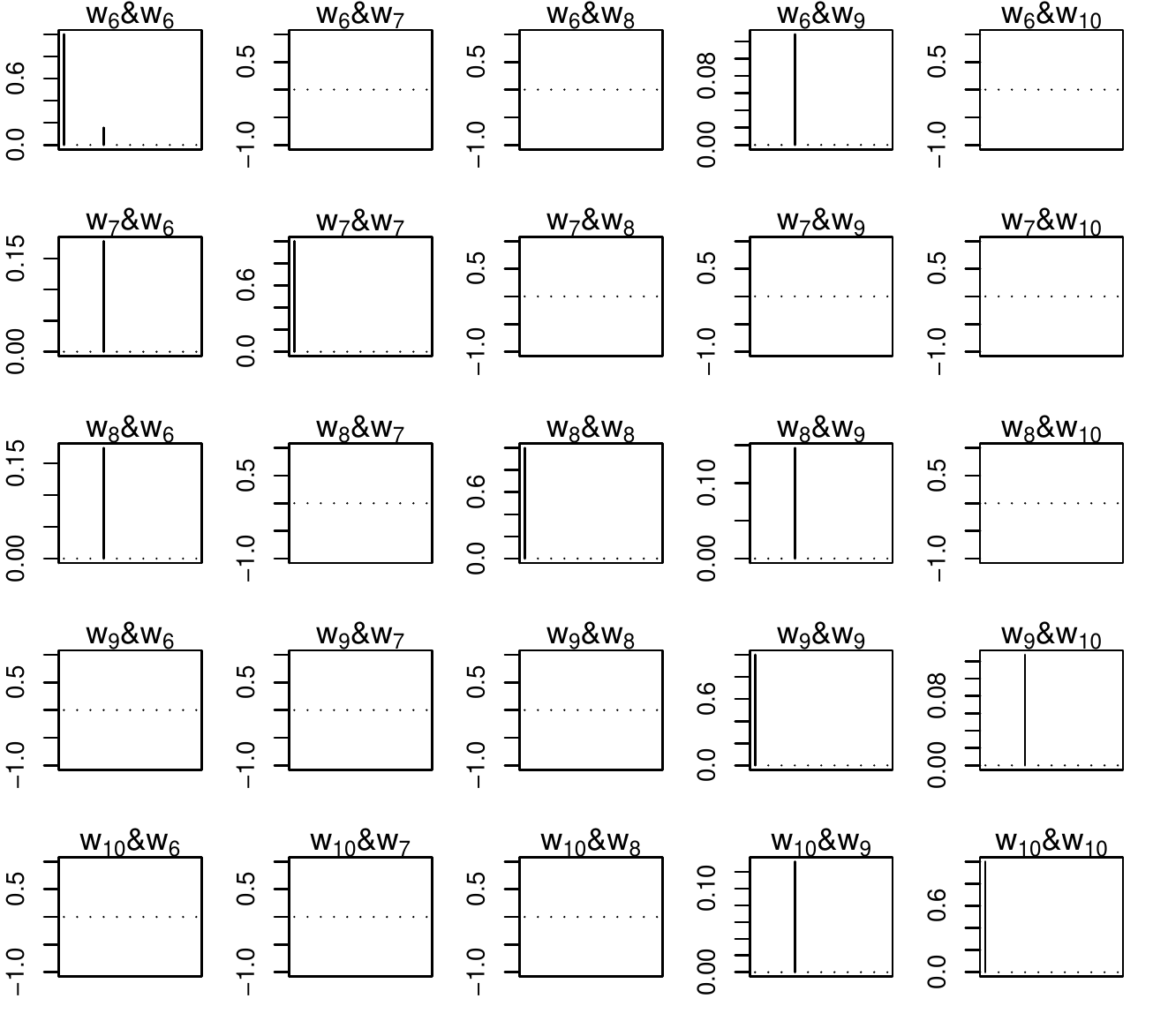}}
\caption{Cross correlogram of the transposed series $\wh \bw_i^t=\bZ_t{\wh \bv_i}$ with thresholding in Example 4. The components of $\wh \bW_t$ can be segmented into 6 groups: $\{1,3,8,10\}$, $\{2,6\}$, $\{4\}$, $\{5\}$, $\{7\}$ and $\{9\}$.}\label{fig12}
\end{center}
\end{figure}

\[
\wh\bW_t=
\left[
\begin{array}{cccc|cc|c|c|c|c}
w_{1,1} & w_{1,2} &w_{1,3} &w_{1,4} &w_{1,5} &w_{1,6} &w_{1,7} &w_{1,8} &w_{1,9} &w_{1,10} \\
w_{2,1} & w_{2,2} &w_{2,3} &w_{2,4} &w_{2,5} &w_{2,6} &w_{2,7} &w_{2,8} &w_{2,9} &w_{2,10} \\
w_{3,1} & w_{3,2} &w_{3,3} &w_{3,4} &w_{3,5} &w_{3,6} &w_{3,7} &w_{3,8} &w_{3,9} &w_{3,10} \\
\hline
w_{4,1} & w_{4,2} &w_{4,3} &w_{4,4} &w_{4,5} &w_{4,6} &w_{4,7} &w_{4,8} &w_{4,9} &w_{4,10}\\
\hline
w_{5,1} & w_{5,2} &w_{5,3} &w_{5,4} &w_{5,5} &w_{5,6} &w_{5,7} &w_{5,8} &w_{5,9} &w_{5,10} \\
\hline
w_{6,1} & w_{6,2} &w_{6,3} &w_{6,4} &w_{6,5} &w_{6,6} &w_{6,7} &w_{6,8} &w_{6,9} &w_{6,10} \\
\hline
w_{7,1} & w_{7,2} &w_{7,3} &w_{7,4} &w_{7,5} &w_{7,6} &w_{7,7} &w_{7,8} &w_{7,9} &w_{7,10}\\
\hline
w_{8,1} & w_{8,2} &w_{8,3} &w_{8,4} &w_{8,5} &w_{8,6} &w_{8,7} &w_{8,8} &w_{8,9} &w_{8,10} \\
\hline
w_{9,1} & w_{9,2} &w_{9,3} &w_{9,4} &w_{9,5} &w_{9,6} &w_{9,7} &w_{9,8} &w_{9,9} &w_{9,10} \\
\hline
w_{10,1} & w_{10,2} &w_{10,3} &w_{10,4} &w_{10,5} &w_{10,6} &w_{10,7} &w_{10,8} &w_{10,9} &w_{10,10} 
\end{array}
\right]_t,
\]
where each block can be modeled separately if their dynamic structures are concerned.
From the above analysis, we can see that our method could provide a substantial dimension reduction for matrix- and tensor-valued time series.


  \section*{Appendix: Proofs}
   \renewcommand{\theequation}{A.\arabic{equation}}
   \setcounter{equation}{0}
We use $C$ as a generic constant whose value may change at different places.
\begin{lemma}\label{lm1}
  Under Assumptions 1-2, if $p$ and $q$ are fixed, then for each $k\leq k_0$, as $n\rightarrow\infty$,
  $$\|\hat{\bSigma}_y(k)-\bSigma_y(k)\|_2=O_p(n^{-1/2})\quad\text{and}\quad \|\hat{\bS}-\bW_y\|_2=O_p(n^{-1/2}),$$
  where $\hat{\bS}$ is defined in (\ref{shat:fixed}).
\end{lemma}

{\bf Proof:} Denote $\hat{\sigma}_{i,j}^{(k)}$ and $\sigma_{i,j}^{(k)}$, respectively, the $(i,j)$-th element of $\hat{\bSigma}_y(k)$ and $\bSigma_y(k)$. Without loss of generality, for each $i,j\in\{1,...,q\}$, we assume $\bmu_i=\bmu_j=\mathbf{0}$. Then
\begin{align}\label{sig:dif}
  \hat{\sigma}_{i,j}^{(k)}-\sigma_{i,j}^{(k)} =& \frac{1}{np}\sum_{t=1}^{n-k}(\by_i^{t+k}-\bar{\by}_i)^\T(\by_j^t-\bar{\by}_j)-\frac{1}{p}E\by_i^{t+k,\T}\by_j^t\notag \\
  =&\frac{1}{np}\sum_{t=1}^{n-k}(\by_i^{t+k,\T}\by_j^t-E\by_i^{t+k,\T}\by_j^t)-\frac{1}{np}\sum_{t=1}^{n-k}\by_i^{t+k,\T}\bar{\by}_j\notag\\
&-\frac{1}{np}\sum_{t=1}^{n-k}\bar{\by}_i^{\T}\by_j^t+\frac{n-k}{np}\bar{\by}_i^{\T}\bar{\by}_j-\frac{k}{np}E\by_i^{t+k,\T}\by_j^t\notag\\
=&I_1+I_2+I_3+I_4+I_5,
\end{align}
where $\bar{\by}_i=n^{-1}\sum_{t=1}^n\by_t^t$ and similarly for $\bar{\by}_j$. Note that, by Minkowski inequality,
\begin{equation}\label{meaneq}
  \left[E\left|\frac{1}{p}\by_i^{t+k,\T}\by_j^t\right|^2\right]^{1/2}=\left[E\left|\frac{1}{p}\sum_{l=1}^py_{l,i}^ty_{l,j}^t\right|^2\right]^{1/2}
  \leq \frac{1}{p}\sum_{l=1}^p(E|y_{l,i}^ty_{l,j}^t|^2)^{1/2}<C,
\end{equation}
where the last inequality follows from Assumption \ref{a1}. By (\ref{meaneq}) and Assumption \ref{a2}, following a similar argument as Lemma A.2 in \cite{changguoyao2015}, we have
$$I_1=O_p(n^{-1/2}),I_2=O_p(n^{-1}), I_3=O_p(n^{-1}),I_3=O_p(n^{-1}),I_4=O_p(n^{-1}),I_5=O_p(n^{-1}).$$
Note that $q$ is fixed, which implies the first one in Lemma \ref{lm1}. For the second one, note that
\begin{equation}\label{sw}
  \hat{\bS}-\bW_y=\sum_{k=1}^{k_0}(\hat{\bSigma}_y(k)-\bSigma_y(k))\hat{\bSigma}_y(k)^\T+\sum_{k=1}^{k_0}\bSigma_y(k)(\hat{\bSigma}_y(k)-\bSigma_y(k))^\T.
\end{equation}
Then the second one follows from the first one. This completes the proof. $\Box$

{\bf Proof of Theorem 1.}  (i). Denote $\hat{\bv}_i$ the $i$-th column of $\hat{\bGamma}_y$, then $\hat{\bz}_i^t=\bY_t\hat{\bv}_i$ for $1\leq i\leq q$. For each pair $(i,j)\in E$ defined in (\ref{graph}), we will show that
\begin{equation}\label{max:dif}
  \max_{|h|\leq m}|\hat{\bGamma}_{i,j}(h)-\bGamma_{i,j}(h)|_{\infty}=O_p(\delta_n),
\end{equation}
where $\delta_n\rightarrow 0$ as $n\rightarrow\infty$ and it will be determined later. Note that the $(k,l)$-th element of $\hat{\bGamma}_{i,j}(h)$ is
\begin{equation}\label{corr:sam}
  \hat{\Corr}(\hat{z}_{k,i}^{t+h},\hat{z}_{l,j}^t):=\frac{\hat{\delta}_{ij,kl}^{(h)}}{\hat{\delta}_{i,k}\hat{\delta}_{j,l}}
\end{equation}
where
$$\hat{\delta}_{ij,kl}^{(h)}=\frac{1}{n}\sum_{t=1}^{n-h}(\hat{z}_{k,i}^{t+h}-\bar{\hat{z}}_{k,i})(\hat{z}_{l,j}^{t}-\bar{\hat{z}}_{l,j}),\quad \hat{\delta}_{i,k}^2=\hat{\delta}_{ii,kk}^{(0)},$$
and $\bar{\hat{z}}_{k,i}$ is the sample mean of $\hat{z}_{k,i}^{t}$. By the algorithm in Section 2, we have $\hat{z}_{k,i}^t=\by_{k:}^t\hat{\bv}_i=\hat{\bv}_i^\T\by_{k:}^{t,\T}$ where $\by_{k:}^t$ is the $k$-th row vector of $\bY_t$. Define $\bSigma_{k,l}^{(h)}=\Cov(\by_{k:}^{t+h},\by_{l:}^t)$ and let $\hat{\bSigma}_{k,l}^{(h)}$ be the sample covariance matrix, then
\begin{align}\label{delta:dif}
  \hat{\delta}_{ij,kl}^{(h)}-\delta_{ij,kl}^{(h)}=&\hat{\bv}_i^\T\hat{\bSigma}_{k,l}^{(h)}\hat{\bv}_j-\bv_i^\T\bSigma_{k,l}^{(h)}\bv_j\notag\\
  =&(\hat{\bv}_i-\bv_i)^\T\hat{\bSigma}_{k,l}^{(h)}\hat{\bv}_j+\bv_i^\T(\hat{\bSigma}_{k,l}^{(h)}-\bSigma_{k,l}^{(h)})\hat{\bv}_j+\bv_i^\T\bSigma_{k,l}^{(h)}(\hat{\bv}_j-\bv_j)
\end{align}
By Theorem 8.1.10 in \cite{golub1996} and a similar argument as the proof in \cite{lamyao2012}, we have
\begin{equation}\label{eigenvec}
  \max_{1\leq i\leq q}\|\hat{\bv}_i-\bv_i\|_2\leq C\frac{\|\hat{\bS}-\bW_y\|_2}{\varpi_n}.
\end{equation}
Since $p$ and $q$ are fixed, by a standard argument as that in Lemma \ref{lm1}, we can show that

\begin{equation}\label{sigmax}
  \max_{|h|\leq m}\max_{1\leq k,l\leq p}\|\hat{\bSigma}_{k,l}^{(h)}-\bSigma_{k,l}^{(h)}\|_2\leq\max_{|h|\leq m}\max_{1\leq k,l\leq p}\|\hat{\bSigma}_{k,l}^{(h)}-\bSigma_{k,l}^{(h)}\|_F=O_p(n^{-1/2}).
\end{equation}
For a positive $\varpi_n$, by Lemma \ref{lm1} and (\ref{delta:dif}), we have
\begin{equation}\label{deltanorm}
  \max_{1\leq i,j\leq q}\max_{1\leq k,l\leq p}|\hat{\delta}_{ij,kl}^{(h)}-\delta_{ij,kl}^{(h)}|=O_p(n^{-1/2}).
\end{equation}
and it also holds for $\max_{1\leq i,j\leq q}\max_{1\leq k,l\leq p}|\hat{\delta}_{i,k}^2-{\delta}_{i,k}^2|$. Note that
\begin{equation}\label{deltaik}
  \max_{1\leq i,j\leq q}\max_{1\leq k,l\leq p}|\hat{\delta}_{i,k}^{-1}\hat{\delta}_{j,l}^{-1}-\delta_{i,k}\delta_{j,l}|\leq (\max_{1\leq i\leq q,1\leq k\leq p}|\hat{\delta}_{i,k}^{-1}-\delta_{i,k}^{-1}|)^2+C\max_{1\leq i\leq q,1\leq k\leq p}|\hat{\delta}_{i,k}^{-1}-\delta_{i,k}^{-1}|
\end{equation}
and
\begin{equation}\label{deltainver}
 \hat{\delta}_{i,k}^{-1}-\delta_{i,k}^{-1}=\frac{\hat{\delta}_{i,k}^2-{\delta}_{i,k}^2}{\hat{\delta}_{i,k}^2-{\delta}_{i,k}^2+\delta_{i,k}(\hat{\delta}_{i,k}+{\delta}_{i,k})}.
\end{equation}
Then,
\begin{equation}\label{max:gam}
\max_{1\leq i,j\leq q}\max_{1\leq k,l\leq p}|\frac{\hat{\delta}_{ij,kl}^{(h)}}{\hat{\delta}_{i,k}\hat{\delta}_{j,l}}
-\frac{{\delta}_{ij,kl}^{(h)}}{{\delta}_{i,k}{\delta}_{j,l}}|=O_p(n^{-1/2}).
\end{equation}
Thus, (\ref{max:dif}) follows from (\ref{max:gam}) with $\delta_n=n^{-1/2}$. By the inequality
$$|\max_{|h|\leq m}|\hat{\bGamma}_{i,j}(h)|_\infty-\max_{|h|\leq m}|\bGamma_{i,j}(h)|_\infty|\leq \max_{|h|\leq m}|\hat{\bGamma}_{i,j}(h)-\bGamma_{i,j}(h)|_\infty,$$
we have proved the fact that if $L_i\geq L_j$, then $\hat{L}_i\geq \hat{L}_j$ with probability tending to 1, where $\hat{L}_i$ is the corresponding estimator of the pair for $L_i$ and $\max_{1\leq i\leq q_0}|\hat{L}_i-L_i|=O_p(n^{-1/2})$, where $q_0$ is defined in Section 2.2.2.

We now prove Theorem 1(i). For $j<d$, we have
\begin{equation*}
\frac{\hat{L}_j+C\delta_n}{\hat{L}_{j+1}+C\delta_n}=\frac{\hat{L}_j-L_j+L_j+C\delta_n}{\hat{L}_{j+1}-L_{j+1}+L_{j+1}+C\delta_n}\leq \max_{j<d}\frac{L_j}{L_{j+1}}
\end{equation*}
with probability tending to 1. When $j=d$,
\begin{equation*}
\frac{\hat{L}_d+C\delta_n}{\hat{L}_{d+1}+C\delta_n}=\frac{\hat{L}_d-L_d+L_d+C\delta_n}{\hat{L}_{j+1}+C\delta_n}\rightarrow\infty,
\end{equation*}
and when $j>d$,
\begin{equation*}
\frac{\hat{L}_j+C\delta_n}{\hat{L}_{j+1}+C\delta_n}\rightarrow C>0.
\end{equation*}
Since $\chi_n\delta_n=o_p(\epsilon_n)$, we have
$$\frac{\hat{L}_j+C\delta_n}{\hat{L}_{j+1}+C\delta_n}<\frac{\hat{L}_d+C\delta_n}{\hat{L}_{d+1}+C\delta_n}\quad\text{for}\quad j<d.$$
Therefore, $P(\hat{E}=E)$ for the $\hat{d}$ defined in (\ref{rhat:md}).

(ii). Note that any two blocks of $\bW_{x,i}$ and $\bW_{x,j}$ do not share the same eigenvalues for $1\leq i<j\leq q_1$. Then, by Theorem 8.1.10 in Golub and Van Loan (1996), see also Lam and Yao (2010) and Chang et al. (2017), we have
$$\sup_{1\leq j\leq q_1}D(\mathcal{M}(\hat{\bA}_j),\mathcal{M}(\bA_j))=O_p(\|\hat{\bS}_y-\bW_y\|)=O_p(n^{-1/2}).$$
This competes the proof. $\Box$

The following lemma is a corollary of Theorem 6.2  of \cite{rio2000}, see also Rio (2017, pp 105-106) for details.
\begin{lemma}\label{lm2}
  Let $\gamma>2$ and $(Z_i)_{i>0}$ be a sequence of real-valued and centered random variables and $(\alpha_k)_{k\geq 0}$ be the sequence of strong mixing coefficients defined via $\{Z_i\}$. Denote $S_k=\sum_{i=1}^kZ_i$. Suppose that the strong mixing coefficients satisfy
  $\alpha_k\leq ck^{-a}$ {for some positive constants} $c\geq 1$ and  $a\geq 1$, and
  $$P(|Z_i|>t)\leq t^{-\gamma}\quad\text{for any} \,\, t>0.$$
  Then, for any $r\geq 1$ and any positive $\lambda$, there exists some positive constant $C(a,\gamma)$ such that
  $$P(\sup_{1\leq k\leq n}|S_k|>4\lambda)\leq 4(1+\frac{\lambda^2}{rn\tilde{s}^2})^{-r/2}+4Cnr^{-1}(r/\lambda)^{(a+1)\gamma/(a+\gamma)},$$
  where $$\tilde{s}^2=\sup_{i>0}(EZ_i^2+2\sum_{j>i}|E(Z_iZ_j)|).$$
\end{lemma}
\begin{remark}\label{rm:lemma}
  If $Z_i$ satisfies Assumptions \ref{a4}-\ref{a5}, we can easily show that $\tilde{s}<\infty$. Selecting $r=\lambda^\tau$ for any $\tau\in(0,1)$, we have
  \begin{align}\label{supen}
  P(\sup_{1\leq k\leq n}|S_k|>4\lambda)
  &\leq4\exp(-\frac{\lambda^\tau}{2}\log(1+\frac{\lambda^{2-\tau}}{n\tilde{s}^2}))+4Cn\lambda^{\tau\frac{a(\gamma-1)}{a+\gamma}-\frac{(a+1)\gamma}{a+\gamma}}\notag\\
  &\leq 4\exp(-\frac{\lambda^\tau}{2}\log(2))+4Cn\lambda^{\tau\frac{a(\gamma-1)}{a+\gamma}-\frac{(a+1)\gamma}{a+\gamma}},
  \end{align}
as long as $\lambda\geq 1\vee(n\tilde{s}^2)^{1/(2-\tau)}$. It follows from (\ref{supen}) that
\begin{equation}\label{concentratioin}
  P(\frac{1}{n}|S_n|\geq x)\leq P(\max_{1\leq k\leq n}|S_k|\geq nx)\leq C\exp(-C(nx)^\tau)+Cn^{1-\beta}x^{-\beta},
\end{equation}
where $\beta=(a+1)\gamma/(a+\gamma)-\tau a(\gamma-1)/(a+\gamma)$. If $a>\gamma/(\gamma-2)$, we can also show that $\beta>2$ if we choose $\tau<[a(\gamma-2)-\gamma]/(a(\gamma-1))$.
\end{remark}

\begin{lemma}\label{lm3}
  Let $\hat{\sigma}_{i,j}^{(k)}$ be the $(i,j)$-th element of $\hat{\bSigma}_y(k)$, $\hat{\gamma}_{ij,kl}^{(h)}$ be the $(k,l)$-th element of $\wh \bSigma_{y,i,j}{(h)}$.
  Then
$$P(\max_{1\leq i,j\leq q}|\hat{\sigma}_{i,j}^{(k)}-{\sigma}_{i,j}^{(k)}|>x)\leq Cq^2\{\exp(-Cn^\tau x^\tau)+n^{1-\beta}x^{-\beta}+p\exp(-Cn^\tau x^{\tau/2})+pn^{1-\beta}x^{-\beta/2}\}$$
and
\begin{align*}
P(\max_{1\leq i,j \leq p}\max_{1\leq k,l\leq q}|\hat{\gamma}_{ij,kl}^{(h)}-{\gamma}_{ij,kl}^{(h)}|>x)\leq& Cp^2q^2\{\exp(-Cn^\tau x^\tau)+n^{1-\beta}x^{-\beta}\\
&+\exp(-Cn^\tau x^{\tau/2})+n^{1-\beta}x^{-\beta/2}\}.
\end{align*}
for any $x>0$ such that $nx\geq 1\vee(n\tilde{s}^2)^{1/(2-\tau)}$, and $1\leq k\leq k_0$, $|h|\leq m$.
\end{lemma}

{\bf Proof.} We only show the first one since the second one is similar. By Assumption \ref{a1}, Minkowski inequality,
\begin{equation}\label{minkov}
  \sup_{t}\sup_{i,j}E|\frac{1}{p}\sum_{l=1}^py_{l,i}^{t+k}y_{l,j}^t-E\frac{1}{p}\sum_{l=1}^py_{l,i}^{t+k}y_{l,j}^t|^{\gamma}
\leq C[\frac{1}{p}\sum_{l=1}^p(E|y_{l,i}^{t+k}y_{l,j}|^{\gamma})^{1/\gamma}]^{\gamma}\leq C.
\end{equation}
Then, by Markov inequality,
\begin{equation}\label{mean:py}
  \sup_t\sup_{i,j}P(|\frac{1}{p}\by_i^{t+k,\T}\by_j^t-E\frac{1}{p}\by_i^{t+k,\T}\by_j^t|\geq x)=O(x^{-\gamma}).
\end{equation}
By Lemma \ref{lm2} and (\ref{concentratioin}), it follows that
\begin{equation}\label{eq1}
  P(|\frac{1}{n}\sum_{t=1}^{n-k}\{\frac{1}{p}\by_i^{t+k,\T}\by_j^t-E\frac{1}{p}\by_i^{t+k,\T}\by_j^t\}|\geq x)\leq C\exp(-C(nx)^\tau)+Cn^{1-\beta}x^{-\beta},
\end{equation}
where $\beta$ and $\tau$ are defined in Remark \ref{rm:lemma}.
Now we consider the second term of (\ref{sig:dif}),
\begin{align*}
  \frac{1}{np}\sum_{t=1}^{n-k}\by_i^{t+k,\T}\bar{\by}_j & =\frac{1}{p}\sum_{l=1}^p(\frac{1}{n}\sum_{t=1}^{n-k}y_{l,i}^{t+k})(\frac{1}{n}\sum_{t=1}^{n}y_{l,j}^t).
\end{align*}
Then,
\begin{align}\label{I2:2}
  P(|\frac{1}{p}\sum_{l=1}^p(\frac{1}{n}\sum_{t=1}^{n-k}y_{l,i}^{t+k})(\frac{1}{n}\sum_{t=1}^{n}y_{l,j}^t)|\geq x) & \leq 2\max_{1\leq i\leq q}\sum_{l=1}^pP(|\frac{1}{n}\sum_{t=1}^{n-k}y_{l,i}^{t+k}|\geq x^{1/2})\notag\\
  &\leq Cp\max_{1\leq i\leq q}\max_{1\leq l\leq p}P(|\frac{1}{n}\sum_{t=1}^{n-k}y_{l,i}^{t+k}|\geq x^{1/2}).
\end{align}
By Assumptions \ref{a1} and \ref{a5}, the conclusion of Lemma \ref{lm2} holds uniformly for all $i$ and $j$ in (\ref{I2:2}), thus, there exists a constant C such that
\begin{equation}\label{I2:re}
  P(|I_2|\geq x)\leq Cp\exp(-Cn^\tau x^{\tau/2})+Cpn^{1-\beta}x^{-\beta/2},
\end{equation}
and we can show it for $I_3$ and $I_4$ in a similar way. The first one of Lemma 3 follows from (\ref{eq1}) and (\ref{I2:re}).

For the proof of the second inequality of Lemma \ref{lm3}, since there is only one term for each time $t$ in $I_2$ instead of $p$ terms in (\ref{I2:2}), the upper bound in the bracket of the second one of Lemma 3 is slightly different from that in the first one. The argument is the same and we omit the details here. This completes the proof. $\Box$

The following three lemmas are similar to Lemmas 6-8 in Chang et al. (2017) and the proofs can be done in a similar way, we therefore omit the details here.
\begin{lemma}\label{lm4}
  Under Assumptions 1, and 3-5, we have
  $$\max_{1\leq j\leq q}\sum_{i=1}^q|\sigma_{i,j}^{(k)}|^\iota\leq C(2S_{\max}+1)s_1s_2,\,\,\max_{1\leq i\leq q}\sum_{j=1}^q|\sigma_{i,j}^{(k)}|^\iota\leq C(2S_{\max}+1)s_1s_2$$
  $$\max_{1\leq i,j\leq p}\max_{1\leq l\leq q}\sum_{k=1}^q|\gamma_{ij,kl}^{(h)}|^\iota\leq C(2S_{\max}+1)s_1s_2, \max_{1\leq i,j\leq p}\max_{1\leq k\leq q}\sum_{l=1}^q|\gamma_{ij,kl}^{(h)}|^\iota\leq C(2S_{\max}+1)s_1s_2,$$
  where $S_{\max}=\max_{1\leq j\leq q_1} l_j$, $\iota$ is defined in Assumption \ref{a3}.

\end{lemma}

\begin{lemma}\label{lm5}
  If Assumptions 1 and 3-5 hold,for any $k\leq k_0$, we have
  $$\|T_u(\hat{\bSigma}_y(k))-\bSigma_y(k)\|_2=O_p\{(q^{2/\beta}n^{-(\beta-1)/\beta})^{(1-\iota)/2}\delta\}$$
  and
  $$\max_{1\leq i,j\leq p}\|T_v(\hat{\bSigma}_{y,i,j}(k))-\bSigma_{y,i,j}(k)\|_2=O_p\{((pq)^{2/\beta}n^{-(\beta-1)/\beta})^{(1-\iota)/2}\delta\} ,$$
  provided that $pq=o(n^{(\beta-1)/2})$, and $\delta$ is defined in (\ref{max:group}).
\end{lemma}
\begin{lemma}\label{lm6}
  If assumptions 1 and 3-5 hold, as $n\rightarrow\infty$ and $pq=o(n^{(\beta-1)/2})$, we have
  $$\|\hat{\bS}-\bW_y\|_2=O_p(\kappa\vartheta_n^{1-\iota}\delta+\vartheta_n^{2(1-\iota)}\delta^2),$$
  where $\delta$ and $\kappa$ are difined in (\ref{max:group}).
\end{lemma}

{\bf Proof of Theorem 2.} (i). By (\ref{delta:dif}),
\begin{align}\label{gammadif}
  |\hat{\bGamma}_{i,j}(h)-\bGamma_{i,j}(h)|_\infty & \leq \max_{1\leq k,l\leq q}|\hat{\delta}_{ij,kl}^{(h)}-\delta_{ij,kl}^{(h)}|\notag\\
  &\leq C\|\hat{\bv}_i-\bv_i\|_2\max_{1\leq k,l\leq q}\|\bSigma_{i,j}^{(h)}\|_2+C\|T_v(\hat{\bSigma}_{i,j}^{(h)})-{\bSigma}_{i,j}^{(h)}\|_2.
\end{align}
By (\ref{eigenvec}) and Lemma 5,
\begin{align}\label{gamma:df}
  \max_{1\leq i,j\leq p}|\hat{\bGamma}_{i,j}(h)-\bGamma_{i,j}(h)|_\infty & =O_p(d_{1n}+d_{2n}).
\end{align}
Let $\delta_n=d_{1n}+d_{2n}$, by a similar argument as the proof of Theorem 1(i),
we can show that $P(\hat{E}=E)\rightarrow1$.

(ii). the proof is similar to Theorem 1(ii). This completes the proof.

\begin{lemma}\label{lm7}
  Let $\gamma_1^{-1}=2r_1^{-1}+r_2^{-1}$ and $\gamma_2^{-1}=r_1^{-1}+r_2^{-1}$, then
  \begin{align*}
    P(\max_{1\leq i,j\leq q}|\hat{\sigma}_{i,j}^{(k)}-\hat{\sigma}_{i,j}^{(k)}|\geq s)\leq& Cq^2n\exp(-Cs^{\gamma_1}n^{\gamma_1})+Cq^2pn\exp(-Cs^{\gamma_2/2}n^{\gamma_2})\\
    &+Cq^2\exp(-Cs^2n)+Cq^2p\exp(-Csn)
  \end{align*}
  and
    \begin{align*}
    P(\max_{1\leq i,j\leq p}\max_{1\leq k,l\leq q}|\hat{\gamma}_{ij,kl}^{(h)}-{\gamma}_{ij,kl}^{(h)}|\geq s)\leq& Cp^2q^2\exp(-Cs^{\gamma_1}n^{\gamma_1})+Cp^2q^2n\exp(-Cs^{\gamma_2/2}n^{\gamma_2})\\
    &+Cp^2q^2\exp(-Cs^2n)+Cp^2q^2\exp(-Csn),
  \end{align*}
  for any $s>0$ such that $ns\rightarrow\infty$.
\end{lemma}

{\bf Proof.} We only prove the first one since the second is similar. It is sufficient to bound $P(|\hat{\sigma}_{i,j}^{(k)}-\hat{\sigma}_{i,j}^{(k)}|\geq s)$ for each $1\leq i,j\leq q$. We consider the first term of (\ref{sig:dif}), by Assumption 6,
\begin{align}\label{mean:poly}
  P(|\frac{1}{p}\by_i^{t+k,\T}\by_j^t-E\frac{1}{p}\by_i^{t+k,\T}\by_j^t|>s) & \leq P(|\frac{1}{p}\by_i^{t+k,\T}\by_j^t|>s/2)\notag\\
  &=P(|\frac{1}{p}\sum_{l=1}^py_{l,i}^{t+k}y_{l,j}^t|>s/2)\notag\\
  &\leq P(\frac{1}{p}\sum_{l=1}^py_{l,i}^{t+k,2}>s/2)+P(\frac{1}{p}\sum_{l=1}^py_{l,j}^{t,2}>s/2)\notag\\
  &\leq C\exp(-Cs^{r_1/2})
\end{align}
for any $s>0$. By the theorem of \cite{merlevede2011}, for any $s>0$ and $ns\rightarrow\infty$,
$$P(|I_1|>s)\leq Cn\exp(-Cs^{\gamma_1}n^{\gamma_1})+C\exp(-Cs^2n).$$
By a similar argument as (\ref{I2:2}) and Assumption 6, we can show the second term of (\ref{sig:dif}),
\begin{equation}\label{I2}
  P(|I_2|>s)\leq Cp\exp(-Cs^{\gamma_2/2}n^{\gamma_2})+Cp\exp(-Csn).
\end{equation}
By a similar argument, we can show that $I_3$ and $I_4$ also satisfy (\ref{I2}) and $I_5$ is a negligible term. By Bonferroni inequality, this proves the first one of Lemma 7. This completes the proof. $\Box$

\begin{lemma}\label{lm8}
  $$\|\hat{\bS}-\bW_y\|=O_p\{\kappa(n^{-1}\log q)^{(1-\iota)/2}\delta+(n^{-1}\log q)^{1-\iota}\delta^2\},$$
\end{lemma}

{\bf Proof.} By a similar argument as Lemma 7 in Chang et al. (2017), we can show that
$$\|T_u(\hat{\bSigma}_y(k))-\bSigma_y(k)\|_2=O_p\{(n^{-1}\log q)^{(1-\iota)/2}\delta\}$$
provided that $\log (pq)=o(n^{\gamma_1/(2-\gamma_1)})$. By a similar argument as Lemma 6, we can obtain the result. This completes the proof. $\Box$

{\bf Proof of Theorem 3.} It is the same as the proof of Theorem 2. $\Box$




\end{document}